\documentclass[12pt]{article}

\usepackage{newtxtext,newtxmath}
\usepackage{graphicx}

\usepackage[letterpaper,margin=1in]{geometry}

\renewenvironment{abstract}
	{\quotation}
	{\endquotation}

\date{}

\makeatletter
\renewcommand{\fnum@figure}{\textbf{Figure \thefigure}}
\renewcommand{\fnum@table}{\textbf{Table \thetable}}
\makeatother

\usepackage{scicite}

\usepackage{url}

\def\scititle{
	Electric polarization induced by magnons and magnon Nernst effects
}
\title{\bfseries \boldmath \scititle}

\author{
	D. Quang To$^{1\ast}$,
	Federico Garcia-Gaitan$^{2}$,
	Yafei Ren$^{2}$,
    Joshua M. O. Zide$^{1}$,\and
    M. Benjamin Jungfleisch$^{2}$,
    John Q. Xiao$^{2}$,
    Branislav K. Nikoli\'{c}$^{2}$, \and
    Garnett W. Bryant$^{3,4}$,
    Matthew F. Doty$^{1\ast}$ \and
	\small$^{1}$Department of Materials Science and Engineering, University of Delaware, Newark, Delaware 19716, USA. \and
	\small$^{2}$Department of Physics and Astronomy, University of Delaware, Newark, Delaware 19716, USA.\and
    \small$^{3}$Nanoscale Device Characterization Division, Joint Quantum Institute, \and \small National Institute of Standards and Technology, Gaithersburg, Maryland 20899-8423, USA.\and
     \small$^{4}$University of Maryland, College Park, Maryland 20742, USA.\and
	\small$^\ast$Corresponding author. Email: quangto@udel.edu; doty@udel.edu\and
}

\begin{document} 

% Insert the title and author list
\maketitle

% Abstract, in bold
% There are strict length limits, and not all formats have abstracts.
% Consult the journal instructions to authors for details.
% Do not cite any references in the abstract.
\begin{abstract} \bfseries \boldmath
Magnons offer a promising path toward energy-efficient information transmission and the development of next-generation classical and quantum computing technologies. However, methods to efficiently excite, manipulate, and detect magnons remain a critical need. Here, we show that magnons, despite their charge-neutrality, can induce electric polarization as a result of both their spin and orbital moments. We demonstrate this by calculating the electric polarization induced by magnons in two-dimensional (2D) honeycomb antiferromagnets. The electric polarization becomes finite when the Dzyaloshinskii-Moriya Interaction (DMI) is present and its magnitude can be increased by symmetries of the system. We illustrate this by computing and comparing the electric polarizations induced by the magnon Nernst effects in 2D materials with N\'eel and Zigzag ordering. Our findings show that in the Zigzag order, where the effect is dominated by the magnon orbital moment, the induced electric polarization is approximately three orders of magnitude greater than in the N\'eel phase. These findings reveal that electric fields could enable both detection and manipulation of magnons under certain conditions by leveraging their spin and orbital angular moment. They also suggest that the discovery or engineering of materials with substantial magnon orbital moments could lead to more practical use of magnons for future computing and information transmission device applications.
\end{abstract}

% The first paragraph of any Science paper does NOT have a heading
% Nor is it indented
\noindent
\section{Introduction}
Magnons are bosonic quasi-particles arising from collective and charge-neutral excitations of localized spins. These quanta obey the Bose-Einstein distribution function at finite temperatures and have zero chemical potential in equilibrium. Because magnons can carry spin information without the need for moving electric charge, they enable low-power data transfer and energy-efficient computing technologies \cite{Kajiwara2010,Chumak2015,Lebrun2018,Xing2019,Pirro2021,flebus2024}. Magnons can hybridize with a variety of other quantum states, including photons \cite{Bhoi2019,Golovchanskiy2021}, electrons \cite{Carpene2008,Rohling2018,Kristian2021}, phonons \cite{Mai2021,Manley2024,To2024}, plasmons \cite{Costa2023,Dyrdal2023}, and excitons \cite{Bae2022,Wang2023,Diederich2023}. For these and other reasons, magnons are also being explored as elements of quantum information processing systems \cite{Tabuchi2015,Dany2020,Wolski2020,Yuan2022,Xu2023,Fukami2024,Bejanaro2024,Dols2024}. 

Realizing the opportunities associated with magnons requires tools for their efficient excitation, manipulation, and detection. The field of spin caloritronics covers a wide range of relevant magnon effects including the magnon Seebeck effect (SE) \cite{Xiao2010}, thermal Hall effect (THE) \cite{Onose2010,Katsura2010, Matsumoto2011,Hirschberger2015,Murakami2017,Zhang2019,Park2020,Neumann2022,Jostein2023}, and spin Nernst effect (SNE) \cite{Cheng2016,Zyuzin2016,Li2020,Bazazzadeh2021,Bazazzadeh2021b,Zhang2022,To2023b}. These effects all involve the generation of longitudinal and transverse heat currents or spin currents, mediated by magnons, in response to a longitudinal applied temperature gradient. However, these effects are difficult to exploit for information processing because it is difficult to apply thermal gradients locally or to modulate them rapidly. 

Electronic systems are easily manipulated by local electric fields and it has been recently recognized that the orbital angular moment (OAM) of electrons can generate a number of previously overlooked transport effects~\cite{Go2018,Go2020,Go2020b,Go2023,Choi2023,Lyalin2023,Seifert2023,Jungfleisch2023}. Magnon OAM is expected to play a similarly important role in the form of a magnon Orbital Nernst Effect (ONE).  While a few studies have recently explored magnon OAM within different crystalline lattices,~\cite{Jia2019,Neumann2020,Fishman2022,Fishman2022b,Fishman2023,Fishman2023b, Go2024intrinsic} these studies have focused on theoretical measures of magnon OAM such as a magnon orbital magnetization~\cite{Neumann2020} or the integral of magnon OAM over the orientation of the magnon wave vector \cite{Fishman2022,Fishman2023}. To relate magnon SNE and ONE, which we collectively refer to as magnon Nernst effects (MNEs), to experimentally measurable quantities,~\cite{Zhang2022} we need a more comprehensive theoretical framework.

In this work, we develop a quantum mechanical formalism that establishes the connection between magnon transport effects (e.g. MNEs) and electric polarization. Specifically, we show that magnons, despite their charge-neutrality, can induce electric polarization in systems exhibiting the Dzyaloshinskii-Moriya Interaction. The formalism reveals that the magnon-induced electric polarization occurs as a result of both magnon spin and orbital moments. Furthermore, the magnitude of this polarization can be significantly enhanced by the system's symmetry properties. We illustrate this by computing and comparing the electric polarizations induced by the magnon Nernst effect in 2D materials with N\'eel and Zigzag ordering. Remarkably, we find that the Zigzag order can induce an electric polarization approximately three orders of magnitude greater than that of the Néel order.

The paper is structured as follows. In Sect.~\ref{MNEs} we describe the origin of magnon Nernst effects in magnetic systems and summarize what material properties are important for maximizing the ONE. In Sect.~\ref{formalism} we introduce the quantum mechanical framework that reveals, and allows us to calculate, the electric polarization induced by the motion of magnon wave-packets. In Sect.~\ref{realization} and \ref{DMI} we demonstrate the value of this formalism by applying it to 2D honeycomb antiferromagnets with Néel and Zigzag order. Through both our formalism and symmetry considerations, we show that spin-orbit coupling (SOC), specifically in the form of Dzyaloshinskii-Moriya Interaction, or SOC-like interactions that couple spins to the lattice (e.g.~ magnon-phonon coupling), is essential for generating a finite electric polarization at the sample edges due to magnon Nernst effects. We note that this conclusion challenges recent predictions that an electric polarization could be induced by magnon ONE in a honeycomb lattice N\'eel ordered 2D AFM even in the absence of DMI~\cite{Go2024intrinsic}.  

While the magnon SNE has been extensively studied in the literature \cite{Cheng2016,Zyuzin2016,Li2020,Bazazzadeh2021,Bazazzadeh2021b,Zhang2022,To2023b}, the magnon ONE was introduced relatively recently using a semiclassical theory \cite{Zhang2019b}. However, reference \cite{Zhang2019b} does not include a clear delineation of the conditions under which semiclassical theory is applicable and, more importantly, does not address the resulting measurable quantities that will be induced by the ONE. The formalism presented in this paper provides a firm quantum mechanical foundation for studying the magnon ONE and its observable manifestation in the form of electric polarization. Taken together, the work reported here suggests a path toward exploiting the orbital degree of freedom of magnons in addition to their spin moment. Specifically, the results reveal that the measurement of induced electric polarization could be used both for the experimental study of magnon OAM and MNEs and as means of detecting the presence of magnon transport within devices. They also suggest that the discovery or engineering of materials with large magnon orbital moments could increase the interactions with electric dipoles, creating more opportunities for electrical read out or control of magnons in future computing and information transmission devices. Finally, the results suggest that it may also be possible to use electromagnetic waves to both manipulate and read out magnons through electric dipoles interaction. 

\section{Origin of Magnon Nernst Effects }\label{MNEs}
\begin{figure}
\centering
    \includegraphics[width= 1\textwidth]{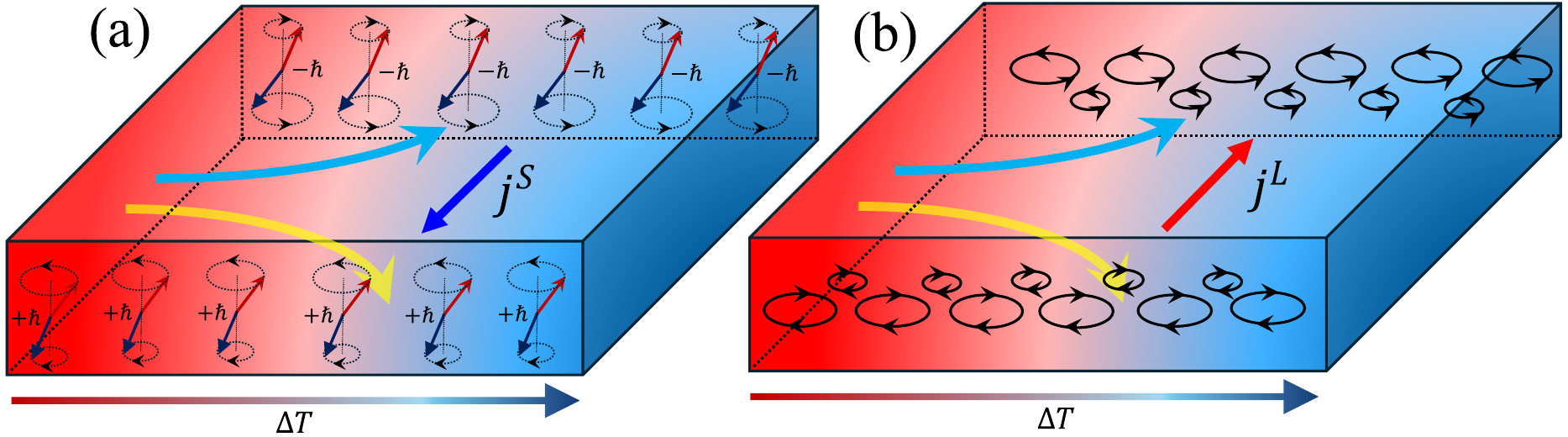}
 \caption{Schematic depiction of the magnon Spin Nernst Effect (a) and magnon Orbital Nernst Effect (b) resulting from an applied temperature gradient $\Delta$T.}
  \label{FigMNEs}
\end{figure}

Magnons are inherently charge-neutral entities and can be generated by the introduction of a temperature gradient that induces a transverse magnon thermal current via the THE and a magnon spin current via the SNE, as schematically depicted in Fig.~\ref{FigMNEs}(a). In this section we show that such thermal gradients can also generate a magnon orbital current (MOC). We call this effect the magnon Orbital Nernst Effect. We then discuss the importance of magnon OAM and Berry curvature to the magnon ONE, which leads to predictions about what classes of materials are most likely to show a large magnon ONE.

In the semiclassical picture, the motion of a magnon wavepacket in the n$^{th}$ band with position $\boldsymbol{r}^{n}_{c}$ and wavevector $\boldsymbol{k}$ subject to an applied temperature gradient is given by \cite{Matsumoto2011b,Xiao2010b}
\begin{align} \label{EqWavePacketMotion}
 \boldsymbol{\dot{r}}^{n}_{c} = \frac{1}{\hbar}\frac{\partial E_{n,\boldsymbol{k}}}{\partial \boldsymbol{k}} - \boldsymbol{\dot{k}} \times \boldsymbol{\Omega}^{n}(\boldsymbol{k})
\end{align}
with $~\hbar \boldsymbol{\dot{k}} = -\nabla U(\boldsymbol{r})$ where $U(\boldsymbol{r})$ is a slowly varying potential experienced by the magnons and $\boldsymbol{\Omega}^{n}\left(\boldsymbol{k} \right)=\sum_{m\neq n }\boldsymbol{\Omega}^{nm}\left(\boldsymbol{k} \right)$ is the Berry curvature with
\begin{align} \label{EqBerry}
   \boldsymbol{\Omega}^{nm}\left(\boldsymbol{k} \right) = i\hbar^{2}\sigma_{3}^{mm}\sigma_{3}^{nn}\frac{\langle n(\boldsymbol{k})\vert \hat{\boldsymbol{v}} \vert m(\boldsymbol{k}) \rangle  \times \langle m(\boldsymbol{k})\vert \hat{\boldsymbol{v}} \vert n(\boldsymbol{k}) \rangle}{\left( \sigma_{3}^{nn}E_{n,\boldsymbol{k}}  - \sigma_{3}^{mm}E_{m,\boldsymbol{k}} \right)^{2}}
\end{align}
the projection of the Berry curvature of the n$^{th}$ band on the m$^{th}$ band. Here, $E_{n,\boldsymbol{k}}$ and $\vert n (\boldsymbol{k})\rangle$ are, respectively, the eigenvectors and eigenvalues of the bosonic Bogoliubov-de Gennes (BdG) Hamiltonian;  $\hat{\boldsymbol{v}}=(\hat{v}_x,\hat{v}_y,\hat{v}_z)$ denotes the velocity vector operator;  and the $\boldsymbol{\sigma}_{3}$ matrix is given by $\boldsymbol{\sigma}_{3} = \begin{pmatrix}
    \boldsymbol{1}_{N \times N} & 0 \\
    0 & -\boldsymbol{1}_{N \times N}
    \end{pmatrix},$
where $\boldsymbol{1}_{N \times N}$ is  a $N \times N$ identity matrix and $\sigma_{3}^{nn}=\langle n(\boldsymbol{k})\vert \boldsymbol{\sigma}_{3}\vert n(\boldsymbol{k})\rangle$ is the n$^{th}$ diagonal element of $\boldsymbol{\sigma}_{3}$.

The key point of Eq.~\ref{EqWavePacketMotion} at this stage of the discussion is the dependence on the Berry Curvature $\boldsymbol{\Omega}^{n}(\boldsymbol{k})$. Berry curvature plays a pivotal role in generating both the magnon THE and the non-trivial topology of magnon bands \cite{Murakami2017, Matsumoto2011b, Go2019,Zhang2020, Diaz2019,Vinas2023}. The non-trivial topology originates from the motion of magnon wavepackets along the edge of the system \cite{Matsumoto2011b}. Magnon wavepackets also exhibit self-rotation, referred to as magnon OAM, as elucidated by Matsumoto using linear response theory \cite{Matsumoto2011b}. As a result, in systems that have both a) non-trivial topological thermal transport of magnons and b) finite self-rotation of magnon wavepackets, the thermal Hall current induced by the magnon bands can also facilitate the transport of magnon OAM. This, in turn, gives rise to a MOC in response to an applied temperature gradient, i.e.~the magnon ONE. The consequence of a finite ONE is the accumulation of the OAM of magnons at the boundaries of the system, as depicted in Fig.~\ref{FigMNEs}(b). In this simple picture, the magnon orbital angular moment population accumulated at the edge of the system due to the ONE is directly proportional to
\begin{align}\label{eq:semiclassicaeta}
    \lambda_{\mu\nu}^{L^{\alpha}} = \frac{k_{B}}{2 \hbar V} &\sum_{k}\sum_{n=1}^{N} \sum_{m\neq n}\left\lbrace \left[\sigma_{3}^{nn}L_{nn}^{\alpha}\left(\boldsymbol{k} \right)+\sigma_{3}^{mm}L_{mm}^{\alpha}\left(\boldsymbol{k} \right) \right] \Omega^{nm}_{\mu\nu}\left(\boldsymbol{k} \right) F(\rho_{n,\boldsymbol{k}} ) \right\rbrace.
\end{align}
where $\lambda_{\mu\nu}^{L^{\alpha}}$ is the Orbital Nernst conductivity (ONC); $V$ is the volume of the system; $\rho_{n,\boldsymbol{k}} = \left(e^{E_{n,\boldsymbol{k}}/k_{B}T}-1\right)^{-1}$ is the Bose-Einstein distribution function; $F\left( \rho_{n,\boldsymbol{k}}\right) = \left(1+ \rho_{n,\boldsymbol{k}} \right)ln\left(1+ \rho_{n,\boldsymbol{k}} \right) -\rho_{n,\boldsymbol{k}}ln\left(\rho_{n,\boldsymbol{k}} \right)$; $\Omega^{nm}_{\mu\nu}\left(\boldsymbol{k} \right)$ is the $\mu\nu$-component of the projected Berry curvature [Eq. \eqref{EqBerry}] and $L^{\alpha}_{nn}$ is the $\alpha$-component of the intra-band magnon OAM in the Bloch wave of the n$^{th}$ band given by
\begin{align} \label{OBM}
    \boldsymbol{L}_{nn}\left(\boldsymbol{k} \right)= -\frac{i\hbar}{2}\sum_{p\neq n}\sigma_{3}^{pp}\frac{ \left\langle n\left(\boldsymbol{k} \right)\left\vert \hat{\boldsymbol{v}}\right\vert p \left( \boldsymbol{k} \right)\right\rangle \times \left\langle p\left(\boldsymbol{k} \right)\left\vert \hat{\boldsymbol{v}}\right\vert n \left( \boldsymbol{k} \right)\right\rangle }{ \sigma_{3}^{nn}E_{n,\boldsymbol{k}}  - \sigma_{3}^{pp}E_{p,\boldsymbol{k}} }.  
\end{align} 
For the detailed derivation of these formulas, please refer to the SM~\footnotemark[2]. 

Several physical consequences related to the ONE can be gleaned from Eqs.~\eqref{eq:semiclassicaeta} and \eqref{OBM}. First, the representation of the magnon intra-band OAM in Bloch states given by Eq.~\eqref{OBM} has deep connections to the magnon Berry curvature given in Eq.~\eqref{EqBerry}. Specifically, both are subject to the constraints of  time reversal symmetry (TRS) and parity-time symmetry (PTS). TRS requires that $\boldsymbol{L}_{nn}\left( \boldsymbol{k} \right) = -\boldsymbol{L}_{nn}\left( - \boldsymbol{k} \right)$. PTS requires the intra-band OAM to be zero, as expressed by $\boldsymbol{L}_{nn}\left( \boldsymbol{k} \right) = 0$ \cite{To2023b}. Thus, non-zero intra-band OAM can only be achieved when PTS is broken. Second, broken TRS is crucial for observing a finite magnon THE. However, Eq.~\eqref{eq:semiclassicaeta} tells us that the ONE can exist even without breaking TRS. The term $\Omega^{L^{\alpha},nm}_{\mu\nu}\left(\boldsymbol{k} \right)=\frac{1}{4}\left[\sigma_{3}^{nn}L_{nn}^{\alpha}\left(\boldsymbol{k} \right)+\sigma_{3}^{mm}L_{mm}^{\alpha}\left(\boldsymbol{k} \right) \right]\Omega^{nm}_{\mu\nu}\left(\boldsymbol{k} \right)$ has even parity with respect to the wavevector because both intra-band OAM and Berry curvature have the same parity with respect to the vector $\boldsymbol{k}$. As a result, $\lambda_{\mu\nu}^{L^{\alpha}}$ can be finite even in the presence of TRS. This observation, in combination with the first point above, suggests that the best strategy for experimentally observing the ONE is to identify material systems exhibiting finite Berry curvature (i.e. breaking PTS) and thus having nonzero intra-band OAM. Collinear 2D honeycomb AFMs with both N\'eel and Zigzag phases are promising in this context. Third, magnon intra-band OAM can have much larger magnitude than magnon spin moment, particularly in the vicinity of the anti-crossing point between two distinct bands. This is similar to the behavior of Berry curvature. Consequently, the ONE should be larger in magnonic crystals in which there are more band crossings for magnons and/or more coupling with other quasiparticles such as phonons \cite{Liu2021,To2023b}. This suggests that the ONE will be more pronounced in systems characterized by strong band hybridization and a narrow energy gap between the bands because the magnitude of both intra-band OAM and Berry curvature are significantly amplified under these conditions \cite{To2023b}.

We note that our discussion here illustrates the ONE using a simple picture in which the topological magnon thermal Hall current gives rise to the magnon ONE through magnon intra-band OAM. However, both magnon intra-band and inter-band OAM can contribute. To treat intra-band and inter-band OAM on an equal footing, we employ linear response theory to derive the linear thermal response mechanism governing the flow of magnon OAM. Please refer to the SM~\footnotemark[2] for the detailed derivation and resulting formulas.

\section{Magnon-induced electric polarization: the role of spin and orbital degrees of freedom}\label{formalism}
Experimental detection of the ONE will be possible when the accumulation of magnon orbital angular moment can be transformed into a magnetization or electric polarization. This is analogous to the way the OHE for electrons can be measured via spin polarizations that are induced by the OHE via spin orbit coupling \cite{Ding2020,Li2023}. Utilizing perturbation theory, Neumann et al.~\cite{Neumann2020} revealed the dependence of the orbital magnetic moment of magnons on the external magnetic field: $\boldsymbol{\mu}_{n,\boldsymbol{k}}^{O} = -\sum_{m=1}^{N} \sum_{\alpha = x,y,z} \frac{\partial E_{n,\boldsymbol{k}}}{\partial \hat{\boldsymbol{\alpha}}_{m}}\cdot \frac{\partial \hat{\boldsymbol{\alpha}}_{m}}{\partial \boldsymbol{B}}$ where $\boldsymbol{B}$ is the applied magnetic field, $\boldsymbol{\alpha}$ is the local spin coordinate system and $E_{n,\boldsymbol{k}}$ is the energy of the magnon in nth state. This dependence is a result of spin orbit coupling or SOC-like interactions that couple spins to the lattice (e.g.~DMI or magnon-phonon coupling). Consequently, the accumulation of the orbital moment of magnons at the boundaries due to the ONE can manifest as a measurable magnetization in systems featuring DMI, magnon-phonon coupling, or similar interactions. 

MNEs may also give rise to a local Electric Polarization (EP) at the boundaries of the system~\cite{Neumann2023}. Employing perturbation theory, we derive a formula to calculate the y-component of the net electric polarization induced by magnon motion in magnetic materials: $P_{y} = \frac{1}{\beta}\int \tilde{P}_{y} d\beta$, where $\beta  = \frac{1}{k_{B}T}$. The full derivation of this formula can be found in the SM~\footnotemark[2]. The resulting formula has two parts: $\tilde{P}_{y} = \tilde{P}_{y}^{S} + \tilde{P}_{y}^{O}$ where

\begin{align} \label{EPG1}
    \tilde{P}_{y}^{S}= -\frac{ g\mu_{B}}{\hbar W_{y}c^{2}}\sum_{n,\boldsymbol{k}}\left[\Omega_{xy}^{S^{z},n}\left( \boldsymbol{k} \right) - \Omega_{zy}^{S^{x},n}\left( \boldsymbol{k} \right) \right]\left[ \boldsymbol{\sigma}_{3}E_{\boldsymbol{k}} \right]_{nn}\rho_{n,\boldsymbol{k}}
\end{align}
and
\begin{align}\label{EPG2}
    \tilde{P}_{y}^{O}=&-\frac{4g\mu_{B}}{W_{y}c^{2}}\sum_{n,\boldsymbol{k}}\left( \rho_{n,\boldsymbol{k}}+\left[ \boldsymbol{\sigma}_{3}E_{\boldsymbol{k}} \right]_{nn}\rho^{\prime}_{n,\boldsymbol{k}}\right)Im\left[\left\langle \partial_{k_{y}}u_{n,\boldsymbol{k}}\left\vert \hat{j}_{x}^{S^{z}}\right\vert u_{n,\boldsymbol{k}}\right\rangle  \left\langle u_{n,\boldsymbol{k}} \right\vert \boldsymbol{\sigma}_{3}\left\vert u_{n,\boldsymbol{k}}\right\rangle +  \right.\notag\\
    &+\left.\left\langle u_{n,\boldsymbol{k}}\left\vert \left(\partial_{k_{y}}\hat{v}_{x} \right)\boldsymbol{\sigma}_{3} \hat{S}^{z}\right\vert u_{n,\boldsymbol{k}}\right\rangle  \left\langle u_{n,\boldsymbol{k}} \right\vert \boldsymbol{\sigma}_{3}\left\vert u_{n,\boldsymbol{k}}\right\rangle + \left\langle u_{n,\boldsymbol{k}}\left\vert \hat{j}_{x}^{S^{z}} \right\vert u_{n,\boldsymbol{k}}\right\rangle  \left\langle u_{n,\boldsymbol{k}} \right\vert \boldsymbol{\sigma}_{3}\left\vert \partial_{k_{y}}u_{n,\boldsymbol{k}}\right\rangle  \right] - \left(x \leftrightarrow z \right)  
\end{align}
with $\hat{j}_{x}^{S^{z}}=\frac{\hat{v}_{x}\boldsymbol{\sigma}_{3} \hat{S}^{z} + \hat{S}^{z}\boldsymbol{\sigma}_{3} \hat{v}_{x}}{4}$ being the spin current operator. 

Equations \eqref{EPG1} and \eqref{EPG2} are important because they provide a microscopic framework for computing the electric polarization generated by magnon transport in an arbitrary magnetic system, even in a case where the spin moment of magnons is not conserved (e.g.~due to magnon-phonon interactions). In principle, these quantities can be numerically computed accounting for both inter- and intra-band contributions to magnon orbital angular moment using relations \eqref{numcalea} and \eqref{numcaleb} reported in the SM~\footnotemark[2]. Analyzing equations \eqref{EPG1} and \eqref{EPG2}, we find that the two parts of $\tilde{P}_{y}$ describe two distinct contributions to the electric polarization induced by the motion of the magnon wave packets: 
\begin{enumerate}
    \item $\tilde{P}_{y}^{S}$ [Eq.~\eqref{EPG1}] describes the electric polarization that arises from the magnon spin current. Specifically, the presence of the spin-Berry curvatures $\Omega_{xy}^{S^{z},n}(\boldsymbol{k})$ and $\Omega_{zy}^{S^{x},n}(\boldsymbol{k})$ result in a spin current in the y direction via the accumulation of spin angular momentum due to, for example, the magnon spin Nernst current carried by magnons under a temperature gradient along the x-direction~\cite{To2023b}. 
    \item $\tilde{P}_{y}^{O}$ [Eq.~\eqref{EPG2}] relates to the orbital angular moment $L^{z}$ of the magnons.  
\end{enumerate}

\noindent In other words, our formula reveals that the net electric polarization induced by magnon motion in magnetic materials has distinct contributions from the magnon spin current [$P_{y}^{S}$, Eq.~\eqref{EPG1}] and the magnon orbital angular moment [$P_{y}^{O}$, Eq.~\eqref{EPG2}].

%MD: everything up until this point is general to both 2D and 3D materials

\section{Magnon Nernst effects and electric polarizations in 2D AFMs}\label{realization}
\begin{figure}[h]
\centering
    \includegraphics[width= 0.7\textwidth]{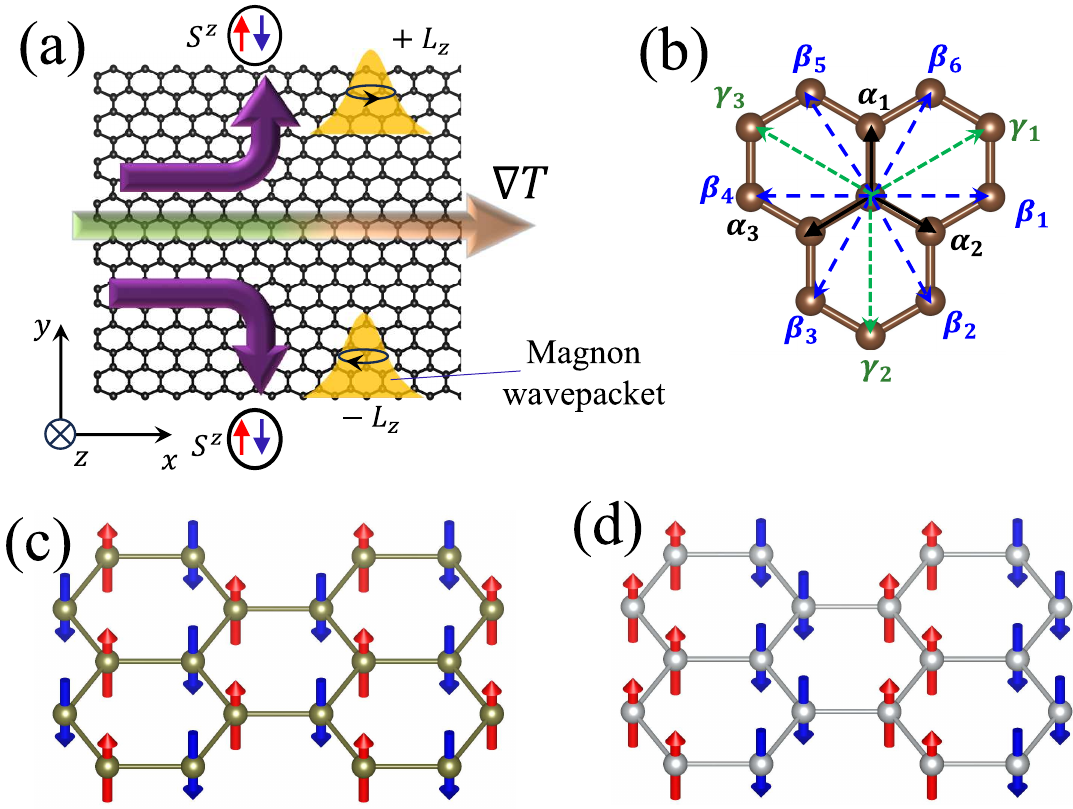}
 \caption{(a) Schematic view of the magnon ONE in a 2D AFM where transverse flow of magnons carrying opposite out-of-the plane orbital moment is induced by temperature gradient $\nabla T$ along the longitudinal direction. (b) The nearest, second nearest and third-nearest neighbor bonds in a honeycomb lattice are denoted by $\alpha_{i}, ~\beta_{i}$ and $\gamma_{i}$, respectively. (c) N\'eel and (d) Zigzag ordering of honeycomb spin lattices.}
  \label{FIG1}
\end{figure}

We now apply our electric polarization formalism to predict the consequences of MNEs in MnPS$_3$ and NiPSe$_3$, which are 2D honeycomb AFMs possessing N\'eel and Zigzag order as shown in Fig.~\ref{FIG1}(c) and (d), respectively. There are two reasons to present these example cases. First, as we show in this section, they allow us to more clearly see and understand the spin and orbital contributions to the resulting electric polarization. Second, as we show in Sect.~\ref{DMI}, they allow us to analyze the importance of symmetry breaking to the emergence of electric polarization. 

Both MnPS$_3$ and NiPSe$_3$ have spins that lie on A and B sublattices with $\boldsymbol{S}_{A}=-\boldsymbol{S}_{B}=S \hat{\boldsymbol{z}}$; the z-axis is out of the plane of the 2D material, as shown in Fig.~\ref{FIG1}(a). The fundamental spin Hamiltonian for this type of system can be expressed as follows~ \cite{Zyuzin2016,Bazazzadeh2021,To2023b}: 
\begin{align}\label{eq:hamiltonian}
    H = \sum_{i,j} J_{ij}\boldsymbol{S}_{i}\cdot \boldsymbol{S}_{j} + \Delta\sum_{i}\left( S_{i}^{z}\right)^{2} + g\mu_{B}B_{z}\sum_{i}S_{i}^{z} + \sum_{\left\langle \left\langle i,j \right\rangle \right\rangle} \boldsymbol{D}_{ij}  \left(\boldsymbol{S}_{i} \times \boldsymbol{S}_{j} \right)
\end{align}    
where $\boldsymbol{S}_{i}=(S_{i}^{x},S_{i}^{y},S_{i}^{z})$ is the operator of total spin localized at a site $i$ of the lattice. The first term represents the exchange energy, with $J_{ij}$ the exchange coupling between spins localized at sites $i$ and $j$. The sum $\sum_{ij}$ runs over all atom pairs in the lattice up to the third-nearest neighbor, as shown in Fig.~\ref{FIG1}b. The second term involves $\Delta$, the easy-axis anisotropy energy. The third term represents the Zeeman energy arising from coupling to the applied magnetic field $B_{z}$ pointing along the $z$-axis, which is perpendicular to the plane. Here $g$ is the Land\'{e} $g$-factor and $\mu_{B}$ is the Bohr magneton. The fourth term captures the Dzyaloshinskii–Moriya Interaction with the DMI vector $\boldsymbol{D}_{ij}$ oriented in the z-direction. The notation $\left\langle \left\langle , \right\rangle \right\rangle$ indicates a summation over second nearest neighbors.

We employ the Holstein-Primakoff (HP) transformation \cite{Holstein1940} that maps spin operators residing on sublattice $A$ or $B$ of a 2D AFM to boson operators. The square roots of operators are expanded into a Taylor series and then truncated ~\cite{Bajpai2021} to linear order to obtain 
\begin{eqnarray}\label{eq:hp}
S_{A}^{+} & = & \sqrt{2S}a_{i},~ S_{A}^{-} = \sqrt{2S}a_{i}^{\dagger},~S_{A}^{z} = S-a_{i}^{\dagger}a_{i},  \nonumber \\ 
\mbox{} S_{B}^{+} & = & \sqrt{2S}b_{j}^{\dagger}, ~S_{B}^{-} = \sqrt{2S}b_{j},~S_{B}^{z} = -S + b_{j}^{\dagger}b_{j}.
\end{eqnarray}
The truncation to linear order is valid as long as the temperature is low, $k_BT \ll J_{ij}$ where $J_{ij}$ is the exchange coupling, and the number of magnons excited is sufficiently small~\cite{Bajpai2021}. This condition is met when the temperature is much lower than the N\'eel temperature of an AFM material, of order T$_N=~150~K$ for the 2D AFM materials considered here. We then recast the spin Hamiltonian as a bosonic Bogoliubov-de Gennes (BdG) Hamiltonian~\cite{Park2020} $H = \sum_{\boldsymbol{k}}\Psi^{\dagger}H(\boldsymbol{k})\Psi$ where $\Psi^{\dagger}= [x_{\boldsymbol{k},1}^{\dagger},x_{\boldsymbol{k},2}^{\dagger},...,x_{\boldsymbol{k},n}^{\dagger}, x_{-\boldsymbol{k},1},x_{-\boldsymbol{k},2},...,x_{-\boldsymbol{k},n}]$ is the Nambu spinor. By using Colpa's method \cite{Colpa1978}, we diagonalize this Hamiltonian to obtain the eigenenergies $E_{n,\boldsymbol{k}}$ and eigenvectors $\vert n(\boldsymbol{k}) \rangle$ of the system, which is what we use in the following to compute the magnon ONE from linear response theory. Please see the Supplementary Material (SM)~\footnote[2]{See Supplemental Material at \url{https://mrsec.udel.edu/about-ud-charm/} which includes Refs. \cite{kubo1957a,kubo1957b,Mahan2013,Smrcka1977,Mook2018,Shi2007,Xiao2005,Thonhauser2005,Fishman2022ITensor,Haldane1988, Chernyshev2016,Habel2024, Zhitomirsky2013, Gohlke2023, Hoyer2024, White2004, Haegeman2016,Chanda2020} along with (1) The BdG Hamiltonian of 2D AFM with N\'eel and Zigzag order (2) Derivations of Eqs.~\eqref{eq:semiclassicaeta}, \eqref{OBM}, \eqref{EPOL}.} for additional information on the construction of the BdG Hamiltonian and the magnetic parameters of the materials employed in this model Hamiltonian.

For the specific case of the 2D AFM materials we are now considering, and assuming the magnon's spin moment to be a well-defined quantum number, we derive:
 \begin{align}
  P_{y}^{S} &=   -\frac{ g\mu_{B}}{\hbar W_{y}c^{2}}\sum_{n,\boldsymbol{k}} \Omega_{xy}^{S^{z},n}\left( \boldsymbol{k} \right)k_{B}T~ ln\left(e^{-\frac{E_{n,\boldsymbol{k}}}{k_{B}T}}-1\right) \label{PS_2D}\\
  P_{y}^{O}  &= -\frac{2g\mu_{B}}{ W_{y}c^{2}}\sum_{n,\boldsymbol{k}} \sigma_{3}^{nn}S^{z}_{nn}L^{z}_{nn}\left(\boldsymbol{k} \right) \rho_{n,\boldsymbol{k}} \label{PO_2D}
\end{align}   
as the finite temperature electric polarization induced by magnon spin Berry curvature and magnon orbital angular moment, respectively. Here $W_{y}$ is the width along the y-direction of the system; $k_{B}$ is the Boltzmann constant, $c$ is the speed of light in vacuum, T is the temperature, $\Omega_{xy}^{S^{z},n}$ is the spin-Berry curvature of nth band, and $S^{z}_{nn}$ and $L^{z}_{nn}\left(\boldsymbol{k} \right)$ are respectively the spin and OAM of the magnon in the nth band. For a more detailed derivation of this formula, please refer to the SM~\footnotemark[2]. Eqs.~\ref{PS_2D} and \ref{PO_2D} allow us to more clearly see that 1) the spin current term arises from the spin Berry curvature [$\Omega_{xy}^{S^{z},n}\left( \boldsymbol{k} \right)$], which is intricately tied to the velocity of the magnon wavepacket's center and 2) the orbital term arises from the intra-band OAM of the magnon [$L^{z}_{nn}\left(\boldsymbol{k} \right)$], which arises from the inherent self-rotation of the magnon wavepacket. 

\begin{figure}
\centering
    \includegraphics[width= 0.5\textwidth]{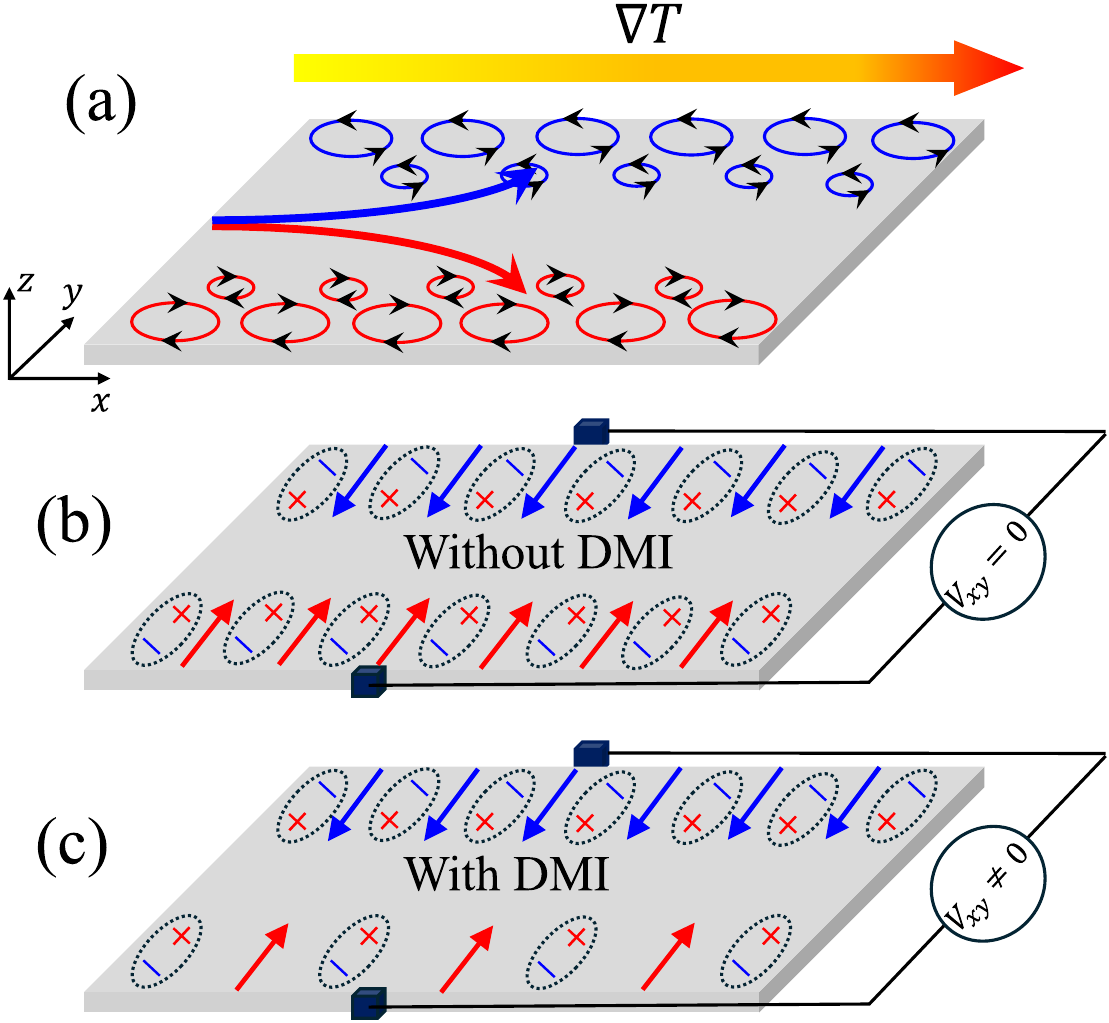}
 \caption{(a) Schematic illustration of magnon orbital accumulation at the edges of a 2D honeycomb AFM with N\'eel order under an applied temperature gradient $\nabla T$. (b) The local electric polarization (red and blue arrow) and local polarized charges induced by magnon spin and magnon orbital angular momentum accumulation due to the magnon ONE in the absence of the DMI. The number of charges on each side is equal, resulting in zero net electric polarization in the absence of DMI. (c) The presence of DMI results in a nonzero net polarized charge due to the difference in the populations of magnons carrying opposite OAM.}
  \label{S9}
\end{figure}

\begin{figure}
\centering
    \includegraphics[width= 1\textwidth]{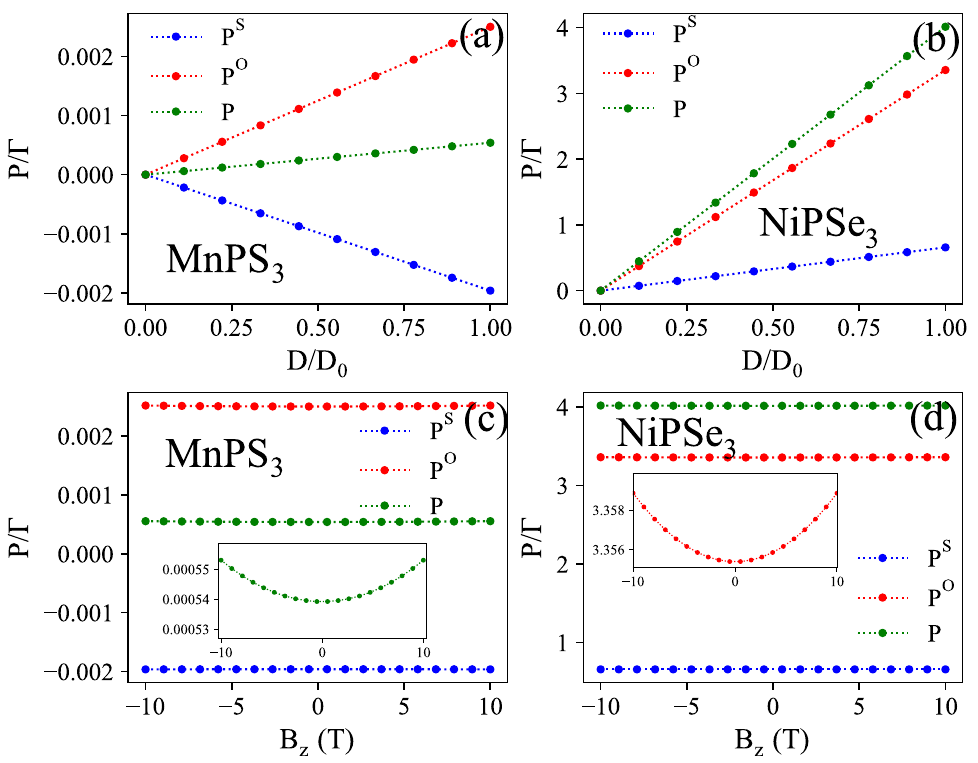}
 \caption{The electric polarization (P) of MnPS$_3$ (Figure a) and NiPSe$_3$ (Figure b) normalized to the factor $\Gamma=-\frac{g\mu_{B}}{\hbar W_{y}c^{2}}$. The polarization is plotted as a function of the Dzyaloshinskii-Moriya Interaction strength, parameterized by the ratio $D/D_{0}$ where $D_{0}$ represents the baseline DMI strength for each material. These calculations were conducted under a constant applied magnetic field of $B_{z}=1~T$ and temperature of 100 K. The electric polarization of MnPS$_3$ (c) and NiPSe$_3$ (d) also varies as a function of the externally applied magnetic field $B_z$. The insets in both figure c and d zoom in on the total electric polarization to reveal a weak dependence on $B_z$. These calculations were conducted for a temperature of 100 K.}
  \label{FIG4}
\end{figure}

We can now apply our theory to explain the electric polarization induced by magnon Nernst effects in MnPS$_3$ and NiPSe$_3$. The presence of a temperature gradient along the x direction leads to the accumulation of magnons with opposite chirality along the $\pm$y boundaries of the system, as schematically depicted in Fig.~\ref{S9}a. A quantitative calculation of the resulting electric polarization for MnPS$_3$ and NiPSe$_3$, as a function of the strength of the DMI interaction, is shown in Fig.~\ref{FIG4}(a) and (b). Remarkably, in the absence of DMI the electric polarization completely vanishes for both N\'eel and Zigzag magnetic orders, even in the presence of an externally applied magnetic field. In the presence of DMI, however, an electric polarization emerges. We will analyze the importance of DMI to the emergence of electric polarization in Sect.~\ref{DMI}. In the remainder of this section we focus on three important observations about the emergent electric polarization.

First, we emphasize that the contribution of magnon OAM to electric polarization should not be viewed as a consequence of magnetic dipole moment current circulation. Unlike electrons, which possess charge and whose motion is governed by electric forces, magnons are charge-neutral and their OAM originates solely from the geometric phase. Therefore, the contributions of magnon OAM to electric polarization should be understood as the changes in the magnon spin current due to the geometric phase of the magnon wave function, which adds a correction term ($P_{y}^{O}$) to the electric polarization induced by the magnon spin current ($P^{S}_{y}$) as described by Eqs.~\eqref{PS_2D} and \eqref{PO_2D}, which are derived quantum mechanically.

Second, the electric polarization $P_{y}$ depends on the width $W_{y}$ of the flake. However, what would be measured in practice is the transverse voltage along the y-direction ($V_{xy}$) induced by the electric polarization ($P_{y}$) under the temperature gradient along x-direction, as shown in Fig.~\ref{S9}(b,c). To derive a prediction of the measurable transverse voltage, we note that the electric field $\Xi_{y}$ along the y-direction induced by $P_{y}$ reads:
\begin{equation}
    \Xi_{y}= \frac{P_{y}}{\varepsilon_{0}\chi} 
\end{equation}
where $\varepsilon_{0}$ is the electric permittivity of free space and $\chi$ is the electric susceptibility. Therefore, the transverse voltage is given by 
\begin{equation}
    V_{xy} = \int_{0}^{W_{y}}\Xi_{y}dy = -\frac{ g\mu_{B}}{\hbar \varepsilon_{0}\chi c^{2}}\sum_{n,\boldsymbol{k}}\left[\Omega_{xy}^{S^{z},n}\left( \boldsymbol{k} \right)k_{B}T~ ln\left(e^{-\frac{E_{n,\boldsymbol{k}}}{k_{B}T}}-1\right) + 2\hbar\sigma_{3}^{nn}S^{z}_{nn}L^{z}_{nn}\left(\boldsymbol{k} \right) \rho_{n,\boldsymbol{k}}\right],
\end{equation}  
which is independent of $W_{y}$. 

\begin{figure}
\centering
    \includegraphics[width= 0.6 \textwidth]{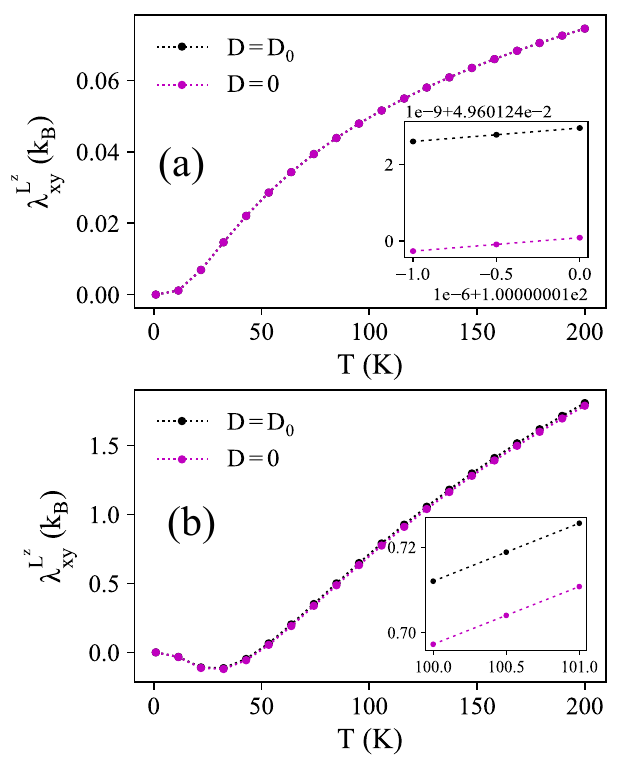}
 \caption{The ONC of MnPS$_3$ (a) and NiPSe$_3$ (b) as a function of temperature at fixed applied magnetic field $B_{z}=1~T$. In figures (a) and (b) the black and pink colors indicate the results calculated with and without DMI, respectively, and insets provide a zoomed-in view of the ONC at approximately T = 100 K.}
  \label{FIG3}
\end{figure}

Third, and perhaps most importantly, we see that the spin ($P^{S}_{y}$) and orbital ($P_{y}^{O}$) contributions to the overall electric polarization ($P$) can differ in both sign and magnitude. For MnPS$_3$ (Fig.~\ref{FIG4}(a)) the magnitudes of the $P^{S}_{y}$ and $P_{y}^{O}$ terms have opposite sign and similar magnitude, leading to a nonzero, but small, net polarization. In contrast, for NiPS$_3$ (Fig.~\ref{FIG4}(b)) the spin and orbital contributions have the same sign, but the orbital contribution ($P_{y}^{O}$) is approximately one order of magnitude larger than the spin contribution ($P^{S}_{y}$). This difference originates in the Orbital Nerst conductivity ($\lambda_{xy}^{L_{z}}$), which determines the the magnitude of the magnon orbital angular moment population that accumulates at the edge of the system due to the ONE as described in Eq.\eqref{eq:semiclassicaeta}. In Fig.~\ref{FIG3} we plot the ONC of MnPS$_3$ (a) and NiPS$_3$ (b) as a function of temperature T with (black) and without (pink) DMI. We find that the ONC are about three orders of magnitude larger than the spin Nernst conductivities predicted for these systems by Bazazzadeh, et al. in Ref.~\cite{Bazazzadeh2021b}. Specifically, the spin Nernst conductivity of NiPSe$_3$ is about 10$^{-3}$ k$_B$ \cite{Bazazzadeh2021b} while the ONC is of order k$_B$, which results in the much larger orbital contribution to the electric polarization reported in Fig.~\ref{FIG4}(b). This striking contrast originates in the fact that the magnon spin angular moment in the systems we consider can only have one of two values: The z-component of spin is locked to the magnon chirality and is independent of the wavevector $\boldsymbol{k}$ \cite{Cheng2016}. In contrast, there is no limit on the maximum OAM of a magnon. This suggests that focusing on materials that can host a large magnon orbital moment may prove advantageous for future device applications.

\section{Symmetry breaking and magnon-induced electric polarization}\label{DMI}

In this section, we investigate the symmetry breaking required for finite magnon-induced electric polarization in 2D honeycomb antiferromagnets. We will demonstrate how interactions like the Dzyaloshinskii-Moriya Interaction and other mechanisms, including magnon-phonon coupling, break the system's symmetry, thereby allowing for a finite electric polarization.

\begin{figure*}[!t]
\centering
    \includegraphics[width= 1 \textwidth]{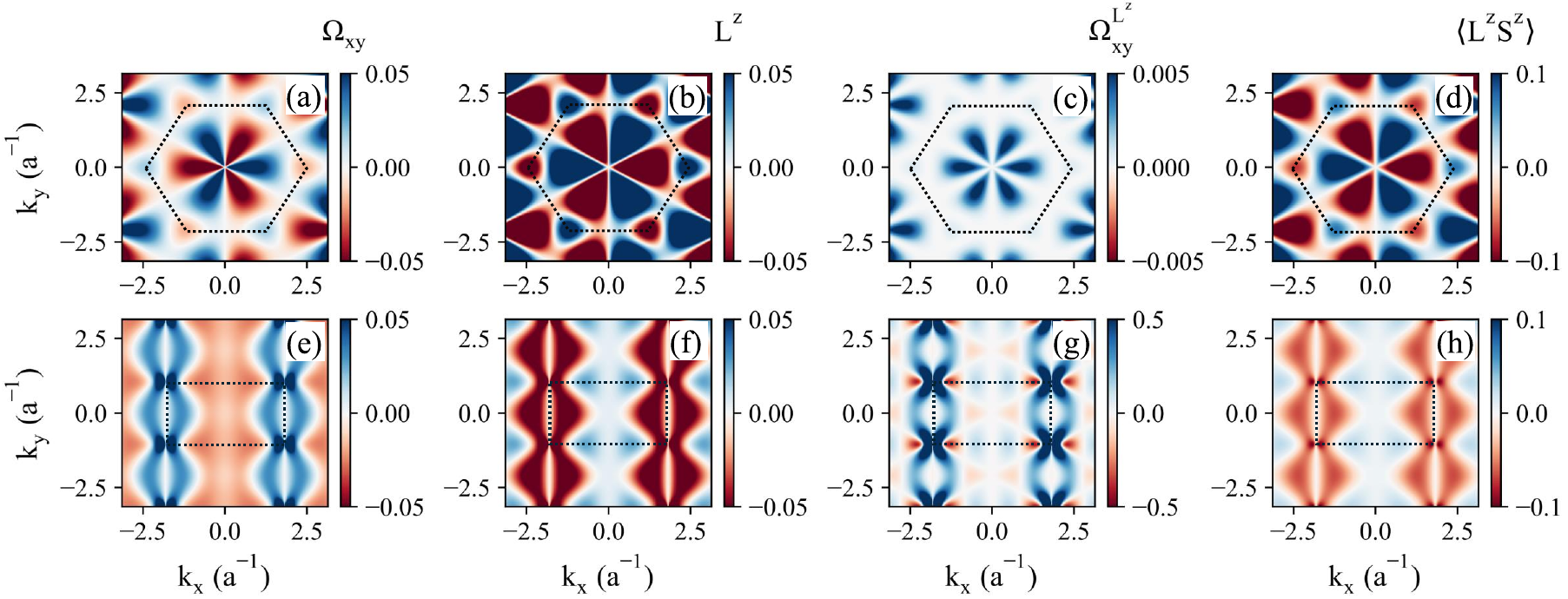}
 \caption{(a,e) The Berry curvature $\Omega_{xy}$, (b,f) the out-of-plane orbital angular moment component L$^{z}$ in the unit of $\hbar$, (c,g) Orbital Berry curvature $\Omega^{L^{z}}_{xy}$, and (d,h) correlation $\langle L^{z} S^{z} \rangle$ as a function of the in-plane wavevector $(k_{x},k_{y})$. Calculations are performed with applied magnetic field $B=1~T$ for the left-handed magnon mode of MnPS$_3$ (a-d), which has N\'eel magnetic order, and NiPSe$_3$ (e-h), which exhibits Zigzag magnetic order.}
  \label{FIG2}
\end{figure*}

Fig.~\ref{FIG4} shows that electric polarization vanishes in the absence of DMI for both N\'eel and Zigzag magnetic orders, but the reasons that EP vanishes are different for the two orders. To understand this more clearly, in Fig.~\ref{FIG2} we plot the Berry curvature, OAM, orbital Berry curvature (OBC), and the correlation $\langle L^{z}S^{z} \rangle^{n} = S^{z}_{nn}L^{z}_{nn}\left(\boldsymbol{k} \right)$ as a function of in-plane wavevector $\left(k_{x},k_{y}\right)$ for both MnPS$_3$ (a-e) and NiPS$_3$ (f-j). As expected, these plots display, respectively, the $C_{3}$ and $C_{2h}$ symmetries of the magnetic structure in the two. The OBC of both the N\'eel and Zigzag phases shows the even parity with respect to the wavevector, which may lead to a finite ONE. 

We first consider MnPS$_3$, which has N\'eel phase. For 2D AFMs in the N\'eel order, a finite magnon ONE may result in the accumulation of magnons carrying opposite intra-band OAM at the system's edges, even in the absence of DMI, while the spin Nernst current vanishes. However, as shown in Eq.~\eqref{PO_2D}, the electric polarization induced by the magnon OAM is governed by the correlation term $S^{z}_{nn}L^{z}_{nn}(\boldsymbol{k})$. This correlation function exhibits odd parity with respect to the wavevector $\boldsymbol{k}$, as shown in Fig.~\ref{FIG2}(d). This occurs because of the odd parity of the magnon intra-band OAM $L^{z}_{nn}\left(\boldsymbol{k} \right)$ even in the presence of DMI, as shown in Fig.~\ref{FIG2}(b). Without DMI, which is not shown here, this correlation function maintains its odd parity and the system possesses effective time reversal symmetry (TRS) \cite{To2023b}, i.e.~$E_{n,\boldsymbol{k}} = E_{n,\boldsymbol{-k}}$. In other words, in the absence of DMI the induced local electric polarizations at the edges are opposite in sign but equal in magnitude, resulting in zero total electric polarization across the sample regardless of the externally applied magnetic field, as schematically illustrated in Fig.~\ref{S9}b. A finite total EP is only observed in the case of N\'eel order when DMI breaks the symmetry between magnons carrying opposite intra-band OAM, as schematically illustrated in Fig.~\ref{S9}c, leading to $E_{n,\boldsymbol{k}} \neq E_{n,\boldsymbol{-k}}$. This disparity in energy levels results in an imbalance in the population at $\boldsymbol{k}$ and $-\boldsymbol{k}$, thereby inducing a finite EP as shown schematically in Fig.~\ref{S9}(c) and quantitatively in Fig.~\ref{FIG4}(a).

In contrast, the correlation function of NiPSe$_3$ with Zigzag order has even parity with respect to the wavevector in the presence of DMI, as shown in Fig.~\ref{FIG2}(h). The even parity would induce substantial EP as a result of ONE, as shown in Fig.~\ref{FIG3}(c). However, in the absence of DMI, the Zigzag phase's correlation function $\langle L^{z}S^{z} \rangle^{n} = S^{z}_{nn}L^{z}_{nn}\left(\boldsymbol{k} \right)$ vanishes because of vanishing magnon intra-band OAM $L^{z}_{nn}\left(\boldsymbol{k} \right)$, resulting in zero magnon OAM's contribution to EP. Nonzero total EP is only observed when DMI induces finite magnon intra-band OAM along with finite magnon SNE in the system. 

To further illustrate the importance of DMI in this picture, we return to Fig.~\ref{FIG3}, which plots the ONC of MnPS$_3$ (a) and NiPS$_3$ (b) as a function of temperature T with (black) and without (pink) DMI. We see that the ONC exists even in the absence of DMI, which indicates that the ONE does not require spin orbit coupling akin to the OHE predicted for electronic systems \cite{Bernevig2005,Go2018,Go2021}. This stands in stark contrast to the SNE, where DMI is required to observe non-zero SNE \cite{Cheng2016}. This disparity arises from the invariance of the systems under the combined $\mathcal{C}_{S}\mathcal{M}_{x}\mathcal{T}_{a}$ symmetry, where $\mathcal{C}_{S}$ is the spin rotation symmetry operation which flips all the spins (and magnetic fields) in the system, $\mathcal{M}_{x}$ represents mirror symmetry with respect to the plane perpendicular to the x-axis, and $\mathcal{T}_{a}$ denotes the translation operator responsible for displacing the system by the vector $\boldsymbol{\beta}_{1}$ or $\boldsymbol{\beta}_{4}$ [see Fig.~\ref{FIG1}(b)]. The ONE does not require the breaking of $\mathcal{C}_{S}\mathcal{M}_{x}\mathcal{T}_{a}$ symmetry, primarily because the OAM is not preserved under non-commutative rotation and translation operations. Consequently, the ONC has non-zero values whether or not DMI is present to break $\mathcal{C}_{S}\mathcal{M}_{x}\mathcal{T}_{a}$ symmetry, as illustrated in Figure~\ref{FIG3}(a) and (b). However, when the $\mathcal{C}_{S}\mathcal{M}_{x}\mathcal{T}_{a}$ symmetry is preserved, the spin polarized current along the y-direction $j_{y}^{S^{z}}$ must vanish [see (SM)~\footnotemark[2] for further detail]. In other words, regardless of the externally applied magnetic field, magnon ONE does not induce a spin-polarized current in the absence of DMI. Only when the presence of DMI breaks the $\mathcal{C}_{S}\mathcal{M}_{x}\mathcal{T}_{a}$ symmetry does a finite spin current emerge within the system as a result of the SNE. The finite SNE in the presence of DMI coupled with finite ONE results in the observable total EP as shown in Figs.~\ref{FIG4}(a) and (b). Meanwhile, as expected, the total EP vanishes when DMI is turned off. 

The dependence of the electric polarization on the magnitude of the DMI interaction strength, which is reported in Figs.~\ref{FIG4}(a) and (b), can also be understood from a symmetry perspective. In the absence of DMI and externally applied magnetic field, the system maintains mirror symmetry ($\mathcal{M}_{y}$) for N\'eel order and glide mirror symmetry ($\mathcal{M}_{y}\tau$) for Zigzag order about the plane normal to the y-direction. Because the systems hold this symmetry, $P_{y} \equiv - P_{y}$ which means that $P_{y}$ must vanish. The presence of an externally applied magnetic field breaks the mirror and glide mirror symmetry, but it does not induce any coupling between different magnon bands and therefore the wavefunctions are unchanged. The relation $E_{n,\boldsymbol{k}} = E_{n,-\boldsymbol{k}}$ is maintained, which means $P_{y}$ remains zero despite the presence of the externally applied magnetic field when DMI is absent ($D/D_{0}=0$ as shown in Fig.~\ref{FIG4}(a,b)). However, the presence of DMI breaks the $\mathcal{M}_{y}$ symmetry and yields $E_{n,\boldsymbol{k}} \neq E_{n,-\boldsymbol{k}}$ \cite{Lee2018,Ghader2019}, thereby allowing the nonzero $P_{y}$ shown in Fig.~\ref{FIG4}(a,b). 

In Fig.~\ref{FIG4}(c,d) we present the dependence of the electric polarization on $B_z$, the magnetic field externally applied along the z-direction. The total electric polarization $P_y$, along with its constituents from spin current $P_{y}^{S}$ and orbital angular moment $P_{y}^{O}$, exhibit relatively small variations as a function of the magnetic field. The magnetic field has minimal impact because it splits the magnon bands corresponding to opposite spins without inducing any coupling between these distinct bands. Consequently, the weak response of the EPs to the magnetic field $B_z$ arises solely from changes in the magnon population $\rho_{n,\boldsymbol{k}}$ and the function $ln\left(e^{-\frac{E_{n,\boldsymbol{k}}}{k_{B}T}}-1\right)$ in Eq.~\eqref{PS_2D}, which are due to the shift in the magnon energy band caused by the Zeeman interaction between local spins and external applied magnetic field. Crucially, all components remain finite even at zero magnetic field, suggesting the potential feasibility of detecting the magnon Orbital Nernst effect even in the absence of an external magnetic field.

We conclude this section with three notes about the possibility of observing the predicted effects in realistic materials. First, we note that the electric polarization in MnPS$_3$ is approximately three orders of magnitude smaller than that in NiPSe$_3$. Additionally, the EP in NiPSe$_3$ is predominantly influenced by magnon orbital angular moment, whereas in MnPS$_3$ both the spin Berry curvature and magnon OAM contribute equally to the EP. This indicates that the measurement of EP in the Zigzag order of NiPSe$_3$ would provide direct evidence of magnon OAM and orbital Nernst effect in this material. Second, in the system considered here the DMI is oriented along the z axis (out of plane), which means it cannot introduce scattering between magnons. In Sect.~\ref{magnon-magnon-interaction} of the SM~\footnotemark[2] we consider what would happen in materials for which this is not the case. Finally, we note that the interaction between magnons and phonons in collinear 2D antiferromagnets and the inherent structure of noncollinear AFMs can break symmetries in a manner similar to the Dzyaloshinskii-Moriya interaction. These effects may also lead to a finite contribution from magnon ONE to the EP in these material systems.

\section{Conclusion and Outlook}\label{conclusion}
In summary, we present a formalism that allows us to connect measurable quantities, namely electric polarization, to magnon spin and orbital moment transport. We introduce the representation of magnon OAM within the Bloch wavefunction, establishing a profound connection between magnon intra-band OAM and magnon Berry curvature. Additionally, we present a full quantum mechanical derivation of electric polarization induced by magnon OAM and magnon spin current. We then apply our formalism to novel phenomena—the Magnon Orbital Nernst Effect and Magnon Spin Nernst Effect—in 2D AFMs exhibiting either Zigzag or N\'eel order on the honeycomb lattice. Through this approach, we demonstrate that materials with N\'eel order and/or Zigzag order in combination with DMI can generate the electric polarization at the edges of the system as a result of magnon Nernst effects. These intriguing findings point the way toward experimental validation of the predicted magnon SNE and magnon ONE phenomenon in 2D honeycomb AFMs with N\'eel and Zigzag order (e.g. MnPS$_3$ and NiPSe$_3$). They may also catalyze further exploration of the emerging field of magnon orbitronics and magnon spinorbitronics. Finally, it is important to highlight that although our work focuses on generating MNEs through thermal effects, our findings extend beyond this scope. Specifically, our results suggest that, despite being electrically neutral, magnons can interact with electromagnetic waves via the electric field component of light. Our theory shows that both the spin and orbital moments of magnons contribute to their electric activity. Consequently, we anticipate that light can transfer its angular momentum into the spin and orbital angular momentum of magnons. This insight suggests the possibility of future research on light-driven magnon dynamics in magnetic materials, leveraging the spin and orbital angular momentum degrees of freedom of magnons.

%%%%%%%%%%%%%%%% REFERENCES %%%%%%%%%%%%%%%

\clearpage % Clear all remaining figures and tables then start a new page

% The list of references goes after the main text and before the acknowledgements
% When preparing an initial submission, we recommend you use BibTeX, like this:
%
\bibliography{Ref} % for a file named science_template.bib
\bibliographystyle{sciencemag}

% After the paper has completed peer review and been revised ready for acceptance,
% you should comment out the lines above and copy-paste the contents of your .bbl
% file here instead. This will help ensure that our conversion software works correctly.
% Remember to re-run BibTeX first - check the timestamp!
%
% Example of the first three entries copy-pasted from science_template.bbl:
%
%\begin{thebibliography}{1}
%
%\bibitem{example}
%A.~N. {Author}, An example reference. \emph{Journal of Improbable Research}
%  \textbf{1}, 67 (2020).
%
%\bibitem{example2}
%F.~M. {Surname}, S.~{Author}, A second example. \emph{Interesting Research
%  Letters} \textbf{32}, 897 (2019).
%
%\bibitem{example_preprint}
%P.~{One}, P.~{Two}, P.~{Three}, {An unpublished preprint}. \emph{preprint}
%  (2021), arXiv:2101.12345.
%
%\end{thebibliography}

%%%%%%%%%%%%%%%% ACKNOWLEDGEMENTS %%%%%%%%%%%%%%%

\section*{Acknowledgments}
This research was primarily supported by NSF through the University of Delaware Materials Research Science and Engineering Center, DMR-2011824.
\paragraph*{Funding:}
DMR-2011824

%%%%%%%%%%%%%%%% SUPPLEMENT LIST %%%%%%%%%%%%%%%

% List the contents of your Supplementary Materials, including the numbers of any
% supplementary figures, tables, external data files etc. and any references that are
% cited only in the supplement. In this example, refs. 7-8 are cited only in the supplement.
% Fill out your numbers accordingly and delete any lines that aren't applicable.
%%%%%%%%%%%%%%%% END OF MAIN TEXT %%%%%%%%%%%%%%%

\newpage

%%%%%%%%%%%%%%%% START OF SUPPLEMENT %%%%%%%%%%%%%%%

% Figures, tables, equations and pages in the supplement are numbered S1, S2 etc.
\renewcommand{\thefigure}{S\arabic{figure}}
\renewcommand{\thetable}{S\arabic{table}}
\renewcommand{\theequation}{S\arabic{equation}}
\renewcommand{\thepage}{S\arabic{page}}
\renewcommand{\thesection}{S\arabic{section}}
\setcounter{section}{0}
\setcounter{figure}{0}
\setcounter{table}{0}
\setcounter{equation}{0}
\setcounter{page}{1} % not 0 as \newpage already started a supplementary page
% References continue the numbering from the main text.

%%%%%%%%%%%%%%%% SUPPLEMENT TITLE PAGE %%%%%%%%%%%%%%%

\begin{center}
\section*{Supplementary Materials for\\ \scititle}

% Author list for the supplement
% Indicate the corresponding authors, but do NOT include institutions here
% It would be nice if the template auto-generated this, but doing so is complicated...
D. Quang To$^{1\ast}$,
	Federico Garcia-Gaitan$^{2}$,
	Yafei Ren$^{2}$,
    Joshua M. O. Zide$^{1}$,\\
    M. Benjamin Jungfleisch$^{2}$,
    John Q. Xiao$^{2}$,
    Branislav K. Nikoli\'{c}$^{2}$, \\
    Garnett W. Bryant$^{3,4}$,
    Matthew F. Doty$^{1\ast}$ \\
	% Additional lines of authors should be inserted using the \and command (not \\)
	% Institution list, in a slightly smaller font
	\small$^{1}$Department of Materials Science and Engineering, University of Delaware, Newark, Delaware 19716, USA. \\
	\small$^{2}$Department of Physics and Astronomy, University of Delaware, Newark, Delaware 19716, USA.\\
    \small$^{3}$Nanoscale Device Characterization Division, Joint Quantum Institute, National Institute of Standards and Technology, Gaithersburg, Maryland 20899-8423, USA.\\
     \small$^{4}$University of Maryland, College Park, Maryland 20742, USA.\\
	% Identify at least one corresponding author, with contact email address
	\small$^\ast$Corresponding author. Email: quangto@udel.edu; doty@udel.edu
\end{center}

% Fill out the numbers for each type of supplementary material,
% and delete any lines that aren't applicable.
% These are just example numbers that don't match the rest of this template.
\subsubsection*{This PDF file includes:}
This PDF file includes\\
Sections S1 to S7\\
Figs. S1 to S8\\
Table S1\\

\newpage

%%%%%%%%%%%%%%%% MATERIALS AND METHODS %%%%%%%%%%%%%%%

\section{Magnon Hamiltonian via Holstein-Primakoff transformation}
\begin{figure}[h]
\centering
    \includegraphics[width= 1\textwidth]{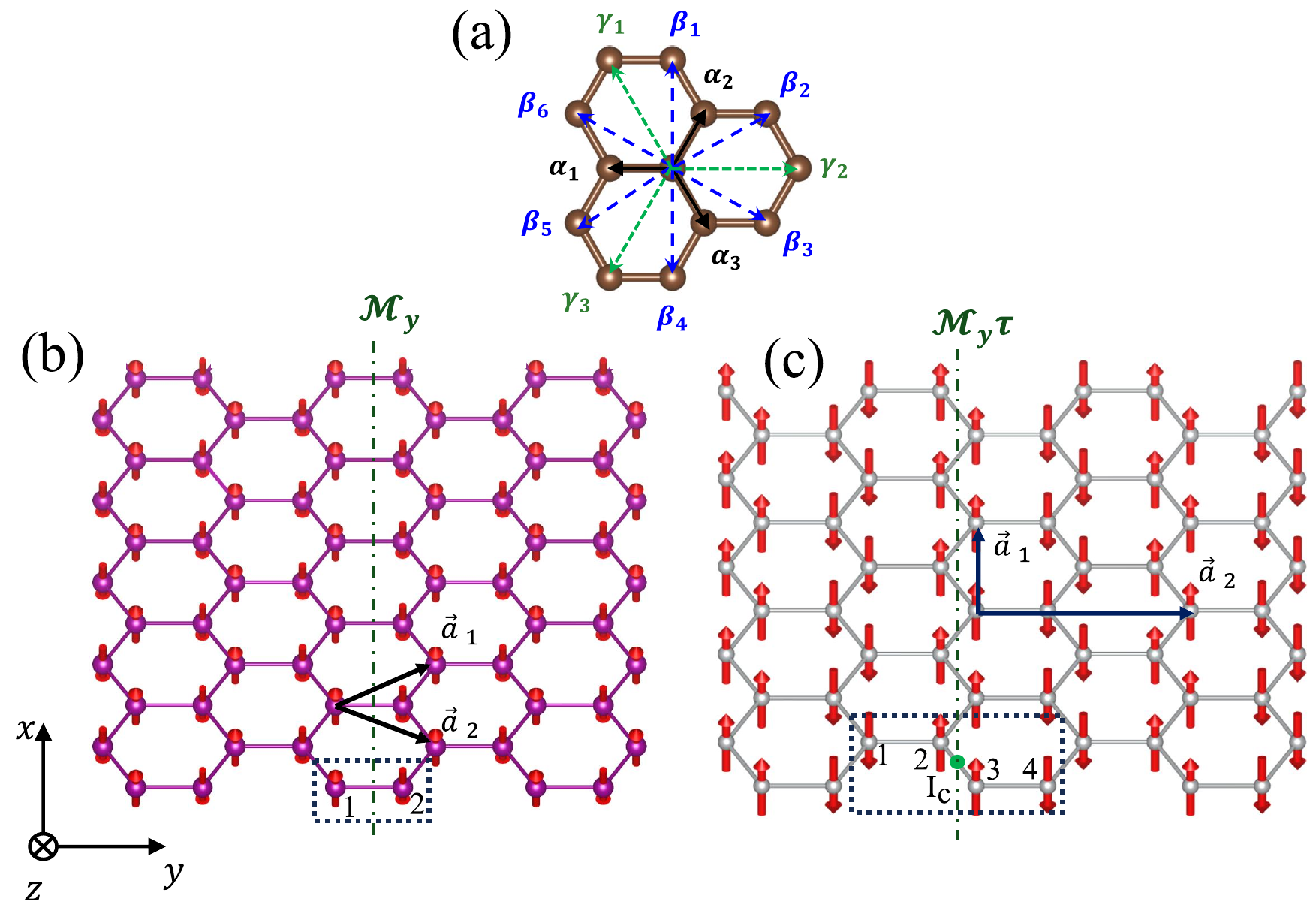}
 \caption{(a) The nearest, second nearest and third nearest neighbor bonds in honeycomb lattice are denoted by $\alpha_{i}, ~\beta_{i}$ and $\gamma_{i}$, respectively. The quasi 2D honeycomb AFM lattice with N\'eel (b) and Zigzag (c) order formed by magnetic atoms. The arrows indicate the  primitive vectors and the dashed rectangular shape shows the unit cell of corresponding lattice. The $\mathcal{M}_{y}$ and $\mathcal{M}_{y}\tau$ represent the mirror and glide mirror symmetry of the N\'eel and Zigzag order, respectively.}
  \label{S1}
\end{figure}

In this work we direct our attention to two distinct types of 2D honeycomb antiferromagnets: those with N\'eel and Zigzag orders whose magnetic structures are illustrated in Fig.~\ref{S1} (b) and (c) respectively. The primitive cell of the N\'eel order comprises two magnetic atoms with opposing spins and lacks an inversion center and, thus, inversion symmetry. Conversely, the Zigzag order's primitive cell accommodates four magnetic atoms, denoted by numbers 1-4 in Fig.~\ref{S1} (c), with the inversion center $I_c$ positioned between two adjacent magnetic atoms exhibiting identical spins. We recall that the fundamental spin Hamiltonian for this type of system can be expressed as follows

\begin{equation} \label{Hamilt}
H = \sum_{i,j} J_{ij}\boldsymbol{S}_{i}\cdot \boldsymbol{S}_{j} + \Delta\sum_{i}\left( S_{i}^{z} \right)^{2} + g\mu_{B}B_{z}\sum_{i}S_{i}^{z} +  \sum_{\left\langle \left\langle i,j \right\rangle \right\rangle} \boldsymbol{D}_{ij}  \left(\boldsymbol{S}_{i} \times \boldsymbol{S}_{j} \right)
\end{equation}

To derive a second-quantization version of Eq.~\eqref{Hamilt} in terms of bosonic operators creating and annihilating magnons, we employ the standard Holstein-Primakoff transformation~\cite{Holstein1940} that maps spin operators residing on sublattice $A$ or $B$ of a two-dimensional antiferromagnet (2D AFM) to bosonic operators whose square root is expanded in a Taylor series and then truncated~\cite{Bajpai2021} to linear order 

\begin{align}\label{eq:hp1}
    S_{A}^{+} = \sqrt{2S}a_{i} ~ ~ ~ ~ ~     S_{A}^{-} = \sqrt{2S}a_{i}^{\dagger} ~ ~ ~ ~ ~ ~ S_{A}^{z} = S-a_{i}^{\dagger}a_{i},
\end{align}
\begin{align}\label{eq:hp2}
    S_{B}^{+} = \sqrt{2S}b_{j}^{\dagger} ~ ~ ~ ~ ~
    S_{B}^{-} = \sqrt{2S}b_{j} ~ ~ ~ ~ ~
    S_{B}^{z} = -S + b_{j}^{\dagger}b_{j}.
\end{align}
Such truncation is valid as long as the temperature is low, $k_BT \ll J_{ij}$, where $J_{ij}$ is the exchange coupling in Eq.~(2) in the main text, and the number of magnons excited is sufficiently small~\cite{Bajpai2021}. Here $a_{i}$ and $b_{j}$ ($a_{i}^{\dagger}$ and $b_{j}^{\dagger}$) are  operators annihilating (creating) magnons at site $i \in A$ or site $j \in B$, respectively. Using the Fourier transform of these operators
\begin{align}
    a_{i} = \frac{1}{\sqrt{N}} \sum_{\boldsymbol{k}} e^{i\boldsymbol{k \cdot r}_{a_{i}}}a_{\boldsymbol{k},i}, ~~~~~~~ a_{i}^{\dagger} = \frac{1}{\sqrt{N}} \sum_{\boldsymbol{k}} e^{-i\boldsymbol{k \cdot r}_{a_{i}}}a_{\boldsymbol{k},i}^{\dagger},
\end{align}
\begin{align}
    b_{i} = \frac{1}{\sqrt{N}} \sum_{\boldsymbol{k}} e^{i\boldsymbol{k \cdot r}_{b_{i}}}b_{\boldsymbol{k},i},  ~~~~~~~  b_{i}^{\dagger} = \frac{1}{\sqrt{N}} \sum_{\boldsymbol{k}} e^{-i\boldsymbol{k \cdot r}_{b_{i}}}b_{\boldsymbol{k},i}^{\dagger},
\end{align}
the Heisenberg Hamiltonian in Eq.~(2) of the main text can be re-written in second-quantization form as
\begin{equation}
    \hat{H} = E^{0} + \hat{H}\left(\boldsymbol{k}\right).
\end{equation}
Here $E^{0}$ is a $k$-independent energy which simply shifts the energy-momentum dispersion of magnons by a constant value and, hence, can be neglected. The $k$-dependent terms, containing operators which create and annihilate magnons in momentum $\hbar\boldsymbol{k}$, are collected into $\hat{H}\left(\boldsymbol{k}\right) = \Psi^{\dagger}\hat{H}_{\boldsymbol{k}}\Psi$ where $\Psi^{\dagger}= \left(x_{\boldsymbol{k},1}^{\dagger},x_{\boldsymbol{k},2}^{\dagger},...,x_{\boldsymbol{k},n}^{\dagger}, x_{-\boldsymbol{k},1},x_{-\boldsymbol{k},2},...,x_{-\boldsymbol{k},n}\right)$ is the Nambu spinor.

For N\'eel order, $\Psi^{\dagger}= \left(a_{\boldsymbol{k}}^{\dagger},b_{\boldsymbol{k}}^{\dagger},a_{-\boldsymbol{k}},b_{-\boldsymbol{k}} \right)$ the bosonic Bogoliubov-de Gennes (BdG) Hamiltonian is \cite{Bazazzadeh2021,Bazazzadeh2021b}
\begin{equation}
\hat{H}_{\boldsymbol{k}} = R_{m}\begin{bmatrix}
    A\left( \boldsymbol{k} \right) &0 \\
    0 & A^{T}\left(- \boldsymbol{k}\right) 
\end{bmatrix}R_{m}^{\dagger}
\label{Hneel}
\end{equation}
where 
\begin{equation}
    A\left( \boldsymbol{k} \right) = S\begin{bmatrix}
        J + J_{2}\zeta_{\boldsymbol{\beta}} -D\zeta_{D} &J_{1}\zeta_{\boldsymbol{\alpha}} + J_{3}\zeta_{\boldsymbol{\gamma}}\\
        J_{1}\zeta_{\boldsymbol{\alpha}}^{*} & J + J_{2}\zeta_{\boldsymbol{\beta}} + D\zeta_{D}
    \end{bmatrix}
\end{equation}
Here $J=3J_{1}-6J_{2}+3J_{3}-\Delta$; $\zeta_{\boldsymbol{\Sigma}} = \sum_{m}\zeta_{\boldsymbol{\Sigma}_{m}}$; $\zeta_{\boldsymbol{\Sigma}_{m}}=e^{i\boldsymbol{k.\Sigma}_{m}}$ ($\boldsymbol{\Sigma}\equiv \boldsymbol{\alpha,\beta,\gamma}$) and $\zeta_{D}=\sum_{m~ \in~odd}2sin\left(\boldsymbol{k.\beta}_{m} \right)$; $R_{m}$ is the rotational matrix given by:
\begin{equation}
    R_{m}=\begin{bmatrix}
        1 &0 &0 &0 \\
        0 &0 &0 &1\\
        0 &0 &1 &0\\
        0 &1 &0 &0
    \end{bmatrix}
\end{equation}
\begin{table}[h] \label{tab1}
\caption{The exchange coupling between localized spins, for 2D AFM used in the main text.}
\small
\begin{tabular}{ccccccccccccccc}
\hline
\hline
Materials &a (\AA) &S($\mu_{B}$) &J$_{1}$ (meV) &J$_{2}$ (meV) &J$_{3}$ (meV)  &$\Delta$ (meV) &D ($\mu eV$) &$M_{eff}$ $\left(\hbar^{2}~ \text{meV}^{-1}\text{Å}^{-2}\right)$ \\
\hline
MnPS$_3$ \cite{Bazazzadeh2021,Bazazzadeh2021b} &5.88 &4.56 &0.527 &0.024 &0.150  &-0.002 &0.39 &0.5\\
NiPSe$_3$ \cite{Bazazzadeh2021,Bazazzadeh2021b}&6.14 &1.56 &-1.131 &-0.069 &3.975 &-0.19 &43.90  &0.1\\
\hline
\hline
\end{tabular}
\end{table}

In the same manner, the BdG Hamiltonian for Zigzag order reads \cite{Bazazzadeh2021,Bazazzadeh2021b}:
\begin{equation}
    \hat{H}_{\boldsymbol{k}} = R_{m}\begin{bmatrix}
    B\left( \boldsymbol{k} \right) &0 \\
    0 & B^{T}\left(- \boldsymbol{k}\right) 
\end{bmatrix}R_{m}^{\dagger}
\label{Hzigzag}
\end{equation}
with $\Psi^{\dagger}= \left(a_{1,\boldsymbol{k}}^{\dagger},b_{1,\boldsymbol{k}}^{\dagger},b_{2,\boldsymbol{k}}^{\dagger},a_{2,\boldsymbol{k}}^{\dagger},a_{1,-\boldsymbol{k}},b_{1,-\boldsymbol{k}},b_{2,-\boldsymbol{k}},a_{2,-\boldsymbol{k}} \right)$. Here 
\begin{equation}
   B\left( \boldsymbol{k} \right) = S\begin{bmatrix}
        C\left( \boldsymbol{k} \right) & D\left( \boldsymbol{k} \right) \\
        D^{\dagger}\left( \boldsymbol{k} \right) &C\left( -\boldsymbol{k} \right)
    \end{bmatrix} 
 \end{equation}
 
\begin{align}
    C\left( \boldsymbol{k} \right)   = &\sigma_{0}\left(-J_{1}+2J_{2}+3J_{3}-\Delta \right) + J_{2}\left[ \sigma_{0}\left(\zeta_{\boldsymbol{\beta}_{1}}+\zeta_{\boldsymbol{\beta}_{4}} \right) + \sigma_{1}\left(\zeta_{\boldsymbol{\beta}_{2}}+\zeta_{\boldsymbol{\beta}_{3}}+\zeta_{\boldsymbol{\beta}_{5}}+\zeta_{\boldsymbol{\beta}_{6}} \right) \right] \\ \notag
    &+iD\left[\sigma_{0}\left(\zeta_{\boldsymbol{\beta}_{1}}-\zeta_{\boldsymbol{\beta}_{4}} \right)+\sigma_{1}\left(-\zeta_{\boldsymbol{\beta}_{2}} + \zeta_{\boldsymbol{\beta}_{3}} + \zeta_{\boldsymbol{\beta}_{5}}-\zeta_{\boldsymbol{\beta}_{6}} \right) \right]
\end{align}

\begin{equation}
    D\left( \boldsymbol{k} \right) = J_{1}\left[\sigma_{0}\left(\zeta_{\boldsymbol{\alpha}_{2}}^{*} + \zeta_{\boldsymbol{\alpha}_{3}}^{*} \right) + \sigma_{1}\zeta_{\boldsymbol{\alpha}_{1}}^{*} \right] + J_{3}\sigma_{1}\left(\zeta_{\boldsymbol{\gamma}_{1}}^{*} +\zeta_{\boldsymbol{\gamma}_{2}}^{*} +\zeta_{\boldsymbol{\gamma}_{3}}^{*} \right)
\end{equation}
Here
\begin{equation}
    \sigma_{0} = \begin{pmatrix}
        \boldsymbol{1}_{N \times N} & 0 \\
        0 &\boldsymbol{1}_{N \times N}
    \end{pmatrix}, ~~~~~ \sigma_{1}= \begin{pmatrix}
        0 &\boldsymbol{1}_{N \times N} \\
        \boldsymbol{1}_{N \times N} &0
    \end{pmatrix}, ~~~~~ \sigma_{2} = \begin{pmatrix}
        0 & -i\boldsymbol{1}_{N \times N} \\
        i\boldsymbol{1}_{N \times N} &0
    \end{pmatrix}, ~~~~~\sigma_{3} = \begin{pmatrix}
        \boldsymbol{1}_{N \times N} & 0 \\
        0 &-\boldsymbol{1}_{N \times N}
    \end{pmatrix}.
\end{equation}
are the Pauli matrices in Bogoliubov space, and the rotation matrix $R_{m}$ is given by
\begin{equation}
    R_{m}=\begin{bmatrix}
        1 &0 &0 &0 &0 &0 &0 &0\\
        0 &0 &0 &0 &0 &0 &0 &1\\
        0 &0 &0 &0 &0 &1 &0 &0\\
        0 &0 &1 &0 &0 &0 &0 &0\\
        0 &0 &0 &0 &1 &0 &0 &0\\
        0 &0 &0 &1 &0 &0 &0 &0\\
        0 &1 &0 &0 &0 &0 &0 &0\\
        0 &0 &0 &0 &0 &0 &1 &0
    \end{bmatrix}
\end{equation}
The material parameters used in our numerical calculations based on these Hamiltonians are listed in Table \ref{tab1}. By using Colpa's method \cite{Colpa1978}, we diagonalize this Hamiltonian to obtain the eigenenergies of the system $E_{\boldsymbol{k}}$ satisfying the following eigenvalue equation 

\begin{equation} \label{EigEq}
\boldsymbol{\sigma}_{3}\hat{H}_{\boldsymbol{k}}T(\boldsymbol{k})=T(\boldsymbol{k})\boldsymbol{\sigma}_{3}E_{\boldsymbol{k}},
\end{equation}
as the generalized eigenvalue problem in which $\boldsymbol{\sigma}_{3}\hat{H}_{\boldsymbol{k}}$ is a non-Hermitian matrix even though $\hat{H}_{\boldsymbol{k}}$ is Hermitian. In other words, the diagonalization of the BdG Hamiltonian deals with non-Hermitian quantum mechanics~\cite{Park2020}, but the eigenvalues $E_{\boldsymbol{k}}$ remain real. In Eq.~\eqref{EigEq}, matrix $T(\boldsymbol{k})$  is ``paraunitiary'' satisfying
\begin{equation}
    T^{\dagger}(\boldsymbol{k}) \boldsymbol{\sigma}_{3}T(\boldsymbol{k})=T(\boldsymbol{k}) \boldsymbol{\sigma}_{3}T^{\dagger}(\boldsymbol{k})=\boldsymbol{\sigma}_{3}
    \label{EQParamatrix},
\end{equation}
The eigenvector $\vert n (\boldsymbol{k})\rangle$ with mth element given by $\vert n (\boldsymbol{k})\rangle_{m}=\left[T(\boldsymbol{k})\right]_{mn}$, with associated eigen-energies $E_{\boldsymbol{k}}$, forms the basis for conducting calculations related to magnon Orbital Angular Moment (OAM), Berry curvature, Orbital Berry curvature, and other relevant quantities discussed in the main text. 

\begin{figure}[h]
\centering
    \includegraphics[width= 0.8\textwidth]{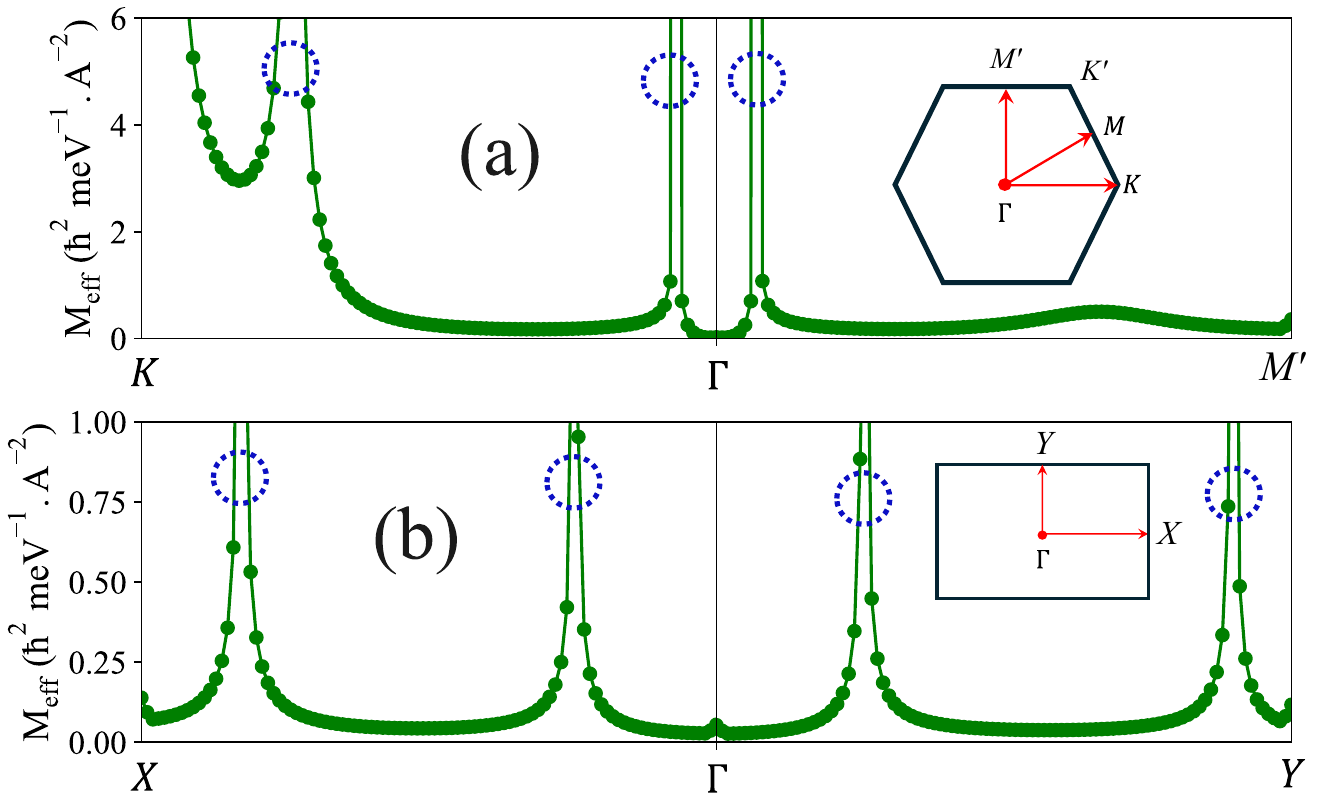}
 \caption{The relative band effective mass $M_{eff}$ calculated for (a) MnPS$_3$ and (b) NiPSe$_3$ along $K-\Gamma-M$ path for MnPS$_3$ and $X-\Gamma-Y$ path for NiPSe$_3$.}
  \label{S2}
\end{figure}

One can observe that in 2D collinear antiferromagnets with out-of-plane (z-direction) magnetic order, both the exchange interaction and the Dzyaloshinskii-Moriya interaction maintain the rotational symmetry about the z-axis. Consequently, the z-component of the total spin, defined as $S^{z}=\sum_{i}\left(S_{iA}^{z}+S_{iB}^{z} \right)$, remains a conserved quantum number \cite{Cheng2016}. By applying the Holstein-Primakoff transformation, one derives the following:
\begin{equation}
    S^{z} = \sum_{\boldsymbol{k}}\left(b_{\boldsymbol{k}}^{\dagger}b_{\boldsymbol{k}}-a_{\boldsymbol{k}}^{\dagger}a_{\boldsymbol{k}} \right) = \sum_{\boldsymbol{k}}S_{\boldsymbol{k}}^{z}
\end{equation}
Note that the $S_{\boldsymbol{k}}^{z}$ commutes with the Hamiltonian, i.e. $\left[H,S_{\boldsymbol{k}}^{z} \right] = 0$. Consequently it is diagonal in the Nambu basis. Employing the Bogoliubov transformation to obtain the normal modes of the magnonic field yields:
\begin{equation}
    S^{z}=\sum_{\alpha,\beta,\boldsymbol{k}}\left(\beta_{\boldsymbol{k}}^{\dagger}\beta_{\boldsymbol{k}} -\alpha_{\boldsymbol{k}}^{\dagger}\alpha_{\boldsymbol{k}}\right)
\end{equation}
where $\alpha_{\boldsymbol{k}}=u_{\boldsymbol{k}}a_{\boldsymbol{k}}-v_{\boldsymbol{k}}b^{\dagger}_{\boldsymbol{k}}$ and $\beta_{\boldsymbol{k}}=u_{\boldsymbol{k}}b_{\boldsymbol{k}}-v_{\boldsymbol{k}}a_{\boldsymbol{k}}^{\dagger}$ with the Bogoliubov coefficients $u_{\boldsymbol{k}}$ and $v_{\boldsymbol{k}}$ being thoughtfully selected to diagonalize the Hamiltonians \eqref{Hneel} and \eqref{Hzigzag}. This demonstrates that $\langle 0 \vert \eta_{\boldsymbol{k}} S^{z} \eta_{\boldsymbol{k}}^{\dagger}\vert 0 \rangle= \pm 1$, where $\eta$ represents either $\beta$ or $\alpha$. Specifically, the $\alpha$ and $\beta$ magnon carry -1 and +1 spin angular momentum along the z-direction, respectively. Consequently, the z-component of spin is closely associated with the magnon chirality, allowing us to interpret $\pm 1$ as the helicity of the magnon bands. Therefore, each magnon band is constrained to one of these two helicity values, which may result in a smaller spin Nernst effect compared to the orbital Nernst effect, as discussed in the main text.

In addition, it is important to note that the approach we have presented here relies on linear spin wave theory, which is applicable only when the temperature is significantly lower than the N\'eel temperature of the antiferromagnetic material.

In Fig.~\ref{S2}, we have plotted the effective mass $M_{eff}=\hbar^{2}\left\vert \partial^{2} E/\partial k^{2} \right\vert^{-1}$ of magnon wavepacket \cite{Matsumoto2011b} of the first magnon band for MnPS$_3$ [Fig.~\ref{S2}(a)] and NiPSe$_3$ [Fig.~\ref{S2}(b)] along respective the $K-\Gamma-M'$ and $X-\Gamma-Y$ path calculated from magnon's band structure by using the Hamiltonian \eqref{Hneel} and \eqref{Hzigzag}. Notably, with the exception of the somewhat ambiguous effective mass behavior near the inflection points as illustrated in Fig.~\ref{S2} by blue dashed circles, the effective mass amplitude for MnPS$_3$ fluctuates between 0.15 to 3 $\left(\hbar^{2}~ \text{meV}^{-1}\text{Å}^{-2}\right)$, while for NiPSe$3$, $M_{eff} \approx 0.1~ \left(\hbar^{2}~ \text{meV}^{-1}\text{Å}^{-2}\right)$. The effective mass of the other magnon bands in both MnPS$_3$ and NiPSe$_3$ exhibits a comparable range of variation. To simplify the analysis, we have chosen to use $M_{eff} = 0.5~\left(\hbar^{2}~ \text{meV}^{-1}\text{Å}^{-2}\right)$ for MnPS$_3$ and $M_{eff} = 0.1~\left(\hbar^{2}~ \text{meV}^{-1}\text{Å}^{-2}\right)$ for NiPSe$_3$ for all magnon bands listed in the Table~\ref{tab1} to perform the calculations of orbital Nernst conductivity presented in the main text. 

\section{Orbital angular moment of magnon}
\subsection{Representations of magnon Orbital angular moment in BdG basis}

In this section, we derive the expression for magnon orbital angular moment in the magnon Bloch representation used in the main text. This approach allows us to treat the intra-band and inter-band orbital angular moment of magnons on an equal footing. 

Magnons exhibit bosonic quasiparticle behavior and can be considered as a motion of a wavepacket. The quasi-velocity operator $\hat{\boldsymbol{v}}$ describing the motion of this wavepacket is given by the Heisenberg equation of motion:
\begin{equation}
    \hat{\boldsymbol{v}} = -\frac{i}{\hbar} \left[\hat{\boldsymbol{r}},\hat{H}_{\boldsymbol{k}} \right]
    \label{PHeis}
\end{equation}

To obtain the representation of magnon orbital angular moment in the magnon Bloch states, we start with the symmetrized expression of the total orbital angular moment operator for the rotation of a magnon wavepacket that can be written as:
\begin{align}
    \hat{\boldsymbol{L}}=\frac{1}{4}\left(\hat{\boldsymbol{r}} \times \hat{\boldsymbol{v}} - \hat{\boldsymbol{v}} \times \hat{\boldsymbol{r}} \right)
    \label{Loper}
\end{align}
where $\hat{\boldsymbol{r}}$ is the position operator, and the velocity operator is given by Eq.~\eqref{PHeis}.

The matrix elements of orbital angular moment in the BdG basis representation read:
\begin{align}
    \left\langle m\left(\boldsymbol{k} \right)\left\vert \hat{L}^{\alpha}  \right\vert  n\left(\boldsymbol{k}\right)\right\rangle &= \frac{\epsilon_{\beta \gamma}}{4}\left\langle m\left(\boldsymbol{k} \right)\left\vert \left(\hat{r}_{\beta}\hat{v}_{\gamma} - \hat{v}_{\beta}\hat{r}_{\gamma} \right)  \right\vert  n\left(\boldsymbol{k}\right)\right\rangle \\
    &= -\frac{i\epsilon_{\beta \gamma} }{4\hbar}\left\langle m\left(\boldsymbol{k} \right)\left\vert \left(\hat{r}_{\beta}\left[\hat{r}_{\gamma},\hat{H}_{\boldsymbol{k}} \right] - \left[\hat{r}_{\beta},\hat{H}_{\boldsymbol{k}} \right]\hat{r}_{\gamma} \right)  \right\vert  n\left(\boldsymbol{k}\right)\right\rangle \\
    &=-\frac{i \epsilon_{\beta \gamma}}{4\hbar}\left\langle m\left(\boldsymbol{k} \right)\left\vert \left(\hat{r}_{\beta}\hat{r}_{\gamma}\hat{H}_{\boldsymbol{k}}- 2\hat{r}_{\beta}\hat{H}_{\boldsymbol{k}}\hat{r}_{\gamma}   + \hat{H}_{\boldsymbol{k}} \hat{r}_{\beta}\hat{r}_{\gamma} \right)  \right\vert  n\left(\boldsymbol{k}\right)\right\rangle
\end{align}
where $\alpha,\beta,\gamma \equiv x,y,z$. Note that:
\begin{align}
    \hat{H}^{\dagger}_{\boldsymbol{k}} \equiv \hat{H}_{\boldsymbol{k}}
\end{align}
and
\begin{align}
    \left\langle m\left(\boldsymbol{k} \right)\left\vert \hat{r}_{\beta}\hat{r}_{\gamma}\hat{H}_{\boldsymbol{k}}  \right\vert  n\left(\boldsymbol{k}\right)\right\rangle = \left[\boldsymbol{\sigma}_{3}E_{\boldsymbol{k}}\right]_{nn}\left\langle m\left(\boldsymbol{k} \right)\left\vert \hat{r}_{\beta}\hat{r}_{\gamma}\boldsymbol{\sigma}_{3}  \right\vert  n\left(\boldsymbol{k}\right)\right\rangle \\
    \left\langle m\left(\boldsymbol{k} \right)\left\vert \hat{H}_{\boldsymbol{k}}\hat{r}_{\beta}\hat{r}_{\gamma}  \right\vert  n\left(\boldsymbol{k}\right)\right\rangle = \left[\boldsymbol{\sigma}_{3}E_{\boldsymbol{k}}\right]_{mm}\left\langle m\left(\boldsymbol{k} \right)\left\vert \boldsymbol{\sigma}_{3} \hat{r}_{\beta}\hat{r}_{\gamma}  \right\vert  n\left(\boldsymbol{k}\right)\right\rangle.
\end{align}
The action of the position operator on the periodic part of the Bloch magnon function is given by
\begin{equation}
    \hat{\boldsymbol{r}} \left\vert n\left( \boldsymbol{k} \right) \right\rangle = i\left\vert \partial_{\boldsymbol{k}} n\left( \boldsymbol{k} \right) \right\rangle
\end{equation}
and because
\begin{equation}
    \left[ \hat{\boldsymbol{r}}, \boldsymbol{\sigma}_{3}\right] = 0
\end{equation}
one obtains
\begin{align}
    \left\langle m\left(\boldsymbol{k} \right)\left\vert \hat{r}_{\beta}\hat{r}_{\gamma}\hat{H}_{\boldsymbol{k}}  \right\vert  n\left(\boldsymbol{k}\right)\right\rangle = \left[\boldsymbol{\sigma}_{3}E _{\boldsymbol{k}}\right]_{nn}\left\langle m\left(\boldsymbol{k} \right)\left\vert \hat{r}_{\beta} \boldsymbol{\sigma}_{3} \hat{r}_{\gamma}  \right\vert  n\left(\boldsymbol{k}\right)\right\rangle = -  \left[\boldsymbol{\sigma}_{3}E _{\boldsymbol{k}}\right]_{nn}\left\langle \partial_{k_{\beta}}m\left(\boldsymbol{k} \right)\left\vert  \boldsymbol{\sigma}_{3}  \right\vert \partial_{k_{\gamma}}  n\left(\boldsymbol{k}\right)\right\rangle\\
    \left\langle m\left(\boldsymbol{k} \right)\left\vert \hat{H}_{\boldsymbol{k}}\hat{r}_{\beta}\hat{r}_{\gamma}  \right\vert  n\left(\boldsymbol{k}\right)\right\rangle = \left[\boldsymbol{\sigma}_{3}E_{\boldsymbol{k}}\right]_{mm}\left\langle m\left(\boldsymbol{k} \right)\left\vert  \hat{r}_{\beta} \boldsymbol{\sigma}_{3} \hat{r}_{\gamma}  \right\vert  n\left(\boldsymbol{k}\right)\right\rangle = -\left[\boldsymbol{\sigma}_{3}E_{\boldsymbol{k}}\right]_{mm}\left\langle \partial_{k_{\beta}}m\left(\boldsymbol{k} \right)\left\vert  \boldsymbol{\sigma}_{3}  \right\vert \partial_{k_{\gamma}}  n\left(\boldsymbol{k}\right)\right\rangle,
\end{align}
which leads to
\begin{align}
    &\left\langle m\left(\boldsymbol{k} \right)\left\vert \hat{L}^{\alpha}  \right\vert  n\left(\boldsymbol{k}\right)\right\rangle =-\frac{i\epsilon_{\beta \gamma}}{4\hbar}\left\langle m\left(\boldsymbol{k} \right)\left\vert \left(\hat{r}_{\beta}\hat{r}_{\gamma}\hat{H}_{\boldsymbol{k}}- 2\hat{r}_{\beta}\hat{H}_{\boldsymbol{k}}\hat{r}_{\gamma}   + \hat{H}_{\boldsymbol{k}} \hat{r}_{\beta}\hat{r}_{\gamma} \right)  \right\vert  n\left(\boldsymbol{k}\right)\right\rangle \\
    &=\frac{i \epsilon_{\beta \gamma}}{4\hbar}\left\lbrace \left( \left[ \boldsymbol{\sigma}_{3}E_{\boldsymbol{k}} \right]_{nn} + \left[ \boldsymbol{\sigma}_{3}E_{\boldsymbol{k}} \right]_{mm}\right) \left\langle \partial_{k_{\beta}}m\left(\boldsymbol{k} \right)\right\vert \boldsymbol{\sigma}_{3}\left\vert \partial_{k_{\gamma}} n\left(\boldsymbol{k} \right) \right\rangle -2\left\langle \partial_{k_{\beta}}m\left(\boldsymbol{k} \right)\right\vert \hat{H}_{\boldsymbol{k}}\left\vert \partial_{k_{\gamma}} n\left(\boldsymbol{k} \right) \right\rangle \right\rbrace.
\end{align}
We recast the above equation as
\begin{align}
    \left\langle m\left(\boldsymbol{k} \right)\left\vert \hat{\boldsymbol{L}}  \right\vert  n\left(\boldsymbol{k}\right)\right\rangle &=\frac{i}{2\hbar}\left\langle \partial_{\boldsymbol{k}}m\left(\boldsymbol{k} \right) \right\vert \times \hat{H}_{\boldsymbol{k}}\left\vert \partial_{\boldsymbol{k}} n\left(\boldsymbol{k} \right)\right\rangle  \\
    &-\frac{i}{4 \hbar} \left( \left[ \boldsymbol{\sigma}_{3}E_{\boldsymbol{k}} \right]_{nn} + \left[ \boldsymbol{\sigma}_{3}E_{\boldsymbol{k}} \right]_{mm} \right) \left\langle \partial_{\boldsymbol{k}}m\left(\boldsymbol{k} \right) \right\vert \times \boldsymbol{\sigma}_{3}\left\vert \partial_{\boldsymbol{k}} n\left(\boldsymbol{k} \right)\right\rangle
\end{align}
For $n \equiv m$ one obtains
\begin{align}
    \left\langle n\left(\boldsymbol{k} \right)\left\vert \hat{\boldsymbol{L}}  \right\vert  n\left(\boldsymbol{k}\right)\right\rangle &=\frac{i}{2\hbar}\left\langle \partial_{\boldsymbol{k}}n\left(\boldsymbol{k} \right) \left\vert \times \left\lbrace\hat{H}_{\boldsymbol{k}}- \left[ \boldsymbol{\sigma}_{3}E_{\boldsymbol{k}} \right]_{nn} \boldsymbol{\sigma}_{3}\right\rbrace\right\vert \partial_{\boldsymbol{k}} n\left(\boldsymbol{k} \right)\right\rangle,
    \label{nthL}
\end{align}
which is the intrinsic orbital angular moment of magnon in the nth-state.

%%%%%%%%%%%%%%%%%%%%%%%%%%%%%%
\subsection{Relationship between magnon orbital angular moment and magnon Berry curvature}

We now discuss the relationship between orbital angular moment in the magnon Bloch representation and the magnon Berry curvature. To do so, we start with the eigenequation for the magnon system, which reads
\begin{align}
\boldsymbol{\sigma}_{3} \hat{H}_{\boldsymbol{k}}\left\vert n\left(\boldsymbol{k} \right)\right\rangle = \left[\boldsymbol{\sigma}_{3} E_{\boldsymbol{k}} \right]_{nn}\left\vert n\left(\boldsymbol{k} \right)\right\rangle.    
\end{align}
Taking the derivative of both sides with respect to $\partial_{\boldsymbol{k}}$, one obtains

\begin{equation}
    \boldsymbol{\sigma}_{3}\partial_{\boldsymbol{k}}\hat{H}_{\boldsymbol{k}} \left\vert n \left( \boldsymbol{k} \right)\right\rangle +  \boldsymbol{\sigma}_{3}\hat{H}_{\boldsymbol{k}} \left\vert \partial_{\boldsymbol{k}} n \left( \boldsymbol{k} \right)\right\rangle = \left[ \boldsymbol{\sigma}_{3}\partial_{\boldsymbol{k}} E_{\boldsymbol{k}} \right]_{nn} \left\vert n\left( \boldsymbol{k}\right)\right\rangle + \left[ \boldsymbol{\sigma}_{3} E_{\boldsymbol{k}} \right]_{nn} \left\vert \partial_{\boldsymbol{k}} n\left( \boldsymbol{k}\right)\right\rangle,
\end{equation}
Multiplying both sides with $\left\langle p \left(\boldsymbol{k}\right)\right\vert \boldsymbol{\sigma}_{3}$ and using $\boldsymbol{\sigma}_{3}\boldsymbol{\sigma}_{3}=\mathcal{I}$, where $\mathcal{I}$ is the identity matrix, one obtains:

\begin{align}
    \left\langle p\left(\boldsymbol{k} \right)\left\vert \partial_{\boldsymbol{k}} \hat{H}_{\boldsymbol{k}}\right\vert n \left( \boldsymbol{k} \right)\right\rangle + \left\langle p\left(\boldsymbol{k} \right)\left\vert  \hat{H}_{\boldsymbol{k}}\right\vert \partial_{\boldsymbol{k}} n \left( \boldsymbol{k} \right)\right\rangle &= \left[\boldsymbol{\sigma}_{3} \partial_{\boldsymbol{k}} E_{ \boldsymbol{k}} \right]_{nn} \left\langle p\left(\boldsymbol{k} \right)\left\vert \boldsymbol{\sigma}_{3}\right\vert n \left( \boldsymbol{k} \right)\right\rangle \notag \\
    &+ \left[\boldsymbol{\sigma}_{3}  E_{ \boldsymbol{k}} \right]_{nn} \left\langle p\left(\boldsymbol{k} \right)\left\vert \boldsymbol{\sigma}_{3}\right\vert \partial_{\boldsymbol{k}} n \left( \boldsymbol{k} \right)\right\rangle.
\end{align}
Note that $\left\langle p\left(\boldsymbol{k} \right)\left\vert \boldsymbol{\sigma}_{3}\right\vert n \left( \boldsymbol{k} \right)\right\rangle= \left\langle n\left(\boldsymbol{k} \right)\left\vert \boldsymbol{\sigma}_{3}\right\vert n \left( \boldsymbol{k} \right)\right\rangle\delta_{p,n}=\sigma_{3}^{nn}\delta_{p,n}$ and

\begin{equation}
    \left\langle p\left(\boldsymbol{k} \right)\left\vert  \hat{H}_{\boldsymbol{k}}\right\vert \partial_{\boldsymbol{k}} n \left( \boldsymbol{k} \right)\right\rangle= \left[ \boldsymbol{\sigma}_{3}E _{\boldsymbol{k}} \right]_{pp}\left\langle p\left(\boldsymbol{k} \right)\left\vert  \boldsymbol{\sigma}_{3}\right\vert \partial_{\boldsymbol{k}} n \left( \boldsymbol{k} \right)\right\rangle
\end{equation}
so that, for $n \equiv p$, one gets:
\begin{align}
    \sigma_{3}^{nn}\left[\boldsymbol{\sigma}_{3} \partial_{\boldsymbol{k}} E_{\boldsymbol{k}} \right]_{nn}  = \left\langle n\left(\boldsymbol{k} \right)\left\vert \partial_{\boldsymbol{k}} \hat{H}_{\boldsymbol{k}}\right\vert n \left( \boldsymbol{k} \right)\right\rangle,
\end{align}
which is the well-known Hellmann-Feynman theorem. For the case $n \neq p$, one has
\begin{equation}
    \left\langle p\left(\boldsymbol{k} \right)\left\vert \partial_{\boldsymbol{k}} \hat{H}_{\boldsymbol{k}}\right\vert n \left( \boldsymbol{k} \right)\right\rangle = \left\lbrace \left[ \boldsymbol{\sigma}_{3}E_{\boldsymbol{k}} \right]_{nn} - \left[ \boldsymbol{\sigma}_{3}E_{\boldsymbol{k}} \right]_{pp} \right\rbrace\left\langle p\left(\boldsymbol{k} \right)\left\vert  \boldsymbol{\sigma}_{3}\right\vert \partial_{\boldsymbol{k}} n \left( \boldsymbol{k} \right)\right\rangle,
\end{equation}
which leads to

\begin{equation}
    \left\langle p\left(\boldsymbol{k} \right)\left\vert  \boldsymbol{\sigma}_{3}\right\vert \partial_{\boldsymbol{k}} n \left( \boldsymbol{k} \right)\right\rangle = \frac{\left\langle p\left(\boldsymbol{k} \right)\left\vert \partial_{\boldsymbol{k}} \hat{H}_{\boldsymbol{k}}\right\vert n \left( \boldsymbol{k} \right)\right\rangle}{ \left[ \boldsymbol{\sigma}_{3}E_{\boldsymbol{k}} \right]_{nn} - \left[ \boldsymbol{\sigma}_{3}E_{\boldsymbol{k}} \right]_{pp}}.
\end{equation}
Multiplying both sides with $\sigma_{3}^{pp}\left\vert m\left(\boldsymbol{k} \right)\right\rangle$ and taking a sum over all $p$ ($p \neq n$), one has:

\begin{equation}
    \sum_{p \neq n}\sigma_{3}^{pp}\left\vert p\left(\boldsymbol{k} \right)\right\rangle \left\langle p\left(\boldsymbol{k} \right)\left\vert \boldsymbol{\sigma}_{3}\right\vert \partial_{\boldsymbol{k}} n \left( \boldsymbol{k} \right)\right\rangle=\sum_{p \neq n}\sigma_{3}^{pp}\left\vert p\left(\boldsymbol{k} \right)\right\rangle\frac{\left\langle p\left(\boldsymbol{k} \right)\left\vert \partial_{\boldsymbol{k}} \hat{H}_{\boldsymbol{k}}\right\vert n \left( \boldsymbol{k} \right)\right\rangle}{ \left[ \boldsymbol{\sigma}_{3}E_{\boldsymbol{k}} \right]_{nn} - \left[ \boldsymbol{\sigma}_{3}E_{\boldsymbol{k}} \right]_{pp}}
\end{equation}

\begin{equation}
    \left[\sum_{p \neq n}\sigma_{3}^{pp}\left\vert p\left(\boldsymbol{k} \right)\right\rangle \left\langle p\left(\boldsymbol{k} \right)\right\vert \boldsymbol{\sigma}_{3} \right]\left\vert \partial_{\boldsymbol{k}} n \left( \boldsymbol{k} \right)\right\rangle=\sum_{p \neq n}\sigma_{3}^{pp}\left\vert p\left(\boldsymbol{k} \right)\right\rangle\frac{\left\langle p\left(\boldsymbol{k} \right)\left\vert \partial_{\boldsymbol{k}} \hat{H}_{\boldsymbol{k}}\right\vert n \left( \boldsymbol{k} \right)\right\rangle}{ \left[ \boldsymbol{\sigma}_{3}E_{\boldsymbol{k}}\right]_{nn} - \left[ \boldsymbol{\sigma}_{3}E_{\boldsymbol{k}} \right]_{pp}}.
\end{equation}
Using the completeness equation of the BdG Hamiltonian $\sum_{p}\sigma_{3}^{pp}\left\vert p\left(\boldsymbol{k} \right)\right\rangle \left\langle p \left(\boldsymbol{k} \right)\right\vert \boldsymbol{\sigma}_{3}=\mathcal{I}$, one obtains:

\begin{equation}
    \left(-i\mathcal{A}_{n} + \partial_{\boldsymbol{k}} \right)\left\vert n \left( \boldsymbol{k} \right)\right\rangle=\sum_{p\neq n}\sigma_{3}^{pp}\frac{\left\langle p\left(\boldsymbol{k} \right)\left\vert \partial_{\boldsymbol{k}} \hat{H}_{\boldsymbol{k}}\right\vert n \left( \boldsymbol{k} \right)\right\rangle}{ \left[ \boldsymbol{\sigma}_{3}E_{\boldsymbol{k}} \right]_{nn} - \left[ \boldsymbol{\sigma}_{3}E_{\boldsymbol{k}} \right]_{pp}}\left\vert p\left(\boldsymbol{k} \right)\right\rangle
\end{equation}
where $\mathcal{A}_{n}= i\sigma_{3}^{nn}\left\langle n \left( \boldsymbol{k}\right)\left\vert \boldsymbol{\sigma}_{3} \right\vert \partial_{\boldsymbol{k}} n \left( \boldsymbol{k} \right)\right\rangle$ is the Berry connection. A gauge choice of $\mathcal{A}_{n}=0$ can be applied, which leads to:
\begin{equation}
      \left\vert \partial_{\boldsymbol{k}} n \left( \boldsymbol{k} \right)\right\rangle=\sum_{p\neq n}\sigma_{3}^{pp}\frac{\left\langle p\left(\boldsymbol{k} \right)\left\vert \partial_{\boldsymbol{k}} \hat{H}_{\boldsymbol{k}}\right\vert n \left( \boldsymbol{k} \right)\right\rangle}{ \left[ \boldsymbol{\sigma}_{3}E_{\boldsymbol{k}} \right]_{nn} - \left[ \boldsymbol{\sigma}_{3}E_{\boldsymbol{k}} \right]_{pp}}\left\vert p\left(\boldsymbol{k} \right)\right\rangle
      \label{EQN:wavefunction}
\end{equation}
In the same manner, one has
\begin{equation}
    \left\langle \partial_{\boldsymbol{k}} m \left(\boldsymbol{k} \right)\right\vert=\sum_{l\neq m}\sigma_{3}^{ll}\frac{\left\langle m\left(\boldsymbol{k} \right)\left\vert \partial_{\boldsymbol{k}} \hat{H}_{\boldsymbol{k}}\right\vert l \left( \boldsymbol{k} \right)\right\rangle}{ \left[ \boldsymbol{\sigma}_{3}E_{\boldsymbol{k}} \right]_{mm} - \left[ \boldsymbol{\sigma}_{3}E_{\boldsymbol{k}} \right]_{ll}}\left\langle l\left(\boldsymbol{k} \right)\right\vert.
    \label{EQN:wavefunction1}
\end{equation}
Inserting the above relations into the expression of the following term:
\begin{align}
&\left\langle \partial_{\boldsymbol{k}}m\left(\boldsymbol{k} \right) \right\vert \times \hat{H}_{\boldsymbol{k}}\left\vert \partial_{\boldsymbol{k}} n\left(\boldsymbol{k} \right)\right\rangle \notag \\
&= \sum_{l\neq m}\sigma_{3}^{ll}\frac{\left\langle m\left(\boldsymbol{k} \right)\left\vert \partial_{\boldsymbol{k}} \hat{H}_{\boldsymbol{k}}\right\vert l \left( \boldsymbol{k} \right)\right\rangle}{ \left[ \boldsymbol{\sigma}_{3}E_{\boldsymbol{k}} \right]_{mm} - \left[ \boldsymbol{\sigma}_{3}E_{\boldsymbol{k}} \right]_{ll}}\left\langle l\left(\boldsymbol{k} \right)\right\vert \times \hat{H}_{\boldsymbol{k}}\sum_{p\neq n}\sigma_{3}^{pp}\frac{\left\langle p\left(\boldsymbol{k} \right)\left\vert \partial_{\boldsymbol{k}} \hat{H}_{\boldsymbol{k}}\right\vert n \left( \boldsymbol{k} \right)\right\rangle}{ \left[ \boldsymbol{\sigma}_{3}E_{\boldsymbol{k}} \right]_{nn} - \left[ \boldsymbol{\sigma}_{3}E_{\boldsymbol{k}} \right]_{pp}}\left\vert p\left(\boldsymbol{k} \right)\right\rangle \\
&= \sum_{l\neq m}\sigma_{3}^{ll}\frac{\left\langle m\left(\boldsymbol{k} \right)\left\vert \partial_{\boldsymbol{k}} \hat{H}_{\boldsymbol{k}}\right\vert l \left( \boldsymbol{k} \right)\right\rangle}{ \left[ \boldsymbol{\sigma}_{3}E_{\boldsymbol{k}} \right]_{mm} - \left[ \boldsymbol{\sigma}_{3}E_{\boldsymbol{k}} \right]_{ll}}\left\langle l\left(\boldsymbol{k} \right)\right\vert \times \sum_{p\neq n}\sigma_{3}^{pp}\frac{\left\langle p\left(\boldsymbol{k} \right)\left\vert \partial_{\boldsymbol{k}} \hat{H}_{\boldsymbol{k}}\right\vert n \left( \boldsymbol{k} \right)\right\rangle}{ \left[ \boldsymbol{\sigma}_{3}E_{\boldsymbol{k}} \right]_{nn} - \left[ \boldsymbol{\sigma}_{3}E_{\boldsymbol{k}} \right]_{pp}}\hat{H}_{\boldsymbol{k}}\left\vert p\left(\boldsymbol{k} \right)\right\rangle \\
&=\sum_{l\neq m}\sigma_{3}^{ll}\frac{\left\langle m\left(\boldsymbol{k} \right)\left\vert \partial_{\boldsymbol{k}} \hat{H}_{\boldsymbol{k}}\right\vert l \left( \boldsymbol{k} \right)\right\rangle}{ \left[ \boldsymbol{\sigma}_{3}E_{\boldsymbol{k}} \right]_{mm} - \left[ \boldsymbol{\sigma}_{3}E_{\boldsymbol{k}} \right]_{ll}}\left\langle l\left(\boldsymbol{k} \right)\right\vert \times \sum_{p\neq n}\sigma_{3}^{pp}\frac{\left\langle p\left(\boldsymbol{k} \right)\left\vert \partial_{\boldsymbol{k}} \hat{H}_{\boldsymbol{k}}\right\vert n \left( \boldsymbol{k} \right)\right\rangle}{ \left[ \boldsymbol{\sigma}_{3}E_{\boldsymbol{k}} \right]_{nn} - \left[ \boldsymbol{\sigma}_{3}E_{\boldsymbol{k}} \right]_{pp}}\boldsymbol{\sigma}_{3} \left[ \boldsymbol{\sigma}_{3}E_{\boldsymbol{k}} \right]_{pp}\left\vert p\left(\boldsymbol{k} \right)\right\rangle \\
&=\sum_{l\neq m}\sigma_{3}^{ll}\frac{\left\langle m\left(\boldsymbol{k} \right)\left\vert \partial_{\boldsymbol{k}} \hat{H}_{\boldsymbol{k}}\right\vert l \left( \boldsymbol{k} \right)\right\rangle}{ \left[ \boldsymbol{\sigma}_{3}E_{\boldsymbol{k}} \right]_{mm} - \left[ \boldsymbol{\sigma}_{3}E_{\boldsymbol{k}} \right]_{ll}}\left\langle l\left(\boldsymbol{k} \right)\right\vert \times \sum_{p\neq n}\sigma_{3}^{pp}\left[ \boldsymbol{\sigma}_{3}E \left(\boldsymbol{k} \right) \right]_{pp}\frac{\left\langle p\left(\boldsymbol{k} \right)\left\vert \partial_{\boldsymbol{k}} \hat{H}_{\boldsymbol{k}}\right\vert n \left( \boldsymbol{k} \right)\right\rangle}{ \left[ \boldsymbol{\sigma}_{3}E_{\boldsymbol{k}} \right]_{nn} - \left[ \boldsymbol{\sigma}_{3}E_{\boldsymbol{k}} \right]_{pp}}\boldsymbol{\sigma}_{3} \left\vert p\left(\boldsymbol{k} \right)\right\rangle \\
&=\sum_{l\neq m}\sum_{p\neq n}\sigma_{3}^{ll}\sigma_{3}^{pp}\left[ \boldsymbol{\sigma}_{3}E_{\boldsymbol{k}} \right]_{pp}\frac{\left\langle m\left(\boldsymbol{k} \right)\left\vert \partial_{\boldsymbol{k}} \hat{H}_{\boldsymbol{k}}\right\vert l \left( \boldsymbol{k} \right)\right\rangle \times \left\langle p\left(\boldsymbol{k} \right)\left\vert \partial_{\boldsymbol{k}} \hat{H}_{\boldsymbol{k}}\right\vert n \left( \boldsymbol{k} \right)\right\rangle}{\left\lbrace \left[ \boldsymbol{\sigma}_{3}E_{\boldsymbol{k}} \right]_{mm} - \left[ \boldsymbol{\sigma}_{3}E_{\boldsymbol{k}} \right]_{ll} \right\rbrace \left\lbrace \left[ \boldsymbol{\sigma}_{3}E_{\boldsymbol{k}} \right]_{nn} - \left[ \boldsymbol{\sigma}_{3}E_{\boldsymbol{k}} \right]_{pp}\right\rbrace }\left\langle l\left(\boldsymbol{k} \right)\right\vert \boldsymbol{\sigma}_{3} \left\vert p\left(\boldsymbol{k} \right)\right\rangle \\
&=\hbar^{2}\sum_{l\neq m}\sum_{p\neq n}\sigma_{3}^{ll}\sigma_{3}^{pp}\left[ \boldsymbol{\sigma}_{3}E_{\boldsymbol{k}} \right]_{pp}\frac{\left\langle m\left(\boldsymbol{k} \right)\left\vert \boldsymbol{v}\right\vert l \left( \boldsymbol{k} \right)\right\rangle \times \left\langle p\left(\boldsymbol{k} \right)\left\vert \boldsymbol{v}\right\vert n \left( \boldsymbol{k} \right)\right\rangle}{\left\lbrace \left[ \boldsymbol{\sigma}_{3}E_{\boldsymbol{k}} \right]_{mm} - \left[ \boldsymbol{\sigma}_{3}E_{\boldsymbol{k}} \right]_{ll} \right\rbrace \left\lbrace \left[ \boldsymbol{\sigma}_{3}E_{\boldsymbol{k}} \right]_{nn} - \left[ \boldsymbol{\sigma}_{3}E_{\boldsymbol{k}} \right]_{pp}\right\rbrace }\left\langle l\left(\boldsymbol{k} \right)\right\vert \boldsymbol{\sigma}_{3} \left\vert p\left(\boldsymbol{k} \right)\right\rangle.
\end{align}

Similarly, one obtains
\begin{align}
 &\left\langle \partial_{\boldsymbol{k}}m\left(\boldsymbol{k} \right) \right\vert \times \boldsymbol{\sigma}_{3}\left\vert \partial_{\boldsymbol{k}} n\left(\boldsymbol{k} \right)\right\rangle \notag\\
 &= \sum_{l\neq m}\sigma_{3}^{ll}\frac{\left\langle m\left(\boldsymbol{k} \right)\left\vert \partial_{\boldsymbol{k}} \hat{H}_{\boldsymbol{k}}\right\vert l \left( \boldsymbol{k} \right)\right\rangle}{ \left[ \boldsymbol{\sigma}_{3}E_{\boldsymbol{k}} \right]_{mm} - \left[ \boldsymbol{\sigma}_{3}E_{\boldsymbol{k}} \right]_{ll}}\left\langle l\left(\boldsymbol{k} \right)\right\vert \times \boldsymbol{\sigma}_{3}\sum_{p\neq n}\sigma_{3}^{pp}\frac{\left\langle p\left(\boldsymbol{k} \right)\left\vert \partial_{\boldsymbol{k}} \hat{H}_{\boldsymbol{k}}\right\vert n \left( \boldsymbol{k} \right)\right\rangle}{ \left[ \boldsymbol{\sigma}_{3}E_{\boldsymbol{k}} \right]_{nn} - \left[ \boldsymbol{\sigma}_{3}E_{\boldsymbol{k}} \right]_{pp}}\left\vert p\left(\boldsymbol{k} \right)\right\rangle \\
 &=\sum_{l\neq m}\sum_{p\neq n}\sigma_{3}^{ll}\sigma_{3}^{pp}\frac{\left\langle m\left(\boldsymbol{k} \right)\left\vert \partial_{\boldsymbol{k}} \hat{H}_{\boldsymbol{k}}\right\vert l \left( \boldsymbol{k} \right)\right\rangle}{ \left[ \boldsymbol{\sigma}_{3}E_{\boldsymbol{k}} \right]_{mm} - \left[ \boldsymbol{\sigma}_{3}E_{\boldsymbol{k}} \right]_{ll}}\left\langle l\left(\boldsymbol{k} \right)\right\vert \times \boldsymbol{\sigma}_{3}\frac{\left\langle p\left(\boldsymbol{k} \right)\left\vert \partial_{\boldsymbol{k}} \hat{H} \left( \boldsymbol{k} \right)\right\vert n \left( \boldsymbol{k} \right)\right\rangle}{ \left[ \boldsymbol{\sigma}_{3}E_{\boldsymbol{k}} \right]_{nn} - \left[ \boldsymbol{\sigma}_{3}E_{\boldsymbol{k}} \right]_{pp}}\left\vert p\left(\boldsymbol{k} \right)\right\rangle \\
 &=\sum_{l\neq m}\sum_{p\neq n}\sigma_{3}^{ll}\sigma_{3}^{pp}\frac{\left\langle m\left(\boldsymbol{k} \right)\left\vert \partial_{\boldsymbol{k}} \hat{H}_{\boldsymbol{k}}\right\vert l \left( \boldsymbol{k} \right)\right\rangle \times \left\langle p\left(\boldsymbol{k} \right)\left\vert \partial_{\boldsymbol{k}} \hat{H}_{\boldsymbol{k}}\right\vert n \left( \boldsymbol{k} \right)\right\rangle}{\left\lbrace \left[ \boldsymbol{\sigma}_{3}E_{\boldsymbol{k}} \right]_{mm} - \left[ \boldsymbol{\sigma}_{3}E_{\boldsymbol{k}} \right]_{ll} \right\rbrace \left\lbrace \left[ \boldsymbol{\sigma}_{3}E_{\boldsymbol{k}} \right]_{nn} - \left[ \boldsymbol{\sigma}_{3}E_{\boldsymbol{k}} \right]_{pp}\right\rbrace }\left\langle l\left(\boldsymbol{k} \right)\right\vert \boldsymbol{\sigma}_{3} \left\vert p\left(\boldsymbol{k} \right)\right\rangle \\
 &=\hbar^{2}\sum_{l\neq m}\sum_{p\neq n}\sigma_{3}^{ll}\sigma_{3}^{pp}\frac{\left\langle m\left(\boldsymbol{k} \right)\left\vert \hat{\boldsymbol{v}}_{\boldsymbol{k}}\right\vert l \left( \boldsymbol{k} \right)\right\rangle \times \left\langle p\left(\boldsymbol{k} \right)\left\vert \hat{\boldsymbol{v}}_{\boldsymbol{k}}\right\vert n \left( \boldsymbol{k} \right)\right\rangle}{\left\lbrace \left[ \boldsymbol{\sigma}_{3}E_{\boldsymbol{k}} \right]_{mm} - \left[ \boldsymbol{\sigma}_{3}E_{\boldsymbol{k}} \right]_{ll} \right\rbrace \left\lbrace \left[ \boldsymbol{\sigma}_{3}E_{\boldsymbol{k}} \right]_{nn} - \left[ \boldsymbol{\sigma}_{3}E_{\boldsymbol{k}} \right]_{pp}\right\rbrace }\left\langle l\left(\boldsymbol{k} \right)\right\vert \boldsymbol{\sigma}_{3} \left\vert p\left(\boldsymbol{k} \right)\right\rangle
\end{align}
where
\begin{equation}
    \hat{\boldsymbol{v}}_{\boldsymbol{k}}=\frac{1}{\hbar} \partial_{\boldsymbol{k}}\hat{H}\left( \boldsymbol{k}\right)
\end{equation}
is the velocity operator. Note that $\left\langle l\left(\boldsymbol{k} \right)\right\vert \boldsymbol{\sigma}_{3} \left\vert p\left(\boldsymbol{k} \right)\right\rangle= \left\langle p\left(\boldsymbol{k} \right)\right\vert \boldsymbol{\sigma}_{3} \left\vert p\left(\boldsymbol{k} \right)\right\rangle\delta_{p,l}$. Therefore, one gets:

\begin{align}
 &\left\langle \partial_{\boldsymbol{k}}m\left(\boldsymbol{k} \right) \right\vert \times \boldsymbol{\sigma}_{3}\left\vert \partial_{\boldsymbol{k}} n\left(\boldsymbol{k} \right)\right\rangle \notag \\
 &= \hbar^{2}\sum_{p\neq m}\sum_{p\neq n}\sigma_{3}^{pp}\sigma_{3}^{pp}\frac{\left\langle m\left(\boldsymbol{k} \right)\left\vert \hat{\boldsymbol{v}}_{\boldsymbol{k}}\right\vert p \left( \boldsymbol{k} \right)\right\rangle \times \left\langle p\left(\boldsymbol{k} \right)\left\vert \hat{\boldsymbol{v}}_{\boldsymbol{k}}\right\vert n \left( \boldsymbol{k} \right)\right\rangle}{\left\lbrace \left[ \boldsymbol{\sigma}_{3}E_{\boldsymbol{k}} \right]_{mm} - \left[ \boldsymbol{\sigma}_{3}E_{\boldsymbol{k}} \right]_{pp} \right\rbrace \left\lbrace \left[ \boldsymbol{\sigma}_{3}E_{\boldsymbol{k}} \right]_{nn} - \left[ \boldsymbol{\sigma}_{3}E_{\boldsymbol{k}} \right]_{pp}\right\rbrace }\left\langle p\left(\boldsymbol{k} \right)\right\vert \boldsymbol{\sigma}_{3} \left\vert p\left(\boldsymbol{k} \right)\right\rangle \\
 & = \hbar^{2}\sum_{p\neq m,n}\sigma_{3}^{pp}\frac{\left\langle m\left(\boldsymbol{k} \right)\left\vert \hat{\boldsymbol{v}}_{\boldsymbol{k}}\right\vert p \left( \boldsymbol{k} \right)\right\rangle \times \left\langle p\left(\boldsymbol{k} \right)\left\vert \hat{\boldsymbol{v}}_{\boldsymbol{k}}\right\vert n \left( \boldsymbol{k} \right)\right\rangle}{\left\lbrace \left[ \boldsymbol{\sigma}_{3}E_{\boldsymbol{k}} \right]_{mm} - \left[ \boldsymbol{\sigma}_{3}E_{\boldsymbol{k}} \right]_{pp} \right\rbrace \left\lbrace \left[ \boldsymbol{\sigma}_{3}E_{\boldsymbol{k}} \right]_{nn} - \left[ \boldsymbol{\sigma}_{3}E_{\boldsymbol{k}} \right]_{pp}\right\rbrace }
\end{align}
leads to
\begin{align}
 \left\langle \partial_{\boldsymbol{k}}m\left(\boldsymbol{k} \right) \right\vert \times \boldsymbol{\sigma}_{3}\left\vert \partial_{\boldsymbol{k}} n\left(\boldsymbol{k} \right)\right\rangle = \hbar^{2}\sum_{p\neq m,n}\sigma_{3}^{pp}\frac{\left\langle m\left(\boldsymbol{k} \right)\left\vert \hat{\boldsymbol{v}}_{\boldsymbol{k}}\right\vert p \left( \boldsymbol{k} \right)\right\rangle \times \left\langle p\left(\boldsymbol{k} \right)\left\vert \hat{\boldsymbol{v}}_{\boldsymbol{k}}\right\vert n \left( \boldsymbol{k} \right)\right\rangle}{\left\lbrace \left[ \boldsymbol{\sigma}_{3}E_{\boldsymbol{k}} \right]_{mm} - \left[ \boldsymbol{\sigma}_{3}E_{\boldsymbol{k}}\right]_{pp} \right\rbrace \left\lbrace \left[ \boldsymbol{\sigma}_{3}E_{\boldsymbol{k}}\right]_{nn} - \left[ \boldsymbol{\sigma}_{3}E_{\boldsymbol{k}}\right]_{pp}\right\rbrace }
\end{align}
where we have used $\sigma_{3}^{pp}\sigma_{3}^{pp}=1$ and $\left\langle p\left(\boldsymbol{k} \right)\right\vert \boldsymbol{\sigma}_{3} \left\vert p\left(\boldsymbol{k} \right)\right\rangle=\sigma_{3}^{pp}$.

Similarly, one obtains:
\begin{align}
\left\langle \partial_{\boldsymbol{k}}m\left(\boldsymbol{k} \right) \right\vert \times \hat{H}_{\boldsymbol{k}}\left\vert \partial_{\boldsymbol{k}} n\left(\boldsymbol{k} \right)\right\rangle =\hbar^{2}\sum_{p\neq m,n}\sigma_{3}^{pp} \left[ \boldsymbol{\sigma}_{3}E_{\boldsymbol{k}} \right]_{pp}\frac{\left\langle m\left(\boldsymbol{k} \right)\left\vert \hat{\boldsymbol{v}}_{\boldsymbol{k}}\right\vert p \left( \boldsymbol{k} \right)\right\rangle \times \left\langle p\left(\boldsymbol{k} \right)\left\vert \hat{\boldsymbol{v}}_{\boldsymbol{k}}\right\vert n \left( \boldsymbol{k} \right)\right\rangle}{\left\lbrace \left[ \boldsymbol{\sigma}_{3}E_{\boldsymbol{k}} \right]_{mm} - \left[ \boldsymbol{\sigma}_{3}E_{\boldsymbol{k}} \right]_{pp} \right\rbrace \left\lbrace \left[ \boldsymbol{\sigma}_{3}E_{\boldsymbol{k}} \right]_{nn} - \left[ \boldsymbol{\sigma}_{3}E_{\boldsymbol{k}}\right]_{pp}\right\rbrace }
\end{align}
Inserting the above relations into the expression of the magnon orbital angular moment, we obtain:
\begin{align}
    &\left\langle m\left(\boldsymbol{k} \right)\left\vert \hat{\boldsymbol{L}}  \right\vert  n\left(\boldsymbol{k}\right)\right\rangle = \frac{i\hbar^{2}}{2\hbar}\left\lbrace\sum_{p\neq m,n}\sigma_{3}^{pp} \left[ \boldsymbol{\sigma}_{3}E_{\boldsymbol{k}} \right]_{pp}\frac{\left\langle m\left(\boldsymbol{k} \right)\left\vert \hat{\boldsymbol{v}}_{\boldsymbol{k}}\right\vert p \left( \boldsymbol{k} \right)\right\rangle \times \left\langle p\left(\boldsymbol{k} \right)\left\vert \hat{\boldsymbol{v}}_{\boldsymbol{k}}\right\vert n \left( \boldsymbol{k} \right)\right\rangle}{\left\lbrace \left[ \boldsymbol{\sigma}_{3}E_{\boldsymbol{k}} \right]_{mm} - \left[ \boldsymbol{\sigma}_{3}E_{\boldsymbol{k}}\right]_{pp} \right\rbrace \left\lbrace \left[ \boldsymbol{\sigma}_{3}E_{\boldsymbol{k}}\right]_{nn} - \left[ \boldsymbol{\sigma}_{3}E_{\boldsymbol{k}}\right]_{pp}\right\rbrace } \right\rbrace \\
    &-\frac{i\hbar^{2}}{4 \hbar} \left\lbrace \sum_{p\neq m,n}\sigma_{3}^{pp} \left( \left[ \boldsymbol{\sigma}_{3}E_{\boldsymbol{k}} \right]_{nn} + \left[ \boldsymbol{\sigma}_{3}E_{\boldsymbol{k}} \right]_{mm} \right)\frac{\left\langle m\left(\boldsymbol{k} \right)\left\vert \hat{\boldsymbol{v}}_{\boldsymbol{k}}\right\vert p \left( \boldsymbol{k} \right)\right\rangle \times \left\langle p\left(\boldsymbol{k} \right)\left\vert \hat{\boldsymbol{v}}_{\boldsymbol{k}}\right\vert n \left( \boldsymbol{k} \right)\right\rangle}{\left\lbrace \left[ \boldsymbol{\sigma}_{3}E_{\boldsymbol{k}}\right]_{mm} - \left[ \boldsymbol{\sigma}_{3}E_{\boldsymbol{k}} \right]_{pp} \right\rbrace \left\lbrace \left[ \boldsymbol{\sigma}_{3}E_{\boldsymbol{k}} \right]_{nn} - \left[ \boldsymbol{\sigma}_{3}E_{\boldsymbol{k}} \right]_{pp}\right\rbrace } \right\rbrace \\
    &= -\frac{i\hbar }{4}\left\lbrace  \sum_{p\neq m,n} \frac{\sigma_{3}^{pp} \left( \left[ \boldsymbol{\sigma}_{3}E_{\boldsymbol{k}} \right]_{nn} + \left[ \boldsymbol{\sigma}_{3}E_{\boldsymbol{k}}\right]_{mm} - 2\left[ \boldsymbol{\sigma}_{3}E_{\boldsymbol{k}} \right]_{pp} \right)\left\langle m\left(\boldsymbol{k} \right)\left\vert \hat{\boldsymbol{v}}_{\boldsymbol{k}}\right\vert p \left( \boldsymbol{k} \right)\right\rangle \times \left\langle p\left(\boldsymbol{k} \right)\left\vert \hat{\boldsymbol{v}}_{\boldsymbol{k}}\right\vert n \left( \boldsymbol{k} \right)\right\rangle}{\left\lbrace \left[ \boldsymbol{\sigma}_{3}E_{\boldsymbol{k}} \right]_{mm} - \left[ \boldsymbol{\sigma}_{3}E_{\boldsymbol{k}} \right]_{pp} \right\rbrace \left\lbrace \left[ \boldsymbol{\sigma}_{3}E_{\boldsymbol{k}}\right]_{nn} - \left[ \boldsymbol{\sigma}_{3}E_{\boldsymbol{k}} \right]_{pp}\right\rbrace } \right\rbrace \\
    &=-\frac{i\hbar }{4}\left\lbrace  \sum_{p\neq m,n}\sigma_{3}^{pp}\left(\frac{1}{\left[ \boldsymbol{\sigma}_{3}E_{\boldsymbol{k}} \right]_{mm} - \left[ \boldsymbol{\sigma}_{3}E_{\boldsymbol{k}} \right]_{pp}} + \frac{1}{\left[ \boldsymbol{\sigma}_{3}E_{\boldsymbol{k}} \right]_{nn} - \left[ \boldsymbol{\sigma}_{3}E_{\boldsymbol{k}} \right]_{pp}}\right)   \left\langle m\left(\boldsymbol{k} \right)\left\vert \hat{\boldsymbol{v}}_{\boldsymbol{k}}\right\vert p \left( \boldsymbol{k} \right)\right\rangle \times \left\langle p\left(\boldsymbol{k} \right)\left\vert \hat{\boldsymbol{v}}_{\boldsymbol{k}}\right\vert n \left( \boldsymbol{k} \right)\right\rangle \right\rbrace
\end{align}
which can be recast as
\begin{align}
    \boldsymbol{L}_{mn}\left(\boldsymbol{k}\right)=\left\langle m\left(\boldsymbol{k} \right)\left\vert \hat{\boldsymbol{L}}  \right\vert  n\left(\boldsymbol{k}\right)\right\rangle =-i\hbar \boldsymbol{\mathcal{N}}_{mn}
\end{align}
where 
\begin{align}
    \boldsymbol{\mathcal{N}}_{mn}=\frac{1}{4} \sum_{p\neq m,n}\sigma_{3}^{pp}&\left(\frac{1}{\left[ \boldsymbol{\sigma}_{3}E \left(\boldsymbol{k} \right) \right]_{mm} - \left[ \boldsymbol{\sigma}_{3}E_{\boldsymbol{k}} \right]_{pp}} + \frac{1}{\left[ \boldsymbol{\sigma}_{3}E_{\boldsymbol{k}} \right]_{nn} - \left[ \boldsymbol{\sigma}_{3}E_{\boldsymbol{k}} \right]_{pp}}\right) \times \notag \\
    &\times \left\langle m\left(\boldsymbol{k} \right)\left\vert \hat{\boldsymbol{v}}_{\boldsymbol{k}}\right\vert p \left( \boldsymbol{k} \right)\right\rangle \times \left\langle p\left(\boldsymbol{k} \right)\left\vert \hat{\boldsymbol{v}}_{\boldsymbol{k}}\right\vert n \left( \boldsymbol{k} \right)\right\rangle. 
    \label{Ntensor}
\end{align}
The intra-band magnon orbital angular moment is given by:
\begin{align}
    \boldsymbol{L}_{nn}\left( \boldsymbol{k}\right)=-i\hbar\boldsymbol{\mathcal{N}}_{nn} 
\end{align}
where 
\begin{equation}
    \boldsymbol{\mathcal{N}}_{nn} = \frac{1}{2}\sum_{p\neq n}\sigma_{3}^{pp}\frac{\left\langle n\left(\boldsymbol{k} \right)\left\vert \hat{\boldsymbol{v}}_{\boldsymbol{k}}\right\vert p \left( \boldsymbol{k} \right)\right\rangle \times \left\langle p\left(\boldsymbol{k} \right)\left\vert \hat{\boldsymbol{v}}_{\boldsymbol{k}}\right\vert n \left( \boldsymbol{k} \right)\right\rangle }{\left[ \boldsymbol{\sigma}_{3}E_{\boldsymbol{k}} \right]_{nn} - \left[ \boldsymbol{\sigma}_{3}E_{\boldsymbol{k}} \right]_{pp}}   
\end{equation}
thereby completing the derivation of Eq.~(4), which shows the deep connections to the magnon Berry curvature in Eq.~(2) in the main text. 

The connection between intra-band magnon orbital angular moment and Berry curvature offers a fascinating shortcut. Much like Berry curvature, intra-band magnon orbital angular moment exhibits analogous behaviors under effective parity time symmetry (PTS) and effective time reversal symmetry (TRS) operations. In this context, PTS imposes a constraint, rendering the intra-band magnon orbital angular moment zero, while TRS introduces an intriguing peculiarity: the intra-band magnon orbital angular moment becomes an odd function with respect to wavevector \cite{To2023b}. This phenomenon is particularly noteworthy in systems such as the 2D honeycomb antiferromagnetic structure with N\'eel order. Here, intra-band magnon orbital angular moment can manifest even in the absence of Dzyaloshinskii-Moriya interaction, courtesy of the self-broken PTS arising from the lack of an inversion center [see Fig.~\ref{S1}(b)]. In contrast, the Zigzag pattern's presence of PTS due to the existence of the inversion center $I_{c}$, as presented in Fig.~\ref{S1}(c), underscores the indispensability of DMI for observing nonvanishing intra-band magnon orbital angular moment. Hence, we see that DMI is essential to the observation of nontrivial intra-band magnon orbital angular moment in systems exhibiting Zigzag order. 

The odd parity of intra-band magnon orbital angular moment with respect to the wavevector in 2D honeycomb systems with N\'eel order results in zero contribution to electric polarization in the absence of DMI. Similarly, in the 2D honeycomb antiferromagnetic structure with Zigzag order, the lack of DMI causes the intra-band magnon orbital angular moment to vanish, thereby leading to zero intra-band magnon orbital angular moment induced electric polarization. Consequently, we see that DMI plays a crucial role in enabling magnon OAM to contribute to electric polarization in 2D honeycomb AFMs with either N\'eel or Zigzag magnetic configuration, as discussed in the main text.

In a continued analogy to Berry curvature, intra-band magnon orbital angular moment increases at anti-crossing points or band extrema. This suggests that materials with enhanced hybridization between different bands would exhibit substantial intra-band magnon orbital angular moment and magnon ONE, consequently resulting in a larger observable quantity such as electric polarization.

Finally, the total magnon orbital angular moment tensor $\boldsymbol{L}$, which encompasses both intra- and inter-band orbital angular momentum, is expressed as
\begin{equation}
    \boldsymbol{L}\left(\boldsymbol{k} \right) = -i\hbar \boldsymbol{\mathcal{N}}
    \label{totalL}
\end{equation}
where $\boldsymbol{\mathcal{N}}$ is a rank 3 tensor whose components and elements are given by Eq.~\eqref{Ntensor}. The expression \eqref{totalL} serves as the foundation for computing the magnon orbital Berry curvature, the magnon orbital Nernst current, and the magnon induced electric polarization, as fully derived in the subsequent sections through the application of linear response theory and perturbation theory.

\section{Magnon-induced electric polarization: A perturbation theory approach}
In this section we provide a comprehensive derivation of the formula used in the main text for electric polarization induced by motions of magnon wave packet. This idea stems from the duality between electric and magnetic fields, offering a shortcut from electron-based mechanisms to those involving magnons \cite{Mook2018,Neumann2023}. Notably, the orbital moment of electrons is known to contribute to magnetization \cite{Thonhauser2005,Xiao2005,Shi2007}. Building upon this understanding, it is reasonable to anticipate that the orbital moment of magnons, along with their finite magnetic dipoles, could similarly influence the electric polarization. This insight opens avenues for detecting the orbital angular moment of magnons and exploring phenomena such as the Orbital Nernst effect of magnons, which are the focus of our study.

We now employ standard perturbation theory to derive the formula for the magnon-induced electric polarization discussed in the main text. To do so, we suppose that the system is in the ground states and calculate the energy correction due to a uniform electric field. We consider noninteracting magnons in which the grand-canonical partition function is given by
\begin{equation}
    \Omega = E-TS + \mu N
\end{equation}
where E is the energy density, T is the temperature and S is the entropy of the system, $\mu$ is the chemical potential and $N$ is the total number of particles. 

The electric polarization is given by
\begin{equation}
    \boldsymbol{P} = -\frac{1}{V_{0}}\left( \frac{\partial \Omega}{\partial \boldsymbol{\Xi}} \right)_{T}.
\end{equation}
Because the zero-temperature electric polarization 
\begin{equation}
    \boldsymbol{\tilde{P}} = -\frac{1}{V_{0}}\left( \frac{\partial E}{\partial \boldsymbol{\Xi}} \right)_{T}
    \label{Porlari}
\end{equation}
is related to the finite temperature electric polarization through the relationship
\begin{equation}
    \boldsymbol{\tilde{P}} = \frac{\partial\left(\beta \boldsymbol{P}\right)}{\partial \beta}
    \label{Rela}
\end{equation}
where $\beta = \frac{1}{k_{B}T}$, in the following we only compute $\boldsymbol{\tilde{P}}$ and then infer $\boldsymbol{P}$ from relationship \eqref{Rela}. To derive the expression for $\boldsymbol{\tilde{P}}$ involving the quantum mechanical wavefunction and energy dispersion, we represent the relation \eqref{Porlari} as
\begin{equation}
    \delta E = \int d\boldsymbol{r} \delta E \left(\boldsymbol{r} \right) = -\int d\boldsymbol{r} \boldsymbol{\tilde{P}} \left( \boldsymbol{r} \right) \boldsymbol{\Xi} \left( \boldsymbol{r}\right)
\end{equation}
where 
\begin{equation}
    \delta E \left(\boldsymbol{r} \right) = \boldsymbol{\tilde{P}} \left( \boldsymbol{r} \right) \boldsymbol{\Xi} \left( \boldsymbol{r}\right)
    \label{localen}
\end{equation} 
is the local change of the energy induced by external electric field.

We start from the single-particle Hamiltonian:
\begin{equation}
    \hat{H} = \hat{H}^{0} + \hat{V}_{\boldsymbol{\Xi}}
\end{equation}
where $\hat{H}_{0}$ is the unperturbed Hamiltonian that yields the band dispersion $E_{n,\boldsymbol{k}}$ and corresponding Bloch wave function $ \psi_{n,\boldsymbol{k}}\left( \boldsymbol{r} \right)  = \left\langle \boldsymbol{r}\left\vert \psi_{n,\boldsymbol{k}}\right.\right\rangle=e^{i\boldsymbol{k.r}}u_{n,\boldsymbol{k}}\left(\boldsymbol{r} \right)$. 

We will now calculate the perturbed electric potential energy of the magnons $\hat{V}_{\boldsymbol{\Xi}}$. The local charge polarization induced by a magnon wavepacket with velocity $\boldsymbol{v}$ carrying magnetic dipole moment $\boldsymbol{m}=g\mu_{B}\boldsymbol{S}$ is described by the following operator \cite{Neumann2023}:
\begin{equation}
    \hat{\boldsymbol{\mathrm{p}}} = \frac{\hat{\boldsymbol{v}}\times \hat{\boldsymbol{m}} - \hat{\boldsymbol{m}}\times \hat{\boldsymbol{v}}}{2V_{0}c^{2}}
\end{equation}
where $V_{0}$ represents the volume of the system, and $c$ is the speed of light in vacuum. Under an externally applied electric field $\boldsymbol{\Xi}$ which is generally non-uniform, the perturbing electric potential energy of the magnon wave packet associated with this charge polarization is expressed as:
\begin{equation}
    \hat{V}_{\boldsymbol{\Xi}}= \frac{g\mu_{B}}{V_{0}c^{2}}\int \left(\hat{\boldsymbol{v}} \times \hat{\boldsymbol{S}} - \hat{\boldsymbol{S}} \times \hat{\boldsymbol{v}} \right) \cdot \boldsymbol{\Xi} dV = \frac{g\mu_{B}}{V_{0}c^{2}}\int\left[\hat{\boldsymbol{v}}\cdot \left(\hat{\boldsymbol{S}} \times \hat{\boldsymbol{e}}_{\boldsymbol{\Xi}} \right) -  \left( \hat{\boldsymbol{e}}_{\boldsymbol{\Xi}}\times \hat{\boldsymbol{S}} \right)\cdot \hat{\boldsymbol{v}} \right]\Xi dV
    \label{pertupoten}
\end{equation}
where $\hat{\boldsymbol{e}}_{\boldsymbol{\Xi}}$ is the unit vector along $\boldsymbol{\Xi}$ direction; $\hat{\boldsymbol{v}}=-\frac{i}{\hbar}\left[\hat{H},\hat{\boldsymbol{r}} \right]$ is the velocity operator, $\hat{\boldsymbol{S}}$ is the spin operator. 

Assuming that the motion of a magnon wave packet is driven by a temperature gradient along the x-axis, we will now calculate the electric polarization induced by magnon transport in the y-direction. In principle, one could evaluate the electric polarization using standard quantum mechanical perturbation theory to calculate the energy correction from a uniform electric field. However, this method encounters difficulties due to the nonlocal nature of the electric polarization operator with Bloch wave functions. To circumvent this issue, we consider an external electric field along the y-direction with an infinitely slow in-plane spatial variation given by:
\begin{equation}
    \boldsymbol{\Xi} = \Xi_{0} ~ cos\left( q_{y} y\right) \hat{\boldsymbol{y}}
    \label{electricfield}
\end{equation}
with $q_{y}$ being small. This leads to

\begin{equation}
    \hat{\boldsymbol{S}} \times \hat{\boldsymbol{e}}_{\boldsymbol{\Xi}} = -\hat{S}^{z}\hat{\boldsymbol{x}} + \hat{S}^{x}\hat{\boldsymbol{z}}
\end{equation}
where $\boldsymbol{\hat{x}}$, $\boldsymbol{\hat{y}}$ and $\boldsymbol{\hat{z}}$ are, respectively, the unit vectors along the x-axis, y-axis and z-axis. 

Since the external electric field varies slowly in space, the corrections due to changes in the wave function under this variation are of a higher order and can be neglected. This allows us to obtain the perturbing potential of the magnon wave packet at position $y$ by inserting \eqref{electricfield} into \eqref{pertupoten} which leads to
\begin{align}
    \hat{V}_{\boldsymbol{\Xi}}&= -\frac{g\mu_{B}\Xi_{0}}{W_{y}c^{2}} \left[\hat{\boldsymbol{x}}\left(\hat{\boldsymbol{v}} \hat{S}^{z} + \hat{S}^{z} \hat{\boldsymbol{v}} \right) - \hat{\boldsymbol{z}}\left(\hat{\boldsymbol{v}} \hat{S}^{x} + \hat{S}^{x} \hat{\boldsymbol{v}} \right)\right]  \int_{0}^{y} cos\left( q_{y} y\right) dy\\
    &= -\frac{g\mu_{B}\Xi_{0} sin\left(q_{y}y \right) \hat{\boldsymbol{x}} }{q_{y}W_{y}c^{2}} \left(\hat{\boldsymbol{v}} \hat{S}^{z} + \hat{S}^{z} \hat{\boldsymbol{v}} \right) + \frac{g\mu_{B}\Xi_{0} sin\left(q_{y}y \right) \hat{\boldsymbol{z}} }{q_{y}W_{y}c^{2}} \left(\hat{\boldsymbol{v}} \hat{S}^{x} + \hat{S}^{x} \hat{\boldsymbol{v}} \right) \\
    &=\hat{V}_{\boldsymbol{\Xi}}^{(1)} +\hat{V}_{\boldsymbol{\Xi}}^{(2)} \label{V12}
    \end{align}  
where 
\begin{align}
    \hat{V}_{\boldsymbol{\Xi}}^{(1)} = -\frac{g\mu_{B}\Xi_{0} sin\left(q_{y}y \right) \hat{\boldsymbol{x}} }{q_{y}W_{y}c^{2}} \left(\hat{\boldsymbol{v}} \hat{S}^{z} + \hat{S}^{z} \hat{\boldsymbol{v}} \right) \\
    \hat{V}_{\boldsymbol{\Xi}}^{(2)} = \frac{g\mu_{B}\Xi_{0} sin\left(q_{y}y \right) \hat{\boldsymbol{z}} }{q_{y}W_{y}c^{2}} \left(\hat{\boldsymbol{v}} \hat{S}^{x} + \hat{S}^{x} \hat{\boldsymbol{v}} \right)
\end{align}
and $W_{y}$ represent the length of the system along the y-direction.

The grand-canonical ensemble energy density
\begin{equation}
    E\left( \boldsymbol{r} \right) = \sum_{n,\boldsymbol{k}}\rho_{n,\boldsymbol{k}} \sigma_{3}^{nn} Re \left\lbrace\psi_{n,\boldsymbol{k}}^{*}\left( \boldsymbol{r} \right)\hat{H}  \psi_{n,\boldsymbol{k}}\left( \boldsymbol{r} \right) \right\rbrace
\end{equation}
where $\rho_{n,\boldsymbol{k}} = [e^{E_{n,\boldsymbol{k}}/k_{B}T}-1]^{-1}$ is the Bose-Einstein distribution function that describes the occupation number of single magnon states of band index n and momentum $\boldsymbol{k}$. The variation in total energy $\delta E$ up to the first order is given by
\begin{align}
    \delta E\left( \boldsymbol{r} \right) = &\sum_{n,\boldsymbol{k}}\left\lbrace \delta\rho_{n,\boldsymbol{k}} \sigma_{3}^{nn}\psi_{n,\boldsymbol{k}}^{*}\left( \boldsymbol{r} \right)\hat{H}^{0}  \psi_{n,\boldsymbol{k}}\left( \boldsymbol{r} \right)  + \rho_{n,\boldsymbol{k}} \sigma_{3}^{nn} \psi_{n,\boldsymbol{k}}^{*}\left( \boldsymbol{r} \right) \hat{V}_{\boldsymbol{\Xi}} \psi_{n,\boldsymbol{k}}\left( \boldsymbol{r} \right) \right. \notag \\
    &+ \left.\rho_{n,\boldsymbol{k}} \sigma_{3}^{nn} \left[ \delta\psi_{n,\boldsymbol{k}}^{*}\left( \boldsymbol{r} \right) \hat{H}^{0}  \psi_{n,\boldsymbol{k}} \left( \boldsymbol{r} \right)  +  \psi_{n,\boldsymbol{k}}^{*}\left( \boldsymbol{r} \right)\hat{H}^{0}  \delta\psi_{n,\boldsymbol{k}} \left( \boldsymbol{r} \right) \right]\right\rbrace.
\end{align}
The first two terms vanish so that 
\begin{align}
    \delta E\left( \boldsymbol{r} \right) = \sum_{n,\boldsymbol{k}}  \rho_{n,\boldsymbol{k}} \sigma_{3}^{nn}\left[ \delta\psi_{n,\boldsymbol{k}}^{*}\left( \boldsymbol{r} \right) \hat{H}^{0}  \psi_{n,\boldsymbol{k}} \left( \boldsymbol{r} \right)  +  \psi_{n,\boldsymbol{k}}^{*}\left( \boldsymbol{r} \right)\hat{H}^{0}  \delta\psi_{n,\boldsymbol{k}} \left( \boldsymbol{r} \right) \right].
\end{align}
in another words, only the change in the wave function will contribute to the electric polarization. 

In the same spirit as Eq.~\eqref{EQN:wavefunction}, one obtains
\begin{align}
    \left\vert \delta \psi_{n,\boldsymbol{k}}\right\rangle &= \sum_{p\neq n,\boldsymbol{k}'} \sigma_{3}^{pp}\frac{\left\langle \psi_{p,\boldsymbol{k}'}\left\vert \delta \hat{H}\right\vert\psi_{n,\boldsymbol{k}}\right\rangle}{\left[ \boldsymbol{\sigma}_{3}E_{\boldsymbol{k}} \right]_{nn} - \left[ \boldsymbol{\sigma}_{3}E_{\boldsymbol{k}'} \right]_{pp}}\left\vert \psi_{p,\boldsymbol{k}'}\right\rangle + \sum_{n,\boldsymbol{k}' \neq \boldsymbol{k}} \sigma_{3}^{nn}\frac{\left\langle \psi_{n,\boldsymbol{k}'}\left\vert \delta \hat{H}\right\vert\psi_{n,\boldsymbol{k}}\right\rangle}{\left[ \boldsymbol{\sigma}_{3}E_{\boldsymbol{k}}  \right]_{nn} - \left[ \boldsymbol{\sigma}_{3}E_{\boldsymbol{k}'} \right]_{nn}}\left\vert \psi_{p,\boldsymbol{k}'}\right\rangle \notag\\
    &=\sum_{p\neq n,\boldsymbol{k}'} \sigma_{3}^{pp}\frac{\left\langle \psi_{p,\boldsymbol{k}'}\left\vert \hat{V}_{\boldsymbol{\Xi}}\right\vert\psi_{n,\boldsymbol{k}}\right\rangle}{\left[ \boldsymbol{\sigma}_{3}E_{\boldsymbol{k}} \right]_{nn} - \left[ \boldsymbol{\sigma}_{3}E_{\boldsymbol{k}'} \right]_{pp}}\left\vert \psi_{p,\boldsymbol{k}'}\right\rangle + \sum_{n,\boldsymbol{k}' \neq \boldsymbol{k}} \sigma_{3}^{nn}\frac{\left\langle \psi_{n,\boldsymbol{k}'}\left\vert \hat{V}_{\boldsymbol{\Xi}}\right\vert\psi_{n,\boldsymbol{k}}\right\rangle}{\left[ \boldsymbol{\sigma}_{3}E_{\boldsymbol{k}} \right]_{nn} - \left[ \boldsymbol{\sigma}_{3}E_{\boldsymbol{k}'} \right]_{nn}}\left\vert \psi_{p,\boldsymbol{k}'}\right\rangle.
\end{align}

The matrix elements of the perturbation $\hat{V}_{\boldsymbol{\Xi}}$ read:
\begin{align}
    &\left\langle \psi_{p,\boldsymbol{k}'}\left\vert \hat{V}_{\boldsymbol{\Xi}}\right\vert \psi_{n,\boldsymbol{k}}\right\rangle = \frac{1}{V_{0}}\int d\boldsymbol{r} \psi_{p,\boldsymbol{k}'}^{*}\left( \boldsymbol{r} \right)\hat{V}_{\boldsymbol{\Xi}} \psi_{n,\boldsymbol{k}}\left( \boldsymbol{r}\right) = \frac{1}{V_{0}}\int d\boldsymbol{r} e^{i\left(\boldsymbol{k+q-k'} \right)\boldsymbol{r}} u_{p,\boldsymbol{k}'}^{*}\left( \boldsymbol{r} \right)\hat{V}_{\boldsymbol{\Xi}} u_{n,\boldsymbol{k}}\left( \boldsymbol{r}\right) \notag\\
    &= -\frac{g\mu_{B}\Xi_{0}\hat{\boldsymbol{x}}}{4iq_{y}V_{0}W_{y}c^{2}}\int d \boldsymbol{r}\left[e^{i\left(\boldsymbol{k+q-k'}\right)\boldsymbol{r}} - e^{-i\left(\boldsymbol{k-q-k'}\right)\boldsymbol{r}} \right]  u_{p,\boldsymbol{k}'}\left(\boldsymbol{r} \right)\left(\hat{\boldsymbol{v}}_{\boldsymbol{k}'} \boldsymbol{\sigma}_{3} \hat{S}^{z} + \hat{S}^{z} \boldsymbol{\sigma}_{3}\hat{\boldsymbol{v}}_{\boldsymbol{k}} \right) u_{n,\boldsymbol{k}} \left(\boldsymbol{r} \right) - \left(x \leftrightarrow z \right) \notag \\
     &= -\frac{g\mu_{B}\Xi_{0}\hat{\boldsymbol{x}}}{4iq_{y}V_{0}W_{y}c^{2}} \int d\boldsymbol{r}\left[e^{i\left(\boldsymbol{k+q-k'}\right)\boldsymbol{r}}  u_{p,\boldsymbol{k}'}\left(\boldsymbol{r} \right) \left(\hat{\boldsymbol{v}}_{\boldsymbol{k}'}\boldsymbol{\sigma}_{3} \hat{S}^{z} + \hat{S}^{z}\boldsymbol{\sigma}_{3}\hat{\boldsymbol{v}}_{\boldsymbol{k}} \right) u_{n,\boldsymbol{k}}\left(\boldsymbol{r} \right) - \left(\boldsymbol{q \rightarrow -q} \right) \right] - \left(x \leftrightarrow z \right)\notag \\
     &= -\frac{g\mu_{B}\Xi_{0}\hat{\boldsymbol{x}}}{4iq_{y}W_{y}c^{2}} \left[\delta_{\boldsymbol{k',k+q}} \left\langle u_{p,\boldsymbol{k}'}\left\vert \left(\hat{\boldsymbol{v}}_{\boldsymbol{k}'} \boldsymbol{\sigma}_{3}\hat{S}^{z} + \hat{S}^{z} \boldsymbol{\sigma}_{3}\hat{\boldsymbol{v}}_{\boldsymbol{k}} \right)\right\vert u_{n,\boldsymbol{k}}\right\rangle - \left(\boldsymbol{q \rightarrow -q} \right) \right] - \left(x \leftrightarrow z \right)
\end{align}
where we have used $sin\left(q_{y}y\right)=sin\left( \boldsymbol{qr}\right)=\frac{e^{i\boldsymbol{qr}} - e^{-i\boldsymbol{qr}}}{2i}$ with $\boldsymbol{q}=\left[0,q_{y},0\right]^{T}$, i.e the wave vector $\boldsymbol{q}$ is along the y-direction and $\left\langle u_{p,\boldsymbol{k}'}\left\vert \left(\hat{\boldsymbol{v}}_{\boldsymbol{k}'}\boldsymbol{\sigma}_{3} \hat{S}^{z} + \hat{S}^{z}\boldsymbol{\sigma}_{3} \hat{\boldsymbol{v}}_{\boldsymbol{k}} \right)\right\vert u_{n,\boldsymbol{k}}\right\rangle = \frac{1}{V_{0}}\int d\boldsymbol{r}~u_{p,\boldsymbol{k}'}\left(\boldsymbol{r} \right) \left(\hat{\boldsymbol{v}}_{\boldsymbol{k}'}\boldsymbol{\sigma}_{3} \hat{S}^{z} + \hat{S}^{z}\boldsymbol{\sigma}_{3} \hat{\boldsymbol{v}}_{\boldsymbol{k}} \right) u_{n,\boldsymbol{k}}\left(\boldsymbol{r} \right)$. From this one obtains
\begin{align}
    &\sum_{p\neq n,\boldsymbol{k}'} \sigma_{3}^{pp}\frac{\left\langle \psi_{p,\boldsymbol{k}'}\left\vert \hat{V}_{\boldsymbol{\Xi}}\right\vert\psi_{n,\boldsymbol{k}}\right\rangle}{\left[ \boldsymbol{\sigma}_{3}E_{\boldsymbol{k}} \right]_{nn} - \left[ \boldsymbol{\sigma}_{3}E_{\boldsymbol{k}'} \right]_{pp}}\left\vert \psi_{p,\boldsymbol{k}'}\right\rangle \notag\\
    &=-\frac{g\mu_{B}\Xi_{0}\hat{\boldsymbol{x}}}{4iq_{y}W_{y}c^{2}}\sum_{p\neq n,\boldsymbol{k}'} \sigma_{3}^{pp}\left[\delta_{\boldsymbol{k',k+q}}\frac{ \left\langle u_{p,\boldsymbol{k}'}\left\vert \left(\hat{\boldsymbol{v}}_{\boldsymbol{k}'}\boldsymbol{\sigma}_{3} \hat{S}^{z} + \hat{S}^{z} \boldsymbol{\sigma}_{3}\hat{\boldsymbol{v}}_{\boldsymbol{k}} \right)\right\vert u_{n,\boldsymbol{k}}\right\rangle  }{\left[ \boldsymbol{\sigma}_{3}E_{\boldsymbol{k}} \right]_{nn} - \left[ \boldsymbol{\sigma}_{3}E_{\boldsymbol{k}'} \right]_{pp}}\left\vert \psi_{p,\boldsymbol{k}'}\right\rangle - \left(\boldsymbol{q \rightarrow -q} \right)\right] - \left(x \leftrightarrow z \right) \notag\\
    &=-\frac{g\mu_{B}\Xi_{0}\hat{\boldsymbol{x}}}{4iq_{y}W_{y}c^{2}}\sum_{p\neq n} \sigma_{3}^{pp}\left[\frac{\left\langle u_{p,\boldsymbol{k+q}}\left\vert \left(\hat{\boldsymbol{v}}_{\boldsymbol{k+q}}\boldsymbol{\sigma}_{3} \hat{S}^{z} + \hat{S}^{z}\boldsymbol{\sigma}_{3} \hat{\boldsymbol{v}}_{\boldsymbol{k}} \right)\right\vert u_{n,\boldsymbol{k}}\right\rangle  }{\left[ \boldsymbol{\sigma}_{3}E_{\boldsymbol{k}} \right]_{nn} - \left[ \boldsymbol{\sigma}_{3}E_{\boldsymbol{k+q}} \right]_{pp}}\left\vert \psi_{p,\boldsymbol{k+q}}\right\rangle - \left(\boldsymbol{q \rightarrow -q} \right)\right] - \left(x \leftrightarrow z \right)
\end{align}
Similarly, one gets
\begin{align}
    &\sum_{n,\boldsymbol{k}'\neq \boldsymbol{k}} \sigma_{3}^{nn}\frac{\left\langle \psi_{n,\boldsymbol{k}'}\left\vert \hat{V}_{\boldsymbol{\Xi}}\right\vert\psi_{n,\boldsymbol{k}}\right\rangle}{\left[ \boldsymbol{\sigma}_{3}E_{\boldsymbol{k}} \right]_{nn} - \left[ \boldsymbol{\sigma}_{3}E_{\boldsymbol{k}'} \right]_{nn}}\left\vert \psi_{n,\boldsymbol{k}'}\right\rangle \notag \\
    &=-\frac{g\mu_{B}\Xi_{0}\hat{\boldsymbol{x}}}{4iq_{y}W_{y}c^{2}}\sum_{n} \sigma_{3}^{nn}\left[\frac{\left\langle u_{n,\boldsymbol{k+q}}\left\vert \left(\hat{\boldsymbol{v}}_{\boldsymbol{k+q}} \boldsymbol{\sigma}_{3}\hat{S}^{z} + \hat{S}^{z} \boldsymbol{\sigma}_{3}\hat{\boldsymbol{v}}_{\boldsymbol{k}} \right)\right\vert u_{n,\boldsymbol{k}}\right\rangle  }{\left[ \boldsymbol{\sigma}_{3}E_{\boldsymbol{k}} \right]_{nn} - \left[ \boldsymbol{\sigma}_{3}E_{\boldsymbol{k+q}} \right]_{nn}}\left\vert \psi_{n,\boldsymbol{k+q}}\right\rangle - \left(\boldsymbol{q \rightarrow -q} \right)\right] - \left(x \leftrightarrow z \right)
\end{align}
so that 
\begin{align}
    \left\vert \delta \psi_{n,\boldsymbol{k}}\right\rangle  &=-\frac{g\mu_{B}\Xi_{0}\hat{\boldsymbol{x}}}{4iq_{y}W_{y}c^{2}}\sum_{p} \sigma_{3}^{pp}\left[\left\vert \psi_{p,\boldsymbol{k+q}}\right\rangle\frac{\left\langle u_{p,\boldsymbol{k+q}}\left\vert \left(\hat{\boldsymbol{v}}_{\boldsymbol{k+q}} \boldsymbol{\sigma}_{3}\hat{S}^{z} + \hat{S}^{z}\boldsymbol{\sigma}_{3} \hat{\boldsymbol{v}}_{\boldsymbol{k}} \right)\right\vert u_{n,\boldsymbol{k}}\right\rangle  }{\left[ \boldsymbol{\sigma}_{3}E_{\boldsymbol{k}} \right]_{nn} - \left[ \boldsymbol{\sigma}_{3}E_{\boldsymbol{k+q}} \right]_{pp}} - \left(\boldsymbol{q \rightarrow -q} \right)\right] - \left(x \leftrightarrow z \right).
\end{align}
The first order perturbation to the wave function now reads:
\begin{align}
    &\delta\psi_{n,\boldsymbol{k}} \left( \boldsymbol{r} \right) = \left\langle \boldsymbol{r}\left\vert \delta \psi_{n,\boldsymbol{k}} \right\rangle\right. \\
    &= -\frac{g\mu_{B}\Xi_{0}\hat{\boldsymbol{x}}}{4iq_{y}W_{y}c^{2}}\sum_{p} \sigma_{3}^{pp}\left[\left\langle \boldsymbol{r} \right.\left\vert \psi_{p,\boldsymbol{k+q}}\right\rangle\frac{\left\langle u_{p,\boldsymbol{k+q}}\left\vert \left(\hat{\boldsymbol{v}}_{\boldsymbol{k+q}} \boldsymbol{\sigma}_{3}\hat{S}^{z} + \hat{S}^{z} \boldsymbol{\sigma}_{3}\hat{\boldsymbol{v}}_{\boldsymbol{k}} \right)\right\vert u_{n,\boldsymbol{k}}\right\rangle  }{\left[ \boldsymbol{\sigma}_{3}E_{\boldsymbol{k}} \right]_{nn} - \left[ \boldsymbol{\sigma}_{3}E_{\boldsymbol{k+q}} \right]_{pp}} - \left(\boldsymbol{q \rightarrow -q} \right)\right]- \left(x \leftrightarrow z \right) \notag\\
    &=-\frac{g\mu_{B}\Xi_{0}\hat{\boldsymbol{x}}}{4iq_{y}W_{y}c^{2}}\sum_{p} \sigma_{3}^{pp}\left[\psi_{p,\boldsymbol{k+q}}\left(\boldsymbol{r} \right)\frac{\left\langle u_{p,\boldsymbol{k+q}}\left\vert \left(\hat{\boldsymbol{v}}_{\boldsymbol{k+q}}\boldsymbol{\sigma}_{3} \hat{S}^{z} + \hat{S}^{z} \boldsymbol{\sigma}_{3}\hat{\boldsymbol{v}}_{\boldsymbol{k}} \right)\right\vert u_{n,\boldsymbol{k}}\right\rangle  }{\left[ \boldsymbol{\sigma}_{3}E_{\boldsymbol{k}} \right]_{nn} - \left[ \boldsymbol{\sigma}_{3}E_{\boldsymbol{k+q}} \right]_{pp}} - \left(\boldsymbol{q \rightarrow -q} \right)\right]- \left(x \leftrightarrow z \right).
\end{align}
The variation of total energy can now be given by
\begin{align}
    &\delta E\left( \boldsymbol{r} \right) \\
    =& -\frac{g\mu_{B}\Xi_{0}\hat{\boldsymbol{x}}}{4iq_{y}W_{y}c^{2}}\sum_{n,p,\boldsymbol{k}}\rho_{n,\boldsymbol{k}}\sigma_{3}^{nn}\sigma_{3}^{pp}\left[\psi_{n,\boldsymbol{k}}^{*} \left(\boldsymbol{r} \right) \hat{H}^{0}\psi_{p,\boldsymbol{k+q}}\left( \boldsymbol{r} \right)\frac{\left\langle u_{p,\boldsymbol{k+q}}\left\vert \left(\hat{\boldsymbol{v}}_{\boldsymbol{k+q}} \boldsymbol{\sigma}_{3}\hat{S}^{z} + \hat{S}^{z} \boldsymbol{\sigma}_{3}\hat{\boldsymbol{v}}_{\boldsymbol{k}} \right)\right\vert u_{n,\boldsymbol{k}}\right\rangle  }{\left[ \boldsymbol{\sigma}_{3}E_{\boldsymbol{k}} \right]_{nn} - \left[ \boldsymbol{\sigma}_{3}E_{\boldsymbol{k+q}} \right]_{pp}} - \left(\boldsymbol{q \rightarrow -q} \right)\right]  \notag \\
    &+c.c. - \left(x \leftrightarrow z \right)
\end{align}
Because 
\begin{align}
   \psi_{n,\boldsymbol{k}}^{*} \left(\boldsymbol{r} \right) \hat{H}^{0} \psi_{p,\boldsymbol{k+q}}\left( \boldsymbol{r} \right) = \left[ \boldsymbol{\sigma}_{3}E_{\boldsymbol{k}} \right]_{nn}  e^{i\boldsymbol{q.r}}u_{n,\boldsymbol{k}}^{*}\left(\boldsymbol{r} \right) \boldsymbol{\sigma}_{3} u_{p,\boldsymbol{k+q}}\left(\boldsymbol{r} \right),
\end{align}
one obtains
\begin{align}
    \delta E \left( \boldsymbol{r} \right) = -\frac{g\mu_{B}\Xi_{0}\hat{\boldsymbol{x}}}{4iq_{y}W_{y}c^{2}}\sum_{n,p,\boldsymbol{k}}\rho_{n,\boldsymbol{k}}\sigma_{3}^{nn}\sigma_{3}^{pp}&\left[e^{i\boldsymbol{q.r}}u_{n,\boldsymbol{k}}^{*}\left(\boldsymbol{r} \right) \boldsymbol{\sigma}_{3} u_{p,\boldsymbol{k+q}}\left(\boldsymbol{r} \right)\left\lbrace\frac{\left[ \boldsymbol{\sigma}_{3}E_{\boldsymbol{k}} \right]_{nn}}{\left[ \boldsymbol{\sigma}_{3}E_{\boldsymbol{k}} \right]_{nn} - \left[ \boldsymbol{\sigma}_{3}E_{\boldsymbol{k+q}} \right]_{pp}}  \right\rbrace \times \right. \notag \\
    & \times \left. \left\langle u_{p,\boldsymbol{k+q}}\left\vert \left(\hat{\boldsymbol{v}}_{\boldsymbol{k+q}}\boldsymbol{\sigma}_{3} \hat{S}^{z} + \hat{S}^{z} \boldsymbol{\sigma}_{3}\hat{\boldsymbol{v}}_{\boldsymbol{k}} \right)\right\vert u_{n,\boldsymbol{k}}\right\rangle- \left(\boldsymbol{q \rightarrow -q} \right)\right] +c.c. - \left(x \leftrightarrow z \right)
    \label{De}
\end{align}
Using Eq.~\eqref{localen}, the zero-temperature total electric polarization is obtained from the Fourier component of the local energy $\delta E \left( \boldsymbol{r} \right)$ taking the limit $q_{y} \rightarrow 0$:
\begin{align}
    \tilde{\boldsymbol{P}} = -\frac{2}{\Xi_{0}V_{0}}\lim_{q_{y} \rightarrow 0} \int \delta E \left( \boldsymbol{r} \right) cos \left(q_{y}y \right) d\boldsymbol{r}.
    \label{Pola}
\end{align}
Inserting Eq.~\eqref{De} into Eq.~\eqref{Pola} and using $\frac{1}{V_{0}}\int u_{n,\boldsymbol{k}}^{*}\left(\boldsymbol{r} \right) \boldsymbol{\sigma}_{3} u_{p,\boldsymbol{k+q}}\left(\boldsymbol{r} \right)d\boldsymbol{r}= \left\langle u_{n,\boldsymbol{k}}\left\vert \boldsymbol{\sigma}_{3}\right\vert u_{p,\boldsymbol{k+q}}\right\rangle$, which is $\boldsymbol{r}$-independent, one can do integration by parts
\begin{align}
    &\frac{1}{V_{0}}\int cos\left(\boldsymbol{q.r} \right)e^{i\boldsymbol{qr}}\left[ u_{n,\boldsymbol{k}}^{*}\left(\boldsymbol{r} \right) \boldsymbol{\sigma}_{3} u_{p,\boldsymbol{k+q}}\left(\boldsymbol{r} \right)\right]d\boldsymbol{r}  \notag\\
    &=cos\left(\boldsymbol{q.r} \right)e^{i\boldsymbol{qr}} \left\langle u_{n,\boldsymbol{k}}\left\vert \boldsymbol{\sigma}_{3}\right\vert u_{p,\boldsymbol{k+q}}\right\rangle-\int \left\langle u_{n,\boldsymbol{k}}\left\vert \boldsymbol{\sigma}_{3}\right\vert u_{p,\boldsymbol{k+q}}\right\rangle \left[q~sin\left(\boldsymbol{qr} \right)+iq ~cos\left( \boldsymbol{qr}\right) \right]e^{i\boldsymbol{qr}} d\boldsymbol{r} \notag\\
    &=cos\left(\boldsymbol{q.r} \right)e^{i\boldsymbol{qr}} \left\langle u_{n,\boldsymbol{k}}\left\vert \boldsymbol{\sigma}_{3}\right\vert u_{p,\boldsymbol{k+q}}\right\rangle-\left\langle u_{n,\boldsymbol{k}}\left\vert \boldsymbol{\sigma}_{3}\right\vert u_{p,\boldsymbol{k+q}}\right\rangle\int  \left[sin\left(\boldsymbol{qr} \right)+i ~cos\left( \boldsymbol{qr}\right) \right]e^{i\boldsymbol{qr}}d\left(\boldsymbol{qr}\right) \notag\\
    &=cos\left(\boldsymbol{q.r} \right)e^{i\boldsymbol{qr}} \left\langle u_{n,\boldsymbol{k}}\left\vert \boldsymbol{\sigma}_{3}\right\vert u_{p,\boldsymbol{k+q}}\right\rangle-i\boldsymbol{qr}\left\langle u_{n,\boldsymbol{k}}\left\vert \boldsymbol{\sigma}_{3}\right\vert u_{p,\boldsymbol{k+q}}\right\rangle
\end{align}
Because $\boldsymbol{q}$ is an infinitesimal vector, 
\begin{equation}
    e^{i\boldsymbol{qr}} \approx 1 + i\boldsymbol{qr}
\end{equation}
and $cos(\boldsymbol{qr}) \approx 1$
leads to
\begin{equation}
     \frac{1}{V_{0}}\int cos\left(\boldsymbol{q.r} \right)e^{i\boldsymbol{qr}}\left[ u_{n,\boldsymbol{k}}^{*}\left(\boldsymbol{r} \right) \boldsymbol{\sigma}_{3} u_{p,\boldsymbol{k+q}}\left(\boldsymbol{r} \right)\right]d\boldsymbol{r}  = \left\langle u_{n,\boldsymbol{k}}\left\vert \boldsymbol{\sigma}_{3}\right\vert u_{p,\boldsymbol{k+q}}\right\rangle.
\end{equation}
Therefore, the electric polarization reads
\begin{align}
    \tilde{P}_{y}&=-\lim_{q_{y} \rightarrow 0}\frac{g\mu_{B}\hat{\boldsymbol{x}}}{2iq_{y}W_{y}c^{2}}\sum_{n,p,\boldsymbol{k}}\sigma_{3}^{nn}\sigma_{3}^{pp}\rho_{n,\boldsymbol{k}}\left[ \boldsymbol{\sigma}_{3}E_{\boldsymbol{k}} \right]_{nn} \times \notag \\
    &\times \left\lbrace\frac{\left\langle u_{p,\boldsymbol{k+q}}\left\vert \left(\hat{\boldsymbol{v}}_{\boldsymbol{k+q}}\boldsymbol{\sigma}_{3} \hat{S}^{z} + \hat{S}^{z}\boldsymbol{\sigma}_{3} \hat{\boldsymbol{v}}_{\boldsymbol{k}} \right)\right\vert u_{n,\boldsymbol{k}}\right\rangle\left\langle u_{n,\boldsymbol{k}} \right\vert \boldsymbol{\sigma}_{3}\left\vert u_{p,\boldsymbol{k+q}}\right\rangle}{\left[ \boldsymbol{\sigma}_{3}E_{\boldsymbol{k}} \right]_{nn} - \left[ \boldsymbol{\sigma}_{3}E_{\boldsymbol{k+q}} \right]_{pp}} - \left(\boldsymbol{q \rightarrow -q} \right)\right\rbrace +c.c. - \left(x \leftrightarrow z \right) \notag\\
    &=-\lim_{q_{y} \rightarrow 0}\frac{g\mu_{B}\hat{\boldsymbol{x}}}{2iq_{y}W_{y}c^{2}}\sum_{n,p,\boldsymbol{k}}\left(\left[ \boldsymbol{\sigma}_{3}E_{\boldsymbol{k}} \right]_{nn}\rho_{n,\boldsymbol{k}}-\left[ \boldsymbol{\sigma}_{3}E_{\boldsymbol{k+q}} \right]_{pp}\rho_{p,\boldsymbol{k+q}}\right)\sigma_{3}^{nn}\sigma_{3}^{pp} \times \notag\\
    & \times \left\lbrace\frac{\left\langle u_{p,\boldsymbol{k+q}}\left\vert \left(\hat{\boldsymbol{v}}_{\boldsymbol{k+q}}\boldsymbol{\sigma}_{3} \hat{S}^{z} + \hat{S}^{z}\boldsymbol{\sigma}_{3} \hat{\boldsymbol{v}}_{\boldsymbol{k}} \right)\right\vert u_{n,\boldsymbol{k}}\right\rangle}{\left[ \boldsymbol{\sigma}_{3}E_{\boldsymbol{k}} \right]_{nn} - \left[ \boldsymbol{\sigma}_{3}E_{\boldsymbol{k+q}} \right]_{pp}}\left\langle u_{n,\boldsymbol{k}} \right\vert \boldsymbol{\sigma}_{3}\left\vert u_{p,\boldsymbol{k+q}}\right\rangle  \right\rbrace +c.c. - \left(x \leftrightarrow z \right) \label{lim}
\end{align}
Using $\left\langle u_{n,\boldsymbol{k}} \right\vert \boldsymbol{\sigma}_{3}\left\vert u_{p,\boldsymbol{k}}\right\rangle = \sigma_{3}^{nn}\delta_{p,n}$ where $\delta_{p,n}$ here is the Kronecker delta function and implementing the limit in Eq~\eqref{lim}, one obtains for $p\neq n$
\begin{align}
    \tilde{P}_{y}^{(1)}&=-\lim_{q_{y} \rightarrow 0}\frac{g\mu_{B}\hat{\boldsymbol{x}}}{2iq_{y}W_{y}c^{2}}\sum_{p\neq n,\boldsymbol{k}}\left(\left[ \boldsymbol{\sigma}_{3}E_{\boldsymbol{k}} \right]_{nn}\rho_{n,\boldsymbol{k}}-\left[ \boldsymbol{\sigma}_{3}E_{\boldsymbol{k+q}} \right]_{pp}\rho_{p,\boldsymbol{k+q}}\right)\sigma_{3}^{nn}\sigma_{3}^{pp} \times \notag\\
    &\times \left\lbrace\frac{\left\langle u_{p,\boldsymbol{k+q}}\left\vert \left(\hat{\boldsymbol{v}}_{\boldsymbol{k+q}}\boldsymbol{\sigma}_{3} \hat{S}^{z} + \hat{S}^{z}\boldsymbol{\sigma}_{3} \hat{\boldsymbol{v}}_{\boldsymbol{k}} \right)\right\vert u_{n,\boldsymbol{k}}\right\rangle}{\left[ \boldsymbol{\sigma}_{3}E_{\boldsymbol{k}} \right]_{nn} - \left[ \boldsymbol{\sigma}_{3}E_{\boldsymbol{k+q}} \right]_{pp}} \left\langle u_{n,\boldsymbol{k}} \right\vert \boldsymbol{\sigma}_{3}\left\vert u_{p,\boldsymbol{k+q}}\right\rangle  \right\rbrace +c.c. - \left(x \leftrightarrow z \right) \\
    &=-\frac{g\mu_{B}\hat{\boldsymbol{x}}}{2iW_{y}c^{2}}\sum_{p\neq n,\boldsymbol{k}}\left(\left[ \boldsymbol{\sigma}_{3}E_{\boldsymbol{k}} \right]_{nn}\rho_{n,\boldsymbol{k}}-\left[ \boldsymbol{\sigma}_{3}E_{\boldsymbol{k}} \right]_{pp}\rho_{p,\boldsymbol{k}}\right)\sigma_{3}^{nn}\sigma_{3}^{pp}  \times \notag\\
    &\times \left\lbrace\frac{\left\langle u_{p,\boldsymbol{k}}\left\vert \left(\hat{\boldsymbol{v}}_{\boldsymbol{k}}\boldsymbol{\sigma}_{3} \hat{S}^{z} + \hat{S}^{z}\boldsymbol{\sigma}_{3} \hat{\boldsymbol{v}}_{\boldsymbol{k}} \right)\right\vert u_{n,\boldsymbol{k}}\right\rangle}{\left[ \boldsymbol{\sigma}_{3}E_{\boldsymbol{k}} \right]_{nn} - \left[ \boldsymbol{\sigma}_{3}E_{\boldsymbol{k}} \right]_{pp}} \left\langle u_{n,\boldsymbol{k}} \right\vert \boldsymbol{\sigma}_{3}\left\vert \partial_{k_{y}}u_{p,\boldsymbol{k}}\right\rangle  \right\rbrace +c.c. - \left(x \leftrightarrow z \right) \\
    &=-\frac{i\hbar g\mu_{B}}{2W_{y}c^{2}}\sum_{n\neq p,\boldsymbol{k}}\sigma_{3}^{nn}\sigma_{3}^{pp}\left(\left[ \boldsymbol{\sigma}_{3}E_{\boldsymbol{k}} \right]_{nn}\rho_{n,\boldsymbol{k}}-\left[ \boldsymbol{\sigma}_{3}E_{\boldsymbol{k}} \right]_{pp}\rho_{p,\boldsymbol{k}}\right) \times \notag\\
    &\times \frac{\left\langle u_{p,\boldsymbol{k}}\left\vert \left(\hat{v}_{x}\boldsymbol{\sigma}_{3} \hat{S}^{z} + \hat{S}^{z}\boldsymbol{\sigma}_{3} \hat{v}_{x} \right)\right\vert u_{n,\boldsymbol{k}}\right\rangle\left\langle  u_{n,\boldsymbol{k}}\left\vert \hat{v}_{y}\right\vert  u_{p,\boldsymbol{k}}\right\rangle}{ \left\lbrace\left[ \boldsymbol{\sigma}_{3}E_{\boldsymbol{k}} \right]_{nn} - \left[ \boldsymbol{\sigma}_{3}E_{\boldsymbol{k}} \right]_{pp} \right\rbrace^{2}}   +c.c. - \left(x \leftrightarrow z \right) \\
    &=-\frac{ g\mu_{B}}{\hbar W_{y}c^{2}}\sum_{n\neq p,\boldsymbol{k}} ~Im\left(2 \hbar^{2} \sigma_{3}^{nn}\sigma_{3}^{pp} \frac{\left\langle u_{p,\boldsymbol{k}}\left\vert \left(\hat{v}_{x}\boldsymbol{\sigma}_{3} \hat{S}^{z} + \hat{S}^{z}\boldsymbol{\sigma}_{3} \hat{v}_{x} \right)\right\vert u_{n,\boldsymbol{k}}\right\rangle\left\langle  u_{n,\boldsymbol{k}}\left\vert \hat{v}_{y}\right\vert  u_{p,\boldsymbol{k}}\right\rangle}{ \left\lbrace\left[ \boldsymbol{\sigma}_{3}E_{\boldsymbol{k}} \right]_{nn} - \left[ \boldsymbol{\sigma}_{3}E_{\boldsymbol{k}} \right]_{pp} \right\rbrace^{2}}\right)\left[ \boldsymbol{\sigma}_{3}E_{\boldsymbol{k}} \right]_{nn}\rho_{n,\boldsymbol{k}} -  \notag \\
    &- \left(x \leftrightarrow z \right)
\end{align}
where we have used 
\begin{equation}
      \left\langle  u_{n,\boldsymbol{k}}\left\vert \hat{v}_{\boldsymbol{k}}\right\vert  u_{p,\boldsymbol{k}}\right\rangle = \frac{1}{\hbar}\left[\left( \boldsymbol{\sigma}_{3}E_{\boldsymbol{k}} \right)_{pp} - \left( \boldsymbol{\sigma}_{3}E_{\boldsymbol{k}} \right)_{nn} \right]\left\langle u_{n,\boldsymbol{k}} \right\vert \boldsymbol{\sigma}_{3}\left\vert \partial_{k_{y}}u_{p,\boldsymbol{k}}\right\rangle
\end{equation}
Noting that
\begin{equation}
    \sum_{p \neq n}Im\left(2 \hbar^{2} \sigma_{3}^{nn}\sigma_{3}^{pp} \frac{\left\langle u_{p,\boldsymbol{k}}\left\vert \left(\hat{v}_{x}\boldsymbol{\sigma}_{3} \hat{S}^{z} + \hat{S}^{z}\boldsymbol{\sigma}_{3} \hat{v}_{x} \right)\right\vert u_{n,\boldsymbol{k}}\right\rangle\left\langle  u_{n,\boldsymbol{k}}\left\vert \hat{v}_{y}\right\vert  u_{p,\boldsymbol{k}}\right\rangle}{ \left\lbrace\left[ \boldsymbol{\sigma}_{3}E_{\boldsymbol{k}} \right]_{nn} - \left[ \boldsymbol{\sigma}_{3}E_{\boldsymbol{k}} \right]_{pp} \right\rbrace^{2}}\right) \equiv \Omega_{xy}^{S^{z},n}\left( \boldsymbol{k} \right) 
\end{equation}
is the magnon spin Berry curvature of $nth$-band \cite{To2023b}, therefore one obtains
\begin{align} 
    \tilde{P}_{y}^{(1)} &=-\frac{ g\mu_{B}}{\hbar W_{y}c^{2}}\sum_{n,\boldsymbol{k}} \Omega_{xy}^{S^{z},n}\left( \boldsymbol{k} \right)\left[ \boldsymbol{\sigma}_{3}E_{\boldsymbol{k}} \right]_{nn}\rho_{n,\boldsymbol{k}} - \left(x \leftrightarrow z \right) \\
    &= -\frac{ g\mu_{B}}{\hbar W_{y}c^{2}}\sum_{n,\boldsymbol{k}}\left[\Omega_{xy}^{S^{z},n}\left( \boldsymbol{k} \right) - \Omega_{zy}^{S^{x},n}\left( \boldsymbol{k} \right) \right]\left[ \boldsymbol{\sigma}_{3}E_{\boldsymbol{k}} \right]_{nn}\rho_{n,\boldsymbol{k}}\label{DF1}
\end{align}

For $p \equiv n$, one gets
\begin{align}
    \tilde{P}_{y}^{(2)}=-\frac{g\mu_{B}\hat{\boldsymbol{x}}}{2iW_{y}c^{2}}\sum_{n,\boldsymbol{k}}\left( \rho_{n,\boldsymbol{k}}+\left[ \boldsymbol{\sigma}_{3}E_{\boldsymbol{k}} \right]_{nn}\rho^{\prime}_{n,\boldsymbol{k}}\right)&\left[\left\langle \partial_{k_{y}}u_{n,\boldsymbol{k}}\left\vert \left(\hat{\boldsymbol{v}}_{\boldsymbol{k}}\boldsymbol{\sigma}_{3} \hat{S}^{z} + \hat{S}^{z}\boldsymbol{\sigma}_{3} \hat{\boldsymbol{v}}_{\boldsymbol{k}} \right)\right\vert u_{n,\boldsymbol{k}}\right\rangle  \left\langle u_{n,\boldsymbol{k}} \right\vert \boldsymbol{\sigma}_{3}\left\vert u_{n,\boldsymbol{k}}\right\rangle + \right.\notag\\
    &+\left\langle u_{n,\boldsymbol{k}}\left\vert \left(\partial_{k_{y}}\hat{\boldsymbol{v}}_{\boldsymbol{k}} \right)\boldsymbol{\sigma}_{3} \hat{S}^{z}\right\vert u_{n,\boldsymbol{k}}\right\rangle  \left\langle u_{n,\boldsymbol{k}} \right\vert \boldsymbol{\sigma}_{3}\left\vert u_{n,\boldsymbol{k}}\right\rangle \notag\\
    &\left. + \left\langle u_{n,\boldsymbol{k}}\left\vert \left(\hat{\boldsymbol{v}}_{\boldsymbol{k}}\boldsymbol{\sigma}_{3} \hat{S}^{z} + \hat{S}^{z}\boldsymbol{\sigma}_{3} \hat{\boldsymbol{v}}_{\boldsymbol{k}} \right)\right\vert u_{n,\boldsymbol{k}}\right\rangle  \left\langle u_{n,\boldsymbol{k}} \right\vert \boldsymbol{\sigma}_{3}\left\vert \partial_{k_{y}}u_{n,\boldsymbol{k}}\right\rangle  \right] \notag \\
    &+c.c. - \left(x \leftrightarrow z \right)\\
    =-\frac{g\mu_{B}}{W_{y}c^{2}}\sum_{n,\boldsymbol{k}}\left( \rho_{n,\boldsymbol{k}}+\left[ \boldsymbol{\sigma}_{3}E_{\boldsymbol{k}} \right]_{nn}\rho^{\prime}_{n,\boldsymbol{k}}\right)&Im\left[\left\langle \partial_{k_{y}}u_{n,\boldsymbol{k}}\left\vert \left(\hat{v}_{x}\boldsymbol{\sigma}_{3} \hat{S}^{z} + \hat{S}^{z}\boldsymbol{\sigma}_{3} \hat{v}_{x} \right)\right\vert u_{n,\boldsymbol{k}}\right\rangle  \left\langle u_{n,\boldsymbol{k}} \right\vert \boldsymbol{\sigma}_{3}\left\vert u_{n,\boldsymbol{k}}\right\rangle +  \right.\notag\\
    &+\left\langle u_{n,\boldsymbol{k}}\left\vert \left(\partial_{k_{y}}\hat{v}_{x} \right)\boldsymbol{\sigma}_{3} \hat{S}^{z}\right\vert u_{n,\boldsymbol{k}}\right\rangle  \left\langle u_{n,\boldsymbol{k}} \right\vert \boldsymbol{\sigma}_{3}\left\vert u_{n,\boldsymbol{k}}\right\rangle \notag\\
    &\left.+ \left\langle u_{n,\boldsymbol{k}}\left\vert \left(\hat{v}_{x}\boldsymbol{\sigma}_{3} \hat{S}^{z} + \hat{S}^{z}\boldsymbol{\sigma}_{3} \hat{v}_{x} \right)\right\vert u_{n,\boldsymbol{k}}\right\rangle  \left\langle u_{n,\boldsymbol{k}} \right\vert \boldsymbol{\sigma}_{3}\left\vert \partial_{k_{y}}u_{n,\boldsymbol{k}}\right\rangle  \right] - \left(x \leftrightarrow z \right) \label{DF2}
\end{align}

Equations \eqref{DF1} and \eqref{DF2} provide a microscopic framework to compute the electric polarization induced by magnon transport in both 3D and 2D magnetic materials. Eq. \eqref{DF1} reveals the contribution of magnon spin current to the electric polarization through the magnon spin Berry curvatures $\Omega_{xy}^{S^{z},n}\left( \boldsymbol{k} \right)$ and $\Omega_{zy}^{S^{x},n}\left( \boldsymbol{k} \right)$ \cite{To2023b}. Meanwhile, the $\tilde{P}_{y}^{(2)}$ term in Eq.~\eqref{DF2} is difficult to analyze at this stage. However, one can infer that this term is related to the orbital angular moment of the magnon, $\hat{\boldsymbol{L}} = \frac{1}{4}\left(\hat{\boldsymbol{r}} \times \boldsymbol{\boldsymbol{v}} - \hat{\boldsymbol{v}}\times \hat{\boldsymbol{r}}\right)$ defined in Eq.~\eqref{Loper}. This is because $\left\langle \partial_{\boldsymbol{k}}u_{n,\boldsymbol{k}}\right\vert \equiv \left\langle u_{n,\boldsymbol{k}}\right\vert\boldsymbol{r}$, ($\boldsymbol{r}$ is the position operator) appears in a multiplication with the velocity operator $\hat{\boldsymbol{v}}_{\boldsymbol{k}}$ as shown in Eq.~\eqref{DF2}. 

For those reasons, we relabel the expressions \eqref{DF1} and \eqref{DF2} as follows:
\begin{align} 
    \tilde{P}_{y}^{S} = -\frac{ g\mu_{B}}{\hbar W_{y}c^{2}}\sum_{n,\boldsymbol{k}}\left[\Omega_{xy}^{S^{z},n}\left( \boldsymbol{k} \right) - \Omega_{zy}^{S^{x},n}\left( \boldsymbol{k} \right) \right]\left[ \boldsymbol{\sigma}_{3}E_{\boldsymbol{k}} \right]_{nn}\rho_{n,\boldsymbol{k}}
    \label{DF1a}
\end{align}
and
\begin{align}
    \tilde{P}_{y}^{O}=-\frac{g\mu_{B}}{W_{y}c^{2}}\sum_{n,\boldsymbol{k}}\left( \rho_{n,\boldsymbol{k}}+\left[ \boldsymbol{\sigma}_{3}E_{\boldsymbol{k}} \right]_{nn}\rho^{\prime}_{n,\boldsymbol{k}}\right)&Im\left[\left\langle \partial_{k_{y}}u_{n,\boldsymbol{k}}\left\vert \left(\hat{v}_{x}\boldsymbol{\sigma}_{3} \hat{S}^{z} + \hat{S}^{z}\boldsymbol{\sigma}_{3} \hat{v}_{x} \right)\right\vert u_{n,\boldsymbol{k}}\right\rangle  \left\langle u_{n,\boldsymbol{k}} \right\vert \boldsymbol{\sigma}_{3}\left\vert u_{n,\boldsymbol{k}}\right\rangle +  \right.\notag\\
    &+\left\langle u_{n,\boldsymbol{k}}\left\vert \left(\partial_{k_{y}}\hat{v}_{x} \right)\boldsymbol{\sigma}_{3} \hat{S}^{z}\right\vert u_{n,\boldsymbol{k}}\right\rangle  \left\langle u_{n,\boldsymbol{k}} \right\vert \boldsymbol{\sigma}_{3}\left\vert u_{n,\boldsymbol{k}}\right\rangle \notag\\
    &\left.+ \left\langle u_{n,\boldsymbol{k}}\left\vert \left(\hat{v}_{x}\boldsymbol{\sigma}_{3} \hat{S}^{z} + \hat{S}^{z}\boldsymbol{\sigma}_{3} \hat{v}_{x} \right)\right\vert u_{n,\boldsymbol{k}}\right\rangle  \left\langle u_{n,\boldsymbol{k}} \right\vert \boldsymbol{\sigma}_{3}\left\vert \partial_{k_{y}}u_{n,\boldsymbol{k}}\right\rangle  \right] - \left(x \leftrightarrow z \right)
    \label{DF2a}
\end{align}
These labels $S$ and $O$ indicate the contributions from the spin Berry curvature and orbital moment to the electric polarization, respectively. Nevertheless, in principle, one can compute numerically the zero-temperature electric polarization $\tilde{P}_{y}=\tilde{P}_{y}^{S}+\tilde{P}_{y}^{O}$ for general 2D or 3D systems by using the following relations:
\begin{align} \label{numcalea}
      \left\vert \partial_{k_{y}} u_{p,\boldsymbol{k}} \right\rangle=\sum_{q\neq p}\sigma_{3}^{qq}\frac{\left\langle  u_{q,\boldsymbol{k}}\left\vert \partial_{k_{y}} \hat{H}^{0}_{\boldsymbol{k}}\right\vert  u_{p,\boldsymbol{k}}\right\rangle}{ \left[ \boldsymbol{\sigma}_{3}E_{\boldsymbol{k}} \right]_{pp} - \left[ \boldsymbol{\sigma}_{3}E_{\boldsymbol{k}} \right]_{qq}}\left\vert  u_{q,\boldsymbol{k}}\right\rangle \\
      \left\langle \partial_{k_{y}} u_{p,\boldsymbol{k}} \right\vert=\sum_{q\neq p}\sigma_{3}^{qq}\frac{\left\langle  u_{p,\boldsymbol{k}}\left\vert \partial_{k_{y}} \hat{H}^{0}_{\boldsymbol{k}}\right\vert  u_{q,\boldsymbol{k}}\right\rangle}{ \left[ \boldsymbol{\sigma}_{3}E_{\boldsymbol{k}} \right]_{pp} - \left[ \boldsymbol{\sigma}_{3}E_{\boldsymbol{k}} \right]_{qq}}\left\langle  u_{q,\boldsymbol{k}}\right\vert
      \label{numcaleb}
\end{align}
This calculation accounts for the contributions from spin Berry curvature and probably both inter- and intra-band magnon orbital angular moment to the $\tilde{P}_{y}$.

Given that our focus in this work is on 2D systems, we set $v_z=0$. Consequently, the contribution of the $S^x$ term and thus the $\hat{V}_{\boldsymbol{\Xi}}^{(2)}$ in Eq.~\eqref{V12} vanishes. This reduction simplifies Eqs. \eqref{DF1a} and \eqref{DF2a} to
\begin{align} 
    \tilde{P}_{y}^{S} &=-\frac{ g\mu_{B}}{\hbar W_{y}c^{2}}\sum_{n,\boldsymbol{k}} \Omega_{xy}^{S^{z},n}\left( \boldsymbol{k} \right)\left[ \boldsymbol{\sigma}_{3}E_{\boldsymbol{k}} \right]_{nn}\rho_{n,\boldsymbol{k}} 
\end{align}
and
\begin{align}
    \tilde{P}_{y}^{O}=-\frac{g\mu_{B}}{W_{y}c^{2}}\sum_{n,\boldsymbol{k}}\left( \rho_{n,\boldsymbol{k}}+\left[ \boldsymbol{\sigma}_{3}E_{\boldsymbol{k}} \right]_{nn}\rho^{\prime}_{n,\boldsymbol{k}}\right)&Im\left[\left\langle \partial_{k_{y}}u_{n,\boldsymbol{k}}\left\vert \left(\hat{v}_{x}\boldsymbol{\sigma}_{3} \hat{S}^{z} + \hat{S}^{z}\boldsymbol{\sigma}_{3} \hat{v}_{x} \right)\right\vert u_{n,\boldsymbol{k}}\right\rangle  \left\langle u_{n,\boldsymbol{k}} \right\vert \boldsymbol{\sigma}_{3}\left\vert u_{n,\boldsymbol{k}}\right\rangle +  \right.\notag\\
    &+\left\langle u_{n,\boldsymbol{k}}\left\vert \left(\partial_{k_{y}}\hat{v}_{x} \right)\boldsymbol{\sigma}_{3} \hat{S}^{z}\right\vert u_{n,\boldsymbol{k}}\right\rangle  \left\langle u_{n,\boldsymbol{k}} \right\vert \boldsymbol{\sigma}_{3}\left\vert u_{n,\boldsymbol{k}}\right\rangle \notag\\
    &\left.+ \left\langle u_{n,\boldsymbol{k}}\left\vert \left(\hat{v}_{x}\boldsymbol{\sigma}_{3} \hat{S}^{z} + \hat{S}^{z}\boldsymbol{\sigma}_{3} \hat{v}_{x} \right)\right\vert u_{n,\boldsymbol{k}}\right\rangle  \left\langle u_{n,\boldsymbol{k}} \right\vert \boldsymbol{\sigma}_{3}\left\vert \partial_{k_{y}}u_{n,\boldsymbol{k}}\right\rangle  \right] \label{DF3}
\end{align}
Futhermore, in order to gain a deeper understanding of the physical implications of magnon orbital moment on polarization, we focus on scenarios where spin is conserved. This approach is valid in this work because the z-component of the total spin should be a good quantum number under the DMI. This allows for simplification, reducing the complexity of relation \eqref{DF3} to:
\begin{align}
    \tilde{P}_{y}^{O}=-\frac{2g\mu_{B}}{W_{y}c^{2}}\sum_{n,\boldsymbol{k}}\left( \rho_{n,\boldsymbol{k}}+\left[ \boldsymbol{\sigma}_{3}E_{\boldsymbol{k}} \right]_{nn}\rho^{\prime}_{n,\boldsymbol{k}}\right)&\sigma_{3}^{nn}S_{nn}^{z}~Im\left[\left\langle \partial_{k_{y}}u_{n,\boldsymbol{k}}\left\vert \hat{v}_{x} \right\vert u_{n,\boldsymbol{k}}\right\rangle  \left\langle u_{n,\boldsymbol{k}} \right\vert \boldsymbol{\sigma}_{3}\left\vert u_{n,\boldsymbol{k}}\right\rangle + \right.\notag\\
    & +\frac{1}{2}\left\langle u_{n,\boldsymbol{k}}\left\vert \left(\partial_{k_{y}}\hat{v}_{x} \right)\right\vert u_{n,\boldsymbol{k}}\right\rangle  \left\langle u_{n,\boldsymbol{k}} \right\vert \boldsymbol{\sigma}_{3}\left\vert u_{n,\boldsymbol{k}}\right\rangle \notag\\
    &\left. +\left\langle u_{n,\boldsymbol{k}}\left\vert \hat{v}_{x}\right\vert u_{n,\boldsymbol{k}}\right\rangle  \left\langle u_{n,\boldsymbol{k}} \right\vert \boldsymbol{\sigma}_{3}\left\vert \partial_{k_{y}}u_{n,\boldsymbol{k}}\right\rangle  \right] 
\end{align}
leads to
\begin{align}
    \tilde{P}_{y}^{O} = -\frac{2g\mu_{B}}{\hbar W_{y}c^{2}}\sum_{n,\boldsymbol{k}}\left( \rho_{n,\boldsymbol{k}}+\left[ \boldsymbol{\sigma}_{3}E_{\boldsymbol{k}} \right]_{nn}\rho^{\prime}_{n,\boldsymbol{k}}\right)\sigma_{3}^{nn}S_{nn}^{z} Im \left\lbrace\left\langle \partial_{k_{y}} u_{n,\boldsymbol{k}}\left\vert\left[ \left(\boldsymbol{\sigma}_{3}E_{\boldsymbol{k}}\right)_{nn}\boldsymbol{\sigma}_{3} - \hat{H}^{0}_{\boldsymbol{k}} \right]\right\vert \partial_{k_{x}} u_{n,\boldsymbol{k}}\right\rangle\right\rbrace
\end{align}
Finally, one obtains
\begin{align}
    &\tilde{P}_{y}=\tilde{P}_{y}^{S} + \tilde{P}_{y}^{O} = -\frac{ g\mu_{B}}{\hbar W_{y}c^{2}}\sum_{n,\boldsymbol{k}}\left[ \Omega_{xy}^{S^{z},n}\left( \boldsymbol{k} \right)\left[ \boldsymbol{\sigma}_{3}E_{\boldsymbol{k}} \right]_{nn}\rho_{n,\boldsymbol{k}} + 2\hbar\sigma_{3}^{nn}S^{z}_{nn}L^{z}_{nn}\left(\boldsymbol{k} \right)\left( \rho_{n,\boldsymbol{k}}+\left[ \boldsymbol{\sigma}_{3}E_{\boldsymbol{k}} \right]_{nn}\rho^{\prime}_{n,\boldsymbol{k}}\right)\right]
\end{align}
where 
\begin{align}
    L^{z}_{nn}\left(\boldsymbol{k} \right) &= \frac{1}{\hbar}Im \left\lbrace \left\langle \partial_{k_{y}} u_{n,\boldsymbol{k}}\left\vert\left[ \left(\boldsymbol{\sigma}_{3}E_{\boldsymbol{k}}\right)_{nn}\boldsymbol{\sigma}_{3} - \hat{H}^{0}_{\boldsymbol{k}} \right]\right\vert \partial_{k_{x}} u_{n,\boldsymbol{k}}\right\rangle\right\rbrace \\
    &\equiv \frac{1}{\hbar}Im \left\lbrace \left\langle \partial_{k_{y}} n\left(\boldsymbol{k}\right)\left\vert\left[ \left(\boldsymbol{\sigma}_{3}E_{\boldsymbol{k}}\right)_{nn}\boldsymbol{\sigma}_{3} - \hat{H}^{0}_{\boldsymbol{k}} \right]\right\vert \partial_{k_{x}} n\left(\boldsymbol{k}\right)\right\rangle\right\rbrace
\end{align}
is the intra-band orbital angular moment of the magnon in the nth state as presented in Eq.~\eqref{nthL}. Here we have used $u_{n,\boldsymbol{k}} \equiv n\left(\boldsymbol{k}\right)$.

At finite temperature, the electric polarization is computed by:
\begin{align} \label{EQelectricpolariz} 
    P_{y} &= \frac{1}{\beta}\int \tilde{P}_{y}d \beta = -\frac{ g\mu_{B}}{\hbar W_{y}c^{2}}\sum_{n,\boldsymbol{k}}\left[ \frac{1}{\beta}\Omega_{xy}^{S^{z},n}\left( \boldsymbol{k} \right) ln\left(e^{-\frac{E_{n,\boldsymbol{k}}}{k_{B}T}}-1\right) + 2\hbar\sigma_{3}^{nn}S^{z}_{nn}L^{z}_{nn}\left(\boldsymbol{k} \right) \rho_{n,\boldsymbol{k}}\right] \\
    & = P_{y}^{S} + P_{y}^{O}
\end{align}
where 
\begin{align}
    P_{y}^{S} = -\frac{ g\mu_{B}}{\hbar W_{y}c^{2}}\sum_{n,\boldsymbol{k}} \frac{1}{\beta}\Omega_{xy}^{S^{z},n}\left( \boldsymbol{k} \right) ln\left(e^{-\frac{E_{n,\boldsymbol{k}}}{k_{B}T}}-1\right)  
\end{align}
and
\begin{align}
    P_{y}^{O} = -\frac{ 2g\mu_{B}}{W_{y}c^{2}}\sum_{n,\boldsymbol{k}} \sigma_{3}^{nn}S^{z}_{nn}L^{z}_{nn}\left(\boldsymbol{k} \right) \rho_{n,\boldsymbol{k}} 
\end{align}
thereby completing the derivation of Eqs.~\ref{PS_2D} and \ref{PO_2D} in the main text. Here we have used the relations: $\frac{d}{d\beta}\left[\beta \rho_{n,\boldsymbol{k}} \right] = \rho_{n,\boldsymbol{k}} +\left[ \boldsymbol{\sigma}_{3}E_{\boldsymbol{k}} \right]_{nn}\rho^{\prime}_{n,\boldsymbol{k}}$.

The electric polarization induced by the motion of the magnon wave packets has two distinct contributions: 
\begin{enumerate}
    \item $P_{y}^{S}$, arising from the spin current through the spin-Berry curvature $\Omega_{xy}^{S^{z},n}$ accounts for the accumulation of spin angular momentum along y-direction due to, for example the magnon spin Nernst current carried by magnons under a temperature gradient along the x-direction.
    \item $P_{y}^{O}$, which results from the intra-band orbital angular moment $L^{z}_{nn}\left(\boldsymbol{k} \right)$ of the magnons. 
\end{enumerate}

We note that our derivation presented in this part is based on perturbation theory within the thermodynamic limit, assuming an infinite x-dimension and the system is not too far away from equilibrium. For the magnon Nernst effect, this condition is satisfied when $\Delta_{x}T \ll T $. The electric polarization $P_{y}$ in Eq.~\eqref{EQelectricpolariz} is the zeroth-order term in the expansion of electric polarization in terms of the temperature gradient $\partial_{x} T$ 
\begin{equation}
    P_{y} = P_{y}^{0} + P_{y}^{1}\partial_{x} T + ...
\end{equation}
The zeroth order $P_{y}^{0}$ as given in Eq.~\eqref{EQelectricpolariz} is associated with the electric dipole moment, which depends on the temperature $T$ but is independent of $\partial_{x} T$. The first-order term $P_{y}^{1}\partial_{x} T$ corresponds to the electric quadrupole moment and indicates the induced electric polarization due to the temperature gradient $\partial_{x} T$. In this work, for the sake of simplification, we focus on the zeroth-order term and leave the contribution from higher-order terms for future investigation. However, we emphasise that the higher-order term respects the system's symmetry. This means that in the absence of DMI, the higher-order term must vanish to ensure that the total net electric polarization is zero, as previously discussed.

\section{Linear response theory of the magnon orbital Nernst effect}
We now turn to the description of the linear response theory for magnon orbital transport discussed in the main text. In the following, we shall elucidate the derivation of the continuity equation for magnon orbital angular moment density and introduce the corresponding orbital angular moment current operator being used in the linear response theory to compute the magnon orbital current in 2D collinear AFMs under a temperature gradient.

\subsection{Orbital angular moment current operator}
We start from the time evolution equation for the local angular moment density $L^{\alpha}\left(\boldsymbol{r}\right) = \frac{1}{2}\Psi^{\dagger}\left( \boldsymbol{r}\right) \hat{L}^{\alpha} \Psi\left( \boldsymbol{r}\right)$  $(\alpha =x,y,z)$, where the following commutators of the bosonic wavefunctions hold:

\begin{align}
    \left[\Psi^{\dagger}_{m}\left( \boldsymbol{r}\right),\Psi_{n}\left( \boldsymbol{r}'\right) \right] &= \sigma_{3}^{mn}\delta_{\boldsymbol{r},\boldsymbol{r}'} \\
    \left[\Psi_{m}\left( \boldsymbol{r}\right),\Psi_{n}\left( \boldsymbol{r}'\right) \right] &= i\sigma_{2}^{mn}\delta_{\boldsymbol{r},\boldsymbol{r}'} \\
    \left[\Psi^{\dagger}_{m}\left( \boldsymbol{r}\right),\Psi^{\dagger}_{n}\left( \boldsymbol{r}'\right) \right] &= -i\sigma_{2}^{mn}\delta_{\boldsymbol{r},\boldsymbol{r}'}
\end{align}
with $\boldsymbol{\sigma}_{i} ~\left(i=1,2,3 \right)$ Pauli matrices acting on particle-hole space. The Heisenberg equation of motion for the angular momentum operator is then:
\begin{equation}
    \frac{\partial L^{\alpha}\left(\boldsymbol{r}\right)}{\partial t} = i \left[ \hat{H}, \hat{L}^{\alpha}\left(\boldsymbol{r}\right) \right]
\end{equation}
where the total Hamiltonian can be expressed as $H = \frac{1}{2} \int d\boldsymbol{r} \tilde{\Psi}^{\dagger}\left(\boldsymbol{r}\right) \hat{H} \tilde{\Psi}\left(\boldsymbol{r}\right)$ with $\hat{H}=\sum_{\boldsymbol{\delta}}\hat{H}_{\boldsymbol{\delta}}e^{i\hat{\boldsymbol{p}}\cdot\boldsymbol{\delta}}$, $\mathcal{T}_{\boldsymbol{\delta}}=e^{i\hat{\boldsymbol{p}}\cdot\boldsymbol{\delta}}$ is the translation operator that satisfies $\mathcal{T}_{\boldsymbol{\delta}}f\left( \boldsymbol{r} \right)=e^{i\hat{\boldsymbol{p}}\cdot\boldsymbol{\delta}}f\left( \boldsymbol{r} \right)=f\left(\boldsymbol{r+\delta} \right)$, $\boldsymbol{\delta}$ is the vector shift between unit cells, and $\tilde{\Psi}\left(\boldsymbol{r}\right)= \left(1+ \boldsymbol{r} . \frac{\nabla \chi}{2} \right) \Psi\left(\boldsymbol{r}\right)=\xi\left( \boldsymbol{r}\right)\Psi\left(\boldsymbol{r}\right)$ with $\nabla \chi$ the temperature gradient. 

To simplify the notation, we adopt the Einstein summation convention in which repeated Roman indices imply summation over the BdG field operator indices, which range from $-N,-N+1,...-1,1,...,N-1,N$. Additionally, we introduce the notation $C_{A,B}=AB-BA=\left[A,B\right]$ representing the commutator of operators A and B and we set the Planck constant $\hbar=1$. With these conventions in place, the continuity equation can be expressed as follows
\begin{align}
    &\frac{\partial L^{\alpha}\left(\boldsymbol{r}\right)}{\partial t} = i \left[ \hat{H}, \hat{L}^{\alpha}\left(\boldsymbol{r}\right) \right] = i \left[ \frac{1}{2}\sum_{\boldsymbol{\delta}}\int d\boldsymbol{r}'\tilde{\Psi}^{\dagger}\left(\boldsymbol{r}' \right) \hat{H}_{\boldsymbol{\delta}}\tilde{\Psi}\left(\boldsymbol{r}'+\boldsymbol{\delta} \right),\frac{1}{2}\Psi^{\dagger}\left(\boldsymbol{r} \right)\hat{L}^{\alpha}\Psi\left(\boldsymbol{r} \right) \right] \\
    &=-\frac{i}{4} \sum_{\boldsymbol{\delta}}\int d\boldsymbol{r}' \left[\Psi^{\dagger}_{n}\left(\boldsymbol{r} \right)\hat{L}^{\alpha}_{nn'}\Psi_{n'}\left(\boldsymbol{r} \right)\tilde{\Psi}^{\dagger}_{m}\left( \boldsymbol{r}'\right)\left(\hat{H}_{\boldsymbol{\delta}} \right)_{mk}\tilde{\Psi}_{k}\left( \boldsymbol{r}'+\boldsymbol{\delta}\right) \right. \notag \\
    &\left.- \tilde{\Psi}^{\dagger}_{m}\left( \boldsymbol{r}'\right)\left(\hat{H}_{\boldsymbol{\delta}} \right)_{mk}\tilde{\Psi}_{k}\left( \boldsymbol{r}'+\boldsymbol{\delta}\right)\Psi^{\dagger}_{n}\left(\boldsymbol{r} \right)\hat{L}^{\alpha}_{nn'}\Psi_{n'}\left(\boldsymbol{r} \right)\right] \\
    &=-\frac{i}{2}\sum_{\boldsymbol{\delta}}\left\lbrace\Psi^{\dagger}_{n}\left(\boldsymbol{r} \right)\hat{L}^{\alpha}_{nn'}\sigma_{3}^{n'm}\left[\xi\left( \boldsymbol{r}\right)\hat{H}_{\boldsymbol{\delta}}\xi\left(\boldsymbol{r+\delta} \right) \right]_{mk}\Psi_{k}\left( \boldsymbol{r+\delta} \right) \right. \notag \\
    &\left.-\Psi^{\dagger}_{m}\left(\boldsymbol{r-\delta} \right)\left[\xi\left( \boldsymbol{r-\delta}\right)\hat{H}_{\boldsymbol{\delta}}\xi\left(\boldsymbol{\delta} \right) \right]_{mk}\sigma_{3}^{kn}\hat{L}^{\alpha}_{nn'}\Psi_{n'}\left(\boldsymbol{r} \right) \right\rbrace.
\end{align}
Using 
\begin{equation}
    \hat{L}^{\alpha}\xi\left(\boldsymbol{r} \right)= C_{\hat{L}^{\alpha},\xi\left(\boldsymbol{r} \right)} +\xi\left(\boldsymbol{r} \right)\hat{L}^{\alpha}
\end{equation}
one has
\begin{align}
    \frac{\partial L^{\alpha}\left(\boldsymbol{r}\right)}{\partial t} =-\frac{i}{2}\sum_{\boldsymbol{\delta}}\left[ \tilde{\Psi}^{\dagger}\left(\boldsymbol{r} \right) \hat{L}^{\alpha} \boldsymbol{\sigma}_{3}\hat{H}_{\boldsymbol{\delta}}\Tilde{\Psi}\left(\boldsymbol{r+\delta} \right)-\tilde{\Psi}^{\dagger}\left(\boldsymbol{r-\delta} \right)\hat{H}_{\boldsymbol{\delta}}\boldsymbol{\sigma}_{3}\hat{L}^{\alpha}\tilde{\Psi}\left( \boldsymbol{r}\right)\right] - \mathcal{O}_{1}
\end{align}
where
\begin{equation}
    \mathcal{O}_{1}=\frac{i}{2}\sum_{\boldsymbol{\delta}}\left\lbrace\Psi^{\dagger}\left(\boldsymbol{r} \right)C_{\hat{L}^{\alpha},\xi\left(\boldsymbol{r} \right)}\boldsymbol{\sigma}_{3}\hat{H}_{\boldsymbol{\delta}}\xi\left(\boldsymbol{r+\delta} \right)\Psi\left( \boldsymbol{r+\delta} \right) + \Psi^{\dagger}\left(\boldsymbol{r-\delta} \right)\xi\left( \boldsymbol{r-\delta}\right)\hat{H}_{\boldsymbol{\delta}}\boldsymbol{\sigma}_{3}C_{\hat{L}^{\alpha},\xi\left(\boldsymbol{r} \right)}\Psi\left(\boldsymbol{r} \right) \right\rbrace.
\end{equation}
Using the approximations $\tilde{\Psi}\left( \boldsymbol{r\pm \delta}\right)=\tilde{\Psi}\left(\boldsymbol{r} \right)\pm \boldsymbol{\delta}\cdot \boldsymbol{\nabla}\tilde{\Psi}\left(\boldsymbol{r} \right)$ and $\tilde{\Psi}\left( \boldsymbol{r}\right)=\tilde{\Psi}\left(\boldsymbol{r\pm \delta} \right)\mp \boldsymbol{\delta}\cdot \boldsymbol{\nabla}\tilde{\Psi}\left(\boldsymbol{r\pm \delta} \right)$, one gets:
\begin{align}
    &-\frac{i}{2}\sum_{\boldsymbol{\delta}} \tilde{\Psi}^{\dagger}\left(\boldsymbol{r} \right) \hat{L}^{\alpha} \boldsymbol{\sigma}_{3}\hat{H}_{\boldsymbol{\delta}}\Tilde{\Psi}\left(\boldsymbol{r+\delta} \right) = -\frac{i}{4}\sum_{\boldsymbol{\delta}} \tilde{\Psi}^{\dagger}\left(\boldsymbol{r} \right) \hat{L}^{\alpha} \boldsymbol{\sigma}_{3}\hat{H}_{\boldsymbol{\delta}}\Tilde{\Psi}\left(\boldsymbol{r+\delta} \right) -\frac{i}{4}\sum_{\boldsymbol{\delta}} \tilde{\Psi}^{\dagger}\left(\boldsymbol{r} \right) \hat{L}^{\alpha} \boldsymbol{\sigma}_{3}\hat{H}_{\boldsymbol{\delta}}\Tilde{\Psi}\left(\boldsymbol{r+\delta} \right) \\
    &=-\frac{i}{4}\sum_{\boldsymbol{\delta}}\left[ \tilde{\Psi}^{\dagger}\left(\boldsymbol{r-\delta} \right) + \boldsymbol{\delta}\cdot \boldsymbol{\nabla}\tilde{\Psi}^{\dagger}\left( \boldsymbol{r}\right)\right]  \hat{L}^{\alpha} \boldsymbol{\sigma}_{3}\hat{H}_{\boldsymbol{\delta}}\Tilde{\Psi}\left(\boldsymbol{r+\delta} \right) -\frac{i}{4}\sum_{\boldsymbol{\delta}} \tilde{\Psi}^{\dagger}\left(\boldsymbol{r} \right) \hat{L}^{\alpha} \boldsymbol{\sigma}_{3}\hat{H}_{\boldsymbol{\delta}} \left[ \tilde{\Psi}\left(\boldsymbol{r} \right)+ \boldsymbol{\delta}\cdot \boldsymbol{\nabla}\Tilde{\Psi}\left(\boldsymbol{r+\delta} \right)\right]\\
    &=-\frac{i}{4}\sum_{\boldsymbol{\delta}}\left[ \boldsymbol{\delta}\cdot \boldsymbol{\nabla}\tilde{\Psi}^{\dagger}\left( \boldsymbol{r}\right)\hat{L}^{\alpha} \boldsymbol{\sigma}_{3}\hat{H}_{\boldsymbol{\delta}}\Tilde{\Psi}\left(\boldsymbol{r+\delta} \right) + \tilde{\Psi}^{\dagger}\left(\boldsymbol{r} \right) \hat{L}^{\alpha} \boldsymbol{\sigma}_{3}\hat{H}_{\boldsymbol{\delta}} \boldsymbol{\delta}\cdot \boldsymbol{\nabla}\Tilde{\Psi}\left(\boldsymbol{r+\delta} \right) \right] \\
    &-\frac{i}{4}\sum_{\boldsymbol{\delta}} \left[\tilde{\Psi}^{\dagger}\left(\boldsymbol{r-\delta} \right)\hat{L}^{\alpha} \boldsymbol{\sigma}_{3}\hat{H}_{\boldsymbol{\delta}}\Tilde{\Psi}\left(\boldsymbol{r+\delta} \right)+ \tilde{\Psi}^{\dagger}\left(\boldsymbol{r} \right) \hat{L}^{\alpha} \boldsymbol{\sigma}_{3}\hat{H}_{\boldsymbol{\delta}}\tilde{\Psi}\left(\boldsymbol{r} \right)\right]\\
    &=-\frac{i}{4}\sum_{\boldsymbol{\delta}}\boldsymbol{\delta}\cdot \boldsymbol{\nabla}\left[\tilde{\Psi}^{\dagger}\left( \boldsymbol{r}\right)\hat{L}^{\alpha} \boldsymbol{\sigma}_{3}\hat{H}_{\boldsymbol{\delta}}\Tilde{\Psi}\left(\boldsymbol{r+\delta} \right) \right] -\frac{i}{2}\sum_{\boldsymbol{\delta}}\tilde{\Psi}^{\dagger}\left(\boldsymbol{r} \right)\hat{L}^{\alpha} \boldsymbol{\sigma}_{3}\hat{H}_{\boldsymbol{\delta}}\Tilde{\Psi}\left(\boldsymbol{r} \right) \notag \\
    &+ \frac{i}{4}\sum_{\boldsymbol{\delta}}\tilde{\Psi}^{\dagger}\left(\boldsymbol{r} \right)C_{\hat{L}^{\alpha},\mathcal{T}_{-\boldsymbol{\delta}}} \boldsymbol{\sigma}_{3}\hat{H}_{\boldsymbol{\delta}}e^{i\hat{\boldsymbol{p}}\cdot \boldsymbol{\delta}}\Tilde{\Psi}\left(\boldsymbol{r} \right)\\
    &=-\boldsymbol{\nabla}\frac{1}{4}\left[\tilde{\Psi}^{\dagger}\left( \boldsymbol{r}\right)\hat{L}^{\alpha} \boldsymbol{\sigma}_{3}\hat{\boldsymbol{v}}\Tilde{\Psi}\left(\boldsymbol{r} \right) \right]-\frac{i}{2}\sum_{\boldsymbol{\delta}}\tilde{\Psi}^{\dagger}\left(\boldsymbol{r} \right)\hat{L}^{\alpha} \boldsymbol{\sigma}_{3}\hat{H}_{\boldsymbol{\delta}}\Tilde{\Psi}\left(\boldsymbol{r} \right)+ \frac{i}{4}\sum_{\boldsymbol{\delta}}\tilde{\Psi}^{\dagger}\left(\boldsymbol{r} \right)C_{\hat{L}^{\alpha},\mathcal{T}_{-\boldsymbol{\delta}}}e^{i\hat{\boldsymbol{p}}\cdot \boldsymbol{\delta}} \boldsymbol{\sigma}_{3}\hat{H}_{\boldsymbol{\delta}}\Tilde{\Psi}\left(\boldsymbol{r} \right)
\end{align}
where $\hat{\boldsymbol{v}} = i\left[\hat{H},\hat{\boldsymbol{r}} \right]=i\sum_{\boldsymbol{\delta}}\boldsymbol{\delta}\cdot \hat{H}_{\boldsymbol{\delta}}e^{i\hat{\boldsymbol{p}}\boldsymbol{\delta}}$ is the velocity operator. Similarly:
\begin{align}
    &\frac{i}{2}\sum_{\boldsymbol{\delta}} \tilde{\Psi}^{\dagger}\left(\boldsymbol{r-\delta} \right)\hat{H}_{\boldsymbol{\delta}}\boldsymbol{\sigma}_{3}\hat{L}^{\alpha}\tilde{\Psi}\left( \boldsymbol{r}\right) =  \frac{i}{4}\sum_{\boldsymbol{\delta}} \tilde{\Psi}^{\dagger}\left(\boldsymbol{r-\delta} \right)\hat{H}_{\boldsymbol{\delta}}\boldsymbol{\sigma}_{3}\hat{L}^{\alpha}\tilde{\Psi}\left( \boldsymbol{r}\right) + \frac{i}{4}\sum_{\boldsymbol{\delta}} \tilde{\Psi}^{\dagger}\left(\boldsymbol{r-\delta} \right)\hat{H}_{\boldsymbol{\delta}}\boldsymbol{\sigma}_{3}\hat{L}^{\alpha}\tilde{\Psi}\left( \boldsymbol{r}\right)\\
    &=\frac{i}{4}\sum_{\boldsymbol{\delta}} \left[\tilde{\Psi}^{\dagger}\left(\boldsymbol{r} \right) - \boldsymbol{\delta}\cdot \boldsymbol{\nabla} \tilde{\Psi}^{\dagger}\left(\boldsymbol{r-\delta} \right) \right]\hat{H}_{\boldsymbol{\delta}}\boldsymbol{\sigma}_{3}\hat{L}^{\alpha}\tilde{\Psi}\left( \boldsymbol{r}\right) + \frac{i}{4}\sum_{\boldsymbol{\delta}} \tilde{\Psi}^{\dagger}\left(\boldsymbol{r-\delta} \right)\hat{H}_{\boldsymbol{\delta}}\boldsymbol{\sigma}_{3}\hat{L}^{\alpha} \left[ \tilde{\Psi}\left( \boldsymbol{r+\delta}\right)- \boldsymbol{\delta}\cdot\boldsymbol{\nabla}\tilde{\Psi}\left( \boldsymbol{r}\right) \right]\\
    &=-\frac{i}{4}\sum_{\boldsymbol{\delta}}\left[\boldsymbol{\delta}\cdot \boldsymbol{\nabla} \tilde{\Psi}^{\dagger}\left(\boldsymbol{r-\delta} \right)\hat{H}_{\boldsymbol{\delta}}\boldsymbol{\sigma}_{3}\hat{L}^{\alpha}\tilde{\Psi}\left( \boldsymbol{r}\right)+\tilde{\Psi}^{\dagger}\left(\boldsymbol{r-\delta} \right)\hat{H}_{\boldsymbol{\delta}}\boldsymbol{\sigma}_{3}\hat{L}^{\alpha}\boldsymbol{\delta}\cdot\boldsymbol{\nabla}\tilde{\Psi}\left( \boldsymbol{r}\right) \right]\\
    &+\frac{i}{4}\left[\tilde{\Psi}^{\dagger}\left(\boldsymbol{r} \right) \hat{H}_{\boldsymbol{\delta}}\boldsymbol{\sigma}_{3}\hat{L}^{\alpha}\tilde{\Psi}\left( \boldsymbol{r}\right)+\tilde{\Psi}^{\dagger}\left(\boldsymbol{r-\delta} \right)\hat{H}_{\boldsymbol{\delta}}\boldsymbol{\sigma}_{3}\hat{L}^{\alpha} \tilde{\Psi}\left( \boldsymbol{r+\delta}\right)\right]\\
    &=-\frac{i}{4}\sum_{\boldsymbol{\delta}}\boldsymbol{\delta}\cdot \boldsymbol{\nabla}\left[ \tilde{\Psi}^{\dagger}\left(\boldsymbol{r-\delta} \right)\hat{H}_{\boldsymbol{\delta}}\boldsymbol{\sigma}_{3}\hat{L}^{\alpha}\tilde{\Psi}\left( \boldsymbol{r}\right)\right]+\frac{i}{2}\tilde{\Psi}^{\dagger}\left(\boldsymbol{r} \right) \hat{H}_{\boldsymbol{\delta}}\boldsymbol{\sigma}_{3}\hat{L}^{\alpha}\tilde{\Psi}\left( \boldsymbol{r}\right) - \frac{i}{4}\sum_{\boldsymbol{\delta}}\tilde{\Psi}^{\dagger}\left(\boldsymbol{r} \right)\hat{H}_{\boldsymbol{\delta}}\boldsymbol{\sigma}_{3}C_{\hat{L}^{\alpha},\mathcal{T}_{-\boldsymbol{\delta}}} e^{i\hat{\boldsymbol{p}}\cdot \boldsymbol{\delta}}\tilde{\Psi}\left( \boldsymbol{r}\right)\\
    &=-\boldsymbol{\nabla}\frac{1}{4}\left[ \tilde{\Psi}^{\dagger}\left(\boldsymbol{r}\right) \hat{\boldsymbol{v}}\boldsymbol{\sigma}_{3}\hat{L}^{\alpha}\tilde{\Psi}\left( \boldsymbol{r}\right)\right]+\frac{i}{2}\tilde{\Psi}^{\dagger}\left(\boldsymbol{r} \right) \hat{H}_{\boldsymbol{\delta}}\boldsymbol{\sigma}_{3}\hat{L}^{\alpha}\tilde{\Psi}\left( \boldsymbol{r}\right) - \frac{i}{4}\sum_{\boldsymbol{\delta}}\tilde{\Psi}^{\dagger}\left(\boldsymbol{r} \right)\hat{H}_{\boldsymbol{\delta}}\boldsymbol{\sigma}_{3}C_{\hat{L}^{\alpha},\mathcal{T}_{-\boldsymbol{\delta}}} e^{i\hat{\boldsymbol{p}}\cdot \boldsymbol{\delta}}\tilde{\Psi}\left( \boldsymbol{r}\right).    
\end{align}
Consequently
\begin{align}\label{eq:OAM}
    \frac{\partial L^{\alpha}\left(\boldsymbol{r}\right)}{\partial t} &= - \boldsymbol{\nabla} \left[\tilde{\Psi}^{\dagger}\left(\boldsymbol{r}\right)\frac{ \left(\hat{L}^{\alpha}\boldsymbol{\sigma}_{3}\hat{\boldsymbol{v}}+\hat{\boldsymbol{v}}\boldsymbol{\sigma}_{3}\hat{L}^{\alpha}  \right)}{4}\tilde{\Psi}\left(\boldsymbol{r}\right)\right]-\frac{1}{2}\tilde{\Psi}^{\dagger}\left(\boldsymbol{r}\right) \left(\hat{L}^{\alpha}\boldsymbol{\sigma}_{3}\hat{H}_{\boldsymbol{\delta}}-\hat{H}_{\boldsymbol{\delta}}\boldsymbol{\sigma}_{3}\hat{L}^{\alpha}  \right)\tilde{\Psi}\left(\boldsymbol{r}\right) + \mathcal{O}_{1} + \mathcal{O}_{2} \notag \\
    & = -\boldsymbol{\nabla}\boldsymbol{j}^{L^{\alpha}} + \boldsymbol{S}^{L^{\alpha}}
\end{align}
where
\begin{equation}
    \mathcal{O}_{2} = \frac{i}{4}\sum_{\boldsymbol{\delta}}\tilde{\Psi}^{\dagger}\left(\boldsymbol{r} \right)C_{\hat{L}^{\alpha},\mathcal{T}_{-\boldsymbol{\delta}}}e^{i\hat{\boldsymbol{p}}\cdot \boldsymbol{\delta}} \boldsymbol{\sigma}_{3}\hat{H}_{\boldsymbol{\delta}}\Tilde{\Psi}\left(\boldsymbol{r} \right)- \frac{i}{4}\sum_{\boldsymbol{\delta}}\tilde{\Psi}^{\dagger}\left(\boldsymbol{r} \right)\hat{H}_{\boldsymbol{\delta}}\boldsymbol{\sigma}_{3}C_{\hat{L}^{\alpha},\mathcal{T}_{-\boldsymbol{\delta}}} e^{i\hat{\boldsymbol{p}}\cdot \boldsymbol{\delta}}\tilde{\Psi}\left( \boldsymbol{r}\right)    
\end{equation}
and
\begin{align}
    \boldsymbol{S}^{L^{\alpha}} = -\frac{1}{2}\tilde{\Psi}^{\dagger}\left(\boldsymbol{r}\right) \left(\hat{L}^{\alpha}\boldsymbol{\sigma}_{3}\hat{H}_{\boldsymbol{\delta}}-\hat{H}_{\boldsymbol{\delta}}\boldsymbol{\sigma}_{3}\hat{L}^{\alpha}  \right)\tilde{\Psi}\left(\boldsymbol{r}\right) + \mathcal{O}_{1} + \mathcal{O}_{2}
    \label{Eqnsourceterm}
\end{align}
is the orbital source density corresponding to the orbital torque density \cite{Go2020} and $\boldsymbol{j}^{L^{\alpha}} = \tilde{\Psi}^{\dagger}\left(\boldsymbol{r}\right)\frac{ \left(\hat{L}^{\alpha}\boldsymbol{\sigma}_{3}\hat{\boldsymbol{v}}+\hat{\boldsymbol{v}}\boldsymbol{\sigma}_{3}\hat{L}^{\alpha}  \right)}{4}\tilde{\Psi}\left(\boldsymbol{r}\right)$ is the local orbital current density. The orbital angular moment current operator can be defined as $\hat{\boldsymbol{j}}^{L^{\alpha}} = \frac{1}{4}\left( \hat{L}^{\alpha}\boldsymbol{\sigma}_{3}\hat{\boldsymbol{v}}+\hat{\boldsymbol{v}}\boldsymbol{\sigma}_{3}\hat{L}^{\alpha}\right)$.

We note that in addition to the conventional source term $-\frac{1}{2}\tilde{\Psi}^{\dagger}\left(\boldsymbol{r}\right) \left(\hat{L}^{\alpha}\boldsymbol{\sigma}_{3}\hat{H}_{\boldsymbol{\delta}}-\hat{H}_{\boldsymbol{\delta}}\boldsymbol{\sigma}_{3}\hat{L}^{\alpha}  \right)\tilde{\Psi}\left(\boldsymbol{r}\right)$, Eq.~\eqref{Eqnsourceterm} also incorporates the terms $\mathcal{O}_{1}$ and $\mathcal{O}_{2}$ arising due to the nonzero commutator $\left[\hat{\boldsymbol{L}},\hat{\boldsymbol{r}} \right] \neq 0$. This indicates that even in a scenario where $-\frac{1}{2}\tilde{\Psi}^{\dagger}\left(\boldsymbol{r}\right) \left(\hat{L}^{\alpha}\boldsymbol{\sigma}_{3}\hat{H}_{\boldsymbol{\delta}}-\hat{H}_{\boldsymbol{\delta}}\boldsymbol{\sigma}_{3}\hat{L}^{\alpha}  \right)\tilde{\Psi}\left(\boldsymbol{r}\right) = 0$, the conservation of Orbital Angular Moment is not guaranteed. Additional considerations such as source density and magnon orbital torque are beyond the scope of this study. Instead, our focus will be exclusively on the orbital current term. The inclusion of a source term will result in some dissipation; nevertheless, our findings remain valid within orbital relaxation time. 

\begin{figure}[h]
\centering
    \includegraphics[width= 1\textwidth]{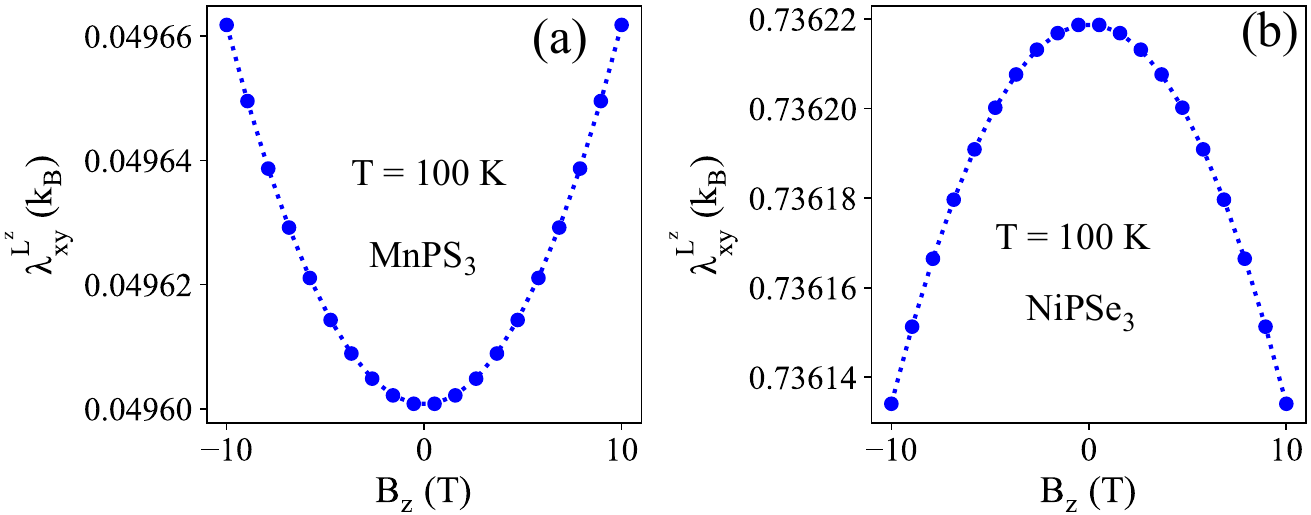}
 \caption{The Orbital Nernst conductivity of MnPS$_3$ (a) and NiPSe$_3$ (b) as a function of externally applied magnetic field along z-direction $B_{z}$ computed at fixed temperature $T=100~K$.}
  \label{S3}
\end{figure}

\subsection{Responses of magnonic system to thermal gradient: Linear response theory}
We are now deriving the expression describing the response of the magnon orbital angular moment current to a temperature gradient within the linear response theory. In this pursuit, we adopt the work of Matsumoto et al. in Ref.~\cite{Matsumoto2014}, Zyuzin et al. in Ref.~\cite{Zyuzin2016}, and Park et al. in Ref.~\cite{Park2020} to deal with magnon orbital angular moment current. Our analysis involves a 2D magnon system exposed to a temperature gradient, and the resulting orbital current can be represented as
\begin{equation}
    \boldsymbol{j}^{L^{\alpha}}\left( \boldsymbol{r} \right) = \tilde{\Psi}^{\dagger}\left(\boldsymbol{r} \right)\left[ \frac{\hat{L}^{\alpha}\boldsymbol{\sigma}_{3}\hat{\boldsymbol{v}}+\hat{\boldsymbol{v}}\boldsymbol{\sigma}_{3}\hat{L}^{\alpha}}{4}\right] \tilde{\Psi}\left(\boldsymbol{r} \right)
\end{equation}
where $\tilde{\Psi}\left(\boldsymbol{r} \right) = \left(1+\boldsymbol{r} \cdot \frac{\boldsymbol{\nabla}\chi}{2} \right)\Psi\left(\boldsymbol{r} \right)$ so that
\begin{equation}
    \boldsymbol{j}^{L^{\alpha}}\left( \boldsymbol{r} \right) = \Psi^{\dagger}\left(\boldsymbol{r} \right)\left(1+\boldsymbol{r} \cdot \frac{\boldsymbol{\nabla}\chi}{2} \right)\left[ \frac{\hat{L}^{\alpha}\boldsymbol{\sigma}_{3}\hat{\boldsymbol{v}}+\hat{\boldsymbol{v}}\boldsymbol{\sigma}_{3}\hat{L}^{\alpha}}{4}\right] \left(1+\boldsymbol{r} \cdot \frac{\boldsymbol{\nabla}\chi}{2} \right)\Psi\left(\boldsymbol{r} \right).
\end{equation}
To linear order in the temperature gradient the Orbital angular moment current is decomposed as
\begin{equation}
    \boldsymbol{j}^{L^{\alpha}}\left( \boldsymbol{r} \right) = \boldsymbol{j}^{L^{\alpha}(0)}\left( \boldsymbol{r} \right) + \boldsymbol{j}^{L^{\alpha}(1)}\left( \boldsymbol{r} \right)
\end{equation}
where
\begin{equation}
    \boldsymbol{j}^{L^{\alpha}(0)}\left( \boldsymbol{r} \right) = \Psi^{\dagger}\left(\boldsymbol{r} \right)\frac{\hat{L}^{\alpha}\boldsymbol{\sigma}_{3}\hat{\boldsymbol{v}}+\hat{\boldsymbol{v}}\boldsymbol{\sigma}_{3}\hat{L}^{\alpha}}{4}\Psi \left(\boldsymbol{r} \right)
\end{equation}
and
\begin{equation}
    \boldsymbol{j}^{L^{\alpha}(1)}\left( \boldsymbol{r} \right) = \frac{1}{2}\Psi^{\dagger}\left(\boldsymbol{r} \right)\left[\left(\frac{\hat{L}^{\alpha}\boldsymbol{\sigma}_{3}\hat{\boldsymbol{v}}+\hat{\boldsymbol{v}}\boldsymbol{\sigma}_{3}\hat{L}^{\alpha}}{4}\right)\hat{\boldsymbol{r}}+\hat{\boldsymbol{r}}\left(\frac{\hat{L}^{\alpha}\boldsymbol{\sigma}_{3}\hat{\boldsymbol{v}}+\hat{\boldsymbol{v}}\boldsymbol{\sigma}_{3}\hat{L}^{\alpha}}{4}\right)\right]\Psi \left(\boldsymbol{r} \right)\boldsymbol{\nabla}\chi.
\end{equation}
Defining $\boldsymbol{J}^{L^{\alpha}}=\int d \boldsymbol{r}~\boldsymbol{j}^{L^{\alpha}}\left(\boldsymbol{r}\right)$, one obtains
\begin{equation}
    \left\langle \boldsymbol{J}^{L^{\alpha}} \right\rangle =  \left\langle \boldsymbol{J}^{L^{\alpha}(0)} \right\rangle_{neq} + \left\langle \boldsymbol{J}^{L^{\alpha}(1)} \right\rangle_{eq}.
\end{equation}

Employing the Kubo formula \cite{kubo1957a,kubo1957b,Mahan2013} we evaluate the first term
\begin{equation}
    \left\langle J_{\mu}^{L^{\alpha}(0)} \right\rangle_{neq} = -\lim_{\omega \rightarrow 0}\frac{\partial}{\partial \omega}\int_{0}^{\beta}d \tau e^{i \omega \tau} \left\langle T_{\tau}J_{\mu}^{L^{\alpha}(0)}\left( \tau\right)J_{\nu}^{(Q)}\left( 0\right) \right\rangle \nabla_{\nu}\chi \equiv -S_{\mu\nu}\nabla_{\nu}\chi
\end{equation}
where 
\begin{align}
    \boldsymbol{J}^{(Q)} = \frac{1}{4}\sum_{\boldsymbol{k}}\Psi^{\dagger}_{\boldsymbol{k}}\left[\hat{H}_{\boldsymbol{k}} \boldsymbol{\sigma}_{3}\hat{\boldsymbol{v}}_{\boldsymbol{k}}+\hat{\boldsymbol{v}}_{\boldsymbol{k}}\boldsymbol{\sigma}_{3}\hat{H}_{\boldsymbol{k}} \right]\Psi_{\boldsymbol{k}}
\end{align}
and
\begin{equation}
    \boldsymbol{J}^{L^{\alpha}(0)} = \frac{1}{4}\sum_{\boldsymbol{k}}\Psi^{\dagger}_{\boldsymbol{k}}\left[\hat{L}^{\alpha} \boldsymbol{\sigma}_{3}\hat{\boldsymbol{v}}_{\boldsymbol{k}}+\hat{\boldsymbol{v}}_{\boldsymbol{k}}\boldsymbol{\sigma}_{3}\hat{L}^{\alpha} \right]\Psi_{\boldsymbol{k}}.
\end{equation}
Introducing the field operator for the energy eigenstates
\begin{equation}
    \Phi_{\boldsymbol{k}} = T_{\boldsymbol{k}}\Psi_{\boldsymbol{k}}
\end{equation}
one has
\begin{align}
    S_{\mu\nu} = \frac{1}{16}\lim_{\omega \rightarrow 0}\frac{\partial}{\partial \omega}\int_{0}^{\beta}d\tau e^{i\omega \tau}\sum_{\boldsymbol{k,k'}}\left\langle \Phi_{\boldsymbol{k}}^{\dagger}\left(\tau \right)T_{\boldsymbol{k}}^{\dagger}\left(\hat{L}^{\alpha}\boldsymbol{\sigma}_{3}\hat{v}_{\boldsymbol{k},\mu}+\hat{v}_{\boldsymbol{k},\mu}\boldsymbol{\sigma}_{3}\hat{L}^{\alpha} \right)T_{\boldsymbol{k}}\Phi_{\boldsymbol{k}}\left( \tau\right) \Phi_{\boldsymbol{k}'}^{\dagger}\left(0 \right) \right. \times \notag \\
    \times \left.T_{\boldsymbol{k}'}^{\dagger}\left(\hat{H}_{\boldsymbol{k}'}\boldsymbol{\sigma}_{3}\hat{v}_{\boldsymbol{k}',\nu}+\hat{v}_{\boldsymbol{k}',\nu}\boldsymbol{\sigma}_{3}\hat{H}_{\boldsymbol{k}'} \right)T_{\boldsymbol{k}'}\Phi_{\boldsymbol{k}'}\left( 0\right)\right\rangle.
\end{align}
Using the identity
\begin{align}
    &\left\langle \Phi_{\boldsymbol{k},m}^{\dagger}\left(\tau \right)\Phi_{\boldsymbol{k},n}\left( \tau\right) \Phi_{\boldsymbol{k}',p}^{\dagger}\left(0 \right)\Phi_{\boldsymbol{k}',q}\left( 0\right)\right\rangle = \left\langle \Phi_{\boldsymbol{k},m}^{\dagger}\left(\tau \right)\Phi_{\boldsymbol{k},n}\left( \tau\right)\right\rangle \left\langle \Phi_{\boldsymbol{k}',p}^{\dagger}\left(0 \right)\Phi_{\boldsymbol{k}',q}\left( 0\right)\right\rangle \notag \\
    &+\left\langle \Phi_{\boldsymbol{k},m}^{\dagger}\left(\tau \right)\Phi_{\boldsymbol{k}',p}^{\dagger}\left(0 \right)\right\rangle \left\langle \Phi_{\boldsymbol{k},n}\left( \tau\right) \Phi_{\boldsymbol{k}',q}\left( 0\right)\right\rangle+\left\langle \Phi_{\boldsymbol{k},m}^{\dagger}\left(\tau \right)\Phi_{\boldsymbol{k}',q}\left( 0\right)\right\rangle \left\langle \Phi_{\boldsymbol{k},n}\left( \tau\right)\Phi_{\boldsymbol{k}',p}^{\dagger}\left(0 \right)\right\rangle
\end{align}
the integral $\int_{0}^{\beta}e^{i\omega \tau}\left\langle \Phi_{\boldsymbol{k},m}^{\dagger}\left(\tau \right)\Phi_{\boldsymbol{k},n}\left( \tau\right)\right\rangle \left\langle \Phi_{\boldsymbol{k}',p}^{\dagger}\left(0 \right)\Phi_{\boldsymbol{k}',q}\left( 0\right)\right\rangle=0$ because $\omega=\frac{2n \pi}{\beta}$ and
\begin{align}
    \left\langle \Phi_{\boldsymbol{k},m}^{\dagger}\left(\tau \right)\Phi_{\boldsymbol{k}',p}^{\dagger}\left(0 \right)\right\rangle &=i\delta_{\boldsymbol{k,-k'}}\sigma_{2}^{mp}\rho\left[\left(\boldsymbol{\sigma}_{3}E_{\boldsymbol{k}} \right)_{mm} \right]e^{\left(\boldsymbol{\sigma}_{3}E_{\boldsymbol{k}}\right)_{mm}\tau}  \\
    \left\langle \Phi_{\boldsymbol{k},n}\left( \tau\right) \Phi_{\boldsymbol{k}',q}\left( 0\right)\right\rangle&=-i\delta_{\boldsymbol{k,-k'}}\sigma_{2}^{nq}\rho\left[-\left(\boldsymbol{\sigma}_{3}E_{\boldsymbol{k}} \right)_{nn} \right]e^{-\left(\boldsymbol{\sigma}_{3}E_{\boldsymbol{k}}\right)_{nn}\tau}\\
    \left\langle \Phi_{\boldsymbol{k},m}^{\dagger}\left(\tau \right)\Phi_{\boldsymbol{k}',q}\left( 0\right)\right\rangle &=\delta_{\boldsymbol{k,k'}}\sigma_{3}^{mq}\rho\left[\left(\boldsymbol{\sigma}_{3}E_{\boldsymbol{k}} \right)_{mm} \right]e^{\left(\boldsymbol{\sigma}_{3}E_{\boldsymbol{k}}\right)_{mm}\tau}\\
    \left\langle \Phi_{\boldsymbol{k},n}\left( \tau\right)\Phi_{\boldsymbol{k}',p}^{\dagger}\left(0 \right)\right\rangle&=-\delta_{\boldsymbol{k,k'}}\sigma_{3}^{np}\rho\left[\left(-\boldsymbol{\sigma}_{3}E_{\boldsymbol{k}} \right)_{nn} \right]e^{-\left(\boldsymbol{\sigma}_{3}E_{\boldsymbol{k}}\right)_{nn}\tau}
\end{align}
where $\rho(x)=\frac{1}{e^{\beta x}-1}$ is the Bose-Einstein distribution with $\beta = k_{B}T$. We define
\begin{align}
    \boldsymbol{\mathcal{V}}_{\boldsymbol{k}}^{L^{\alpha}} &= T_{\boldsymbol{k}}^{\dagger}\left[\hat{L}^{\alpha} \boldsymbol{\sigma}_{3}\hat{\boldsymbol{v}}_{\boldsymbol{k}}+\hat{\boldsymbol{v}}_{\boldsymbol{k}}\boldsymbol{\sigma}_{3}\hat{L}^{\alpha} \right]T_{\boldsymbol{k}},\\
    \boldsymbol{\mathcal{V}}_{\boldsymbol{k}}&=T_{\boldsymbol{k}}^{\dagger}\hat{\boldsymbol{v}}_{\boldsymbol{k}}T_{\boldsymbol{k}},
\end{align}
which leads to
\begin{align}
    S_{\mu\nu}=\frac{1}{16}\lim_{\omega \rightarrow 0}\frac{\partial}{\partial \omega}\int_{0}^{\beta}&d\tau e^{i\omega \tau}\sum_{\boldsymbol{k,k'}}\left[ \boldsymbol{\mathcal{V}}_{\boldsymbol{k},\mu}^{L^{\alpha}}\right]_{mn}\left[E_{\boldsymbol{k}'}\boldsymbol{\sigma}_{3}\boldsymbol{\mathcal{V}}_{\boldsymbol{k}',\nu} +\boldsymbol{\mathcal{V}}_{\boldsymbol{k}',\nu} \boldsymbol{\sigma}_{3}E_{\boldsymbol{k}'}\right]_{pq} \left[\sigma_{2}^{mp}\sigma_{2}^{nq}\delta_{\boldsymbol{k},-\boldsymbol{k}'} -\sigma_{3}^{mq}\sigma_{3}^{np}\delta_{\boldsymbol{k,k'}}\right] \notag \\
    & \times\rho\left[\left(\boldsymbol{\sigma}_{3}E_{\boldsymbol{k}} \right)_{mm} \right]\rho\left[-\left(\boldsymbol{\sigma}_{3}E_{\boldsymbol{k}} \right)_{nn} \right]e^{\left[\left(\boldsymbol{\sigma}_{3}E_{\boldsymbol{k}} \right)_{mm} - \left(\boldsymbol{\sigma}_{3}E_{\boldsymbol{k}} \right)_{nn} \right]\tau}.
\end{align}
Using the integral
\begin{equation}
    \lim_{\omega \rightarrow 0}\frac{\partial}{\partial \omega}\int_{0}^{\beta}d\tau e^{\left[i\omega + \left(\boldsymbol{\sigma}_{3}E_{\boldsymbol{k}} \right)_{mm} - \left(\boldsymbol{\sigma}_{3}E_{\boldsymbol{k}} \right)_{nn}  \right]\tau} = \lim_{\omega \rightarrow 0}\frac{\partial}{\partial \omega} \frac{e^{\beta \left(\boldsymbol{\sigma}_{3}E_{\boldsymbol{k}}\right)_{mm}-\beta \left(\boldsymbol{\sigma}_{3}E_{\boldsymbol{k}} \right)_{nn}}-1}{i\omega + \left(\boldsymbol{\sigma}_{3}E_{\boldsymbol{k}}\right)_{mm}- \left(\boldsymbol{\sigma}_{3}E_{\boldsymbol{k}} \right)_{nn}}=-i\frac{e^{\beta \left(\boldsymbol{\sigma}_{3}E_{\boldsymbol{k}}\right)_{mm}-\beta \left(\boldsymbol{\sigma}_{3}E_{\boldsymbol{k}} \right)_{nn}}-1}{\left[\left(\boldsymbol{\sigma}_{3}E_{\boldsymbol{k}}\right)_{mm}- \left(\boldsymbol{\sigma}_{3}E_{\boldsymbol{k}} \right)_{nn}\right]^{2}}
\end{equation}
together with
\begin{equation}
    \rho(x)-\rho(-y)=-\rho(x)\rho(-y)\left( e^{\beta x -\beta y} -1 \right)
\end{equation}
one obtains
\begin{align}
    S_{\mu\nu} &= \frac{i}{16}\sum_{\boldsymbol{k,k'}}\left[ \boldsymbol{\mathcal{V}}_{\boldsymbol{k},\mu}^{L^{\alpha}}\right]_{mn}\left[E_{\boldsymbol{k}'}\boldsymbol{\sigma}_{3}\boldsymbol{\mathcal{V}}_{\boldsymbol{k}',\nu} +\boldsymbol{\mathcal{V}}_{\boldsymbol{k}',\nu} \boldsymbol{\sigma}_{3}E_{\boldsymbol{k}'}\right]_{pq} \left[\sigma_{2}^{mp}\sigma_{2}^{nq}\delta_{\boldsymbol{k},-\boldsymbol{k}'} -\sigma_{3}^{mq}\sigma_{3}^{np}\delta_{\boldsymbol{k,k'}}\right] \times \notag\\ 
    &\times \frac{\rho\left[ \left( \boldsymbol{\sigma}_{3}E_{\boldsymbol{k}}\right)_{mm}\right]-\rho\left[ \left( \boldsymbol{\sigma}_{3}E_{\boldsymbol{k}}\right)_{nn}\right]}{\left[\left(\boldsymbol{\sigma}_{3}E_{\boldsymbol{k}}\right)_{mm}-\left(\boldsymbol{\sigma}_{3}E_{\boldsymbol{k}}\right)_{nn} \right]^{2}} \\
    & = -\frac{i}{8}\sum_{\boldsymbol{k}}\left[ \boldsymbol{\mathcal{V}}_{\boldsymbol{k},\mu}^{L^{\alpha}}\right]_{mn}\left[\boldsymbol{\sigma}_{3}\left(T_{\boldsymbol{k}}^{\dagger}\hat{v}_{\boldsymbol{k},\nu} T_{\boldsymbol{k}}\boldsymbol{\sigma}_{3}E_{\boldsymbol{k}} + E_{\boldsymbol{k}}\boldsymbol{\sigma}_{3}T_{\boldsymbol{k}}^{\dagger}\hat{v}_{\boldsymbol{k},\nu} T_{\boldsymbol{k}}\right)\boldsymbol{\sigma}_{3} \right]_{nm}\frac{\rho\left[ \left( \boldsymbol{\sigma}_{3}E_{\boldsymbol{k}}\right)_{mm}\right]-\rho\left[ \left( \boldsymbol{\sigma}_{3}E_{\boldsymbol{k}}\right)_{nn}\right]}{\left[\left(\boldsymbol{\sigma}_{3}E_{\boldsymbol{k}}\right)_{mm}-\left(\boldsymbol{\sigma}_{3}E_{\boldsymbol{k}}\right)_{nn} \right]^{2}}\\
    &=-\frac{i}{8}\sum_{\boldsymbol{k}}\langle n \left(\boldsymbol{k} \right) \vert \hat{L}^{\alpha}\boldsymbol{\sigma}_{3}\hat{v}_{\boldsymbol{k},\mu}  +   \hat{v}_{\boldsymbol{k},\mu}\boldsymbol{\sigma}_{3}\hat{L}^{\alpha}\vert m\left(\boldsymbol{k} \right) \rangle \langle m\left(\boldsymbol{k} \right) \vert \hat{v}_{\boldsymbol{k},\nu}\vert n\left(\boldsymbol{k} \right)\rangle \times \notag \\
    &\times \frac{\left(E_{\boldsymbol{k}}^{nn} \sigma_{3}^{mm} + E_{\boldsymbol{k}}^{mm}\sigma_{3}^{nn} \right)\left\lbrace\rho\left[ \left( \boldsymbol{\sigma}_{3}E_{\boldsymbol{k}}\right)_{mm}\right]-\rho\left[ \left( \boldsymbol{\sigma}_{3}E_{\boldsymbol{k}}\right)_{nn}\right]\right\rbrace}{\left[\left(\boldsymbol{\sigma}_{3}E_{\boldsymbol{k}}\right)_{mm}-\left(\boldsymbol{\sigma}_{3}E_{\boldsymbol{k}}\right)_{nn} \right]^{2}}.
\end{align}
Consequently,
\begin{align}
    &S_{\mu\nu} = -\frac{i}{8}\sum_{\boldsymbol{k}}\langle n \left(\boldsymbol{k} \right) \vert \hat{L}^{\alpha}\boldsymbol{\sigma}_{3}\hat{v}_{\boldsymbol{k},\mu}  +   \hat{v}_{\boldsymbol{k},\mu}\boldsymbol{\sigma}_{3}\hat{L}^{\alpha}\vert m\left(\boldsymbol{k} \right) \rangle \langle m \left(\boldsymbol{k} \right)\vert \hat{v}_{\boldsymbol{k},\nu}\vert n\left(\boldsymbol{k} \right)\rangle \times \notag \\
    &\times \frac{\left(E_{\boldsymbol{k}}^{nn} \sigma_{3}^{mm} + E_{\boldsymbol{k}}^{mm}\sigma_{3}^{nn} \right)\left\lbrace\rho\left[ \left( \boldsymbol{\sigma}_{3}E_{\boldsymbol{k}}\right)_{mm}\right]-\rho\left[ \left( \boldsymbol{\sigma}_{3}E_{\boldsymbol{k}}\right)_{nn}\right]\right\rbrace}{\left[\left(\boldsymbol{\sigma}_{3}E_{\boldsymbol{k}}\right)_{mm}-\left(\boldsymbol{\sigma}_{3}E_{\boldsymbol{k}}\right)_{nn} \right]^{2}}
\end{align}

In our next step we are going to evaluate
\begin{equation}
    \left\langle J_{\mu}^{\hat{L}^{\alpha}(1)} \right\rangle_{eq}=-M_{\mu\nu}\nabla_{\nu}\chi
\end{equation}
by using the Smrcka and Streda method \cite{Smrcka1977,Matsumoto2014,Zyuzin2016}, where
\begin{equation}
    M_{\mu\nu} = -\frac{1}{8}\sum_{\boldsymbol{k}}\int d \eta \rho\left(\eta \right)Tr\left\lbrace\boldsymbol{\sigma}_{3}\left[\left(\hat{L}^{\alpha}\boldsymbol{\sigma}_{3}\hat{v}_{\mu,\boldsymbol{k}}+\hat{v}_{\mu,\boldsymbol{k}}\boldsymbol{\sigma}_{3}\hat{L}^{\alpha}\right)\hat{r}_{\nu}+\hat{r}_{\nu}\left(\hat{L}^{\alpha}\boldsymbol{\sigma}_{3}\hat{v}_{\mu,\boldsymbol{k}}+\hat{v}_{\mu,\boldsymbol{k}}\boldsymbol{\sigma}_{3}\hat{L}^{\alpha}\right)\right]\delta\left(\eta -\boldsymbol{\sigma}_{3}\hat{H}_{\boldsymbol{k}} \right) \right\rbrace.
\end{equation}
We first define
\begin{align}
    \hat{\boldsymbol{w}}_{\boldsymbol{k}} = \hat{L}^{\alpha}\boldsymbol{\sigma}_{3}\hat{\boldsymbol{v}}_{\boldsymbol{k}}, ~~~~~~~~\hat{\boldsymbol{u}}_{\boldsymbol{k}} = \hat{\boldsymbol{v}}_{\boldsymbol{k}}\boldsymbol{\sigma}_{3}\hat{L}^{\alpha}.
\end{align}
For the sake of simplicity, we temporarily omit the notation dependent on the wave vector $\boldsymbol{k}$ and will reintroduce it at the conclusion of our discussion. We introduce
\begin{align}
    A_{\mu\nu} = \frac{i}{2}Tr \left[\boldsymbol{\sigma}_{3}\hat{w}_{\mu}\frac{dG^{+}}{d \eta}\boldsymbol{\sigma}_{3}\hat{v}_{\nu}\delta\left( \eta -\boldsymbol{\sigma}_{3}\hat{H}\right) - \boldsymbol{\sigma}_{3}\hat{w}_{\mu}\delta\left(\eta -\boldsymbol{\sigma}_{3}\hat{H} \right)\boldsymbol{\sigma}_{3}\hat{v}_{\nu}\frac{dG^{-}}{d\eta} \right]\\
    B_{\mu\nu}=\frac{i}{2}Tr \left[\boldsymbol{\sigma}_{3}\hat{w}_{\mu}G^{+}\boldsymbol{\sigma}_{3}\hat{v}_{\nu}\delta\left( \eta -\boldsymbol{\sigma}_{3}\hat{H}\right) - \boldsymbol{\sigma}_{3}\hat{w}_{\mu}\delta\left(\eta -\boldsymbol{\sigma}_{3}\hat{H} \right)\boldsymbol{\sigma}_{3}\hat{v}_{\nu}G^{-} \right]
\end{align}
where 
\begin{equation}
    G^{\pm} = \frac{1}{\eta \pm i\epsilon-\sigma_{3}\hat{H}}
\end{equation}
is the Green's function that satisfies:
\begin{align} \label{Greenre}
    i\delta\left(\eta - \boldsymbol{\sigma}_{3}\hat{H} \right)=-\frac{1}{2 \pi}\left(G^{+}-G^{-} \right), ~~~~~ \frac{dG^{\pm}}{d\eta} = -\left( G^{\pm}\right)^{2},~~~~  i\frac{d}{d\eta}\delta\left(\eta - \boldsymbol{\sigma}_{3} \hat{H}\right) = \frac{1}{2\pi}\left[\left( G^{+}\right)^{2}-\left( G^{-}\right)^{2} \right]
\end{align}
so that one has
\begin{align}
    \frac{dB_{\mu\nu}}{d\eta} &= \frac{i}{2}Tr \left[\boldsymbol{\sigma}_{3}\hat{w}_{\mu}\frac{dG^{+}}{d\eta}\boldsymbol{\sigma}_{3}\hat{v}_{\nu}\delta\left( \eta -\boldsymbol{\sigma}_{3}\hat{H}\right) + \boldsymbol{\sigma}_{3}\hat{w}_{\mu}G^{+}\boldsymbol{\sigma}_{3}\hat{v}_{\nu}\frac{d\delta\left( \eta -\boldsymbol{\sigma}_{3}\hat{H}\right)}{d\eta}\right. \\
    &- \left.\boldsymbol{\sigma}_{3}\hat{w}_{\mu}\frac{d\delta\left(\eta -\boldsymbol{\sigma}_{3}\hat{H} \right)}{d\eta}\boldsymbol{\sigma}_{3}\hat{v}_{\nu}G^{-} - \boldsymbol{\sigma}_{3}\hat{w}_{\mu}\delta\left(\eta -\boldsymbol{\sigma}_{3}\hat{H} \right)\boldsymbol{\sigma}_{3}\hat{v}_{\nu}\frac{dG^{-}}{d\eta} \right].
\end{align}
Consequently,
\begin{align}
    &A_{\mu\nu}-\frac{1}{2}\frac{dB_{\mu\nu}}{d\eta} =  \frac{i}{4}Tr \left[\boldsymbol{\sigma}_{3}\hat{w}_{\mu}\frac{dG^{+}}{d\eta}\boldsymbol{\sigma}_{3}\hat{v}_{\nu}\delta\left( \eta -\boldsymbol{\sigma}_{3}\hat{H}\right) - \boldsymbol{\sigma}_{3}\hat{w}_{\mu}G^{+}\boldsymbol{\sigma}_{3}\hat{v}_{\nu}\frac{d\delta\left( \eta -\boldsymbol{\sigma}_{3}\hat{H}\right)}{d\eta}\right. \notag \\
    &+ \left.\boldsymbol{\sigma}_{3}\hat{w}_{\mu}\frac{d\delta\left(\eta -\boldsymbol{\sigma}_{3}\hat{H} \right)}{d\eta}\boldsymbol{\sigma}_{3}\hat{v}_{\nu}G^{-} - \boldsymbol{\sigma}_{3}\hat{w}_{\mu}\delta\left(\eta -\boldsymbol{\sigma}_{3}\hat{H} \right)\boldsymbol{\sigma}_{3}\hat{v}_{\nu}\frac{dG^{-}}{d\eta} \right] \\
    & = \frac{1}{8\pi}Tr \left[\boldsymbol{\sigma}_{3}\hat{w}_{\mu}\left( G^{+}\right)^{2}\boldsymbol{\sigma}_{3}\hat{v}_{\nu}\left(G^{+}-G^{-} \right) - \boldsymbol{\sigma}_{3}\hat{w}_{\mu}G^{+}\boldsymbol{\sigma}_{3}\hat{v}_{\nu}\left[\left( G^{+}\right)^{2}-\left( G^{-}\right)^{2} \right]\right. \notag \\
    &+ \left.\boldsymbol{\sigma}_{3}\hat{w}_{\mu}\left[\left( G^{+}\right)^{2}-\left( G^{-}\right)^{2} \right]\boldsymbol{\sigma}_{3}\hat{v}_{\nu}G^{-} - \boldsymbol{\sigma}_{3}\hat{w}_{\mu}\left(G^{+}-G^{-} \right)\boldsymbol{\sigma}_{3}\hat{v}_{\nu}\left( G^{-}\right)^{2} \right]  \\
    &= \frac{1}{8\pi}Tr\left[\boldsymbol{\sigma}_{3}\hat{w}_{\mu}\left(G^{+}\right)^{2}\boldsymbol{\sigma}_{3}\hat{v}_{\nu}G^{+}+\boldsymbol{\sigma}_{3}\hat{w}_{\mu}G^{-}\boldsymbol{\sigma}_{3}\hat{v}_{\nu}\left(G^{-}\right)^{2} -\boldsymbol{\sigma}_{3}\hat{w}_{\mu}G^{+}\boldsymbol{\sigma}_{3}\hat{v}_{\nu}\left(G^{+}\right)^{2}-\boldsymbol{\sigma}_{3}\hat{w}_{\mu}\left(G^{-}\right)^{2}\boldsymbol{\sigma}_{3}\hat{v}_{\nu}G^{-}\right].
\end{align}
Note that
\begin{align}
    \hat{\boldsymbol{v}} = i\left[\hat{\boldsymbol{r}},\boldsymbol{\sigma}_{3}\left(G^{\pm}\right)^{-1} \right], ~~~~\hat{\boldsymbol{w}}=i\boldsymbol{\sigma}_{3}\hat{L}^{\alpha}\left[\hat{\boldsymbol{r}},\boldsymbol{\sigma}_{3}\left(G^{\pm}\right)^{-1} \right],~~~~\hat{\boldsymbol{u}}=i\left[\hat{\boldsymbol{r}},\boldsymbol{\sigma}_{3}\left(G^{\pm}\right)^{-1} \right]\boldsymbol{\sigma}_{3}\hat{L}^{\alpha}.
\end{align}
Using those relations, we obtain
\begin{align}
    &A_{\mu\nu}-\frac{1}{2}\frac{dB_{\mu\nu}}{d\eta} \notag \\
    &= \frac{1}{8\pi}Tr\left[\boldsymbol{\sigma}_{3}\hat{w}_{\mu}\left(G^{+}\right)^{2}\boldsymbol{\sigma}_{3}\hat{v}_{\nu}G^{+}+\boldsymbol{\sigma}_{3}\hat{w}_{\mu}G^{-}\boldsymbol{\sigma}_{3}\hat{v}_{\nu}\left(G^{-}\right)^{2} -\boldsymbol{\sigma}_{3}\hat{w}_{\mu}G^{+}\boldsymbol{\sigma}_{3}\hat{v}_{\nu}\left(G^{+}\right)^{2}-\boldsymbol{\sigma}_{3}\hat{w}_{\mu}\left(G^{-}\right)^{2}\boldsymbol{\sigma}_{3}\hat{v}_{\nu}G^{-}\right]\\
    & = \frac{i}{8\pi}Tr\left\lbrace\boldsymbol{\sigma}_{3}\hat{w}_{\mu}\left(G^{+}\right)^{2}\left[\left( G^{+}\right)^{-1}\hat{r}_{\nu}-\hat{r}_{\nu}\left( G^{+}\right)^{-1} \right]G^{+}+\boldsymbol{\sigma}_{3}\hat{w}_{\mu}G^{-}\left[\left( G^{-}\right)^{-1}\hat{r}_{\nu}-\hat{r}_{\nu}\left( G^{-}\right)^{-1} \right]\left(G^{-}\right)^{2} \right. \notag \\
    &\left.-\boldsymbol{\sigma}_{3}\hat{w}_{\mu}G^{+}\left[\left( G^{+}\right)^{-1}\hat{r}_{\nu}-\hat{r}_{\nu}\left( G^{+}\right)^{-1} \right]\left(G^{+}\right)^{2}-\boldsymbol{\sigma}_{3}\hat{w}_{\mu}\left(G^{-}\right)^{2}\left[\left( G^{-}\right)^{-1}\hat{r}_{\nu}-\hat{r}_{\nu}\left( G^{-}\right)^{-1} \right]G^{-}\right\rbrace
\end{align}
where we have used $\boldsymbol{\sigma}_{3}\boldsymbol{\sigma}_{3} \equiv \mathcal{I}$, the identity matrix. Finally, we arrive at
\begin{align} \label{ABmn}
    &A_{\mu\nu}-\frac{1}{2}\frac{dB_{\mu\nu}}{d\eta} \notag \\
    &= \frac{i}{8\pi}Tr\left\lbrace\boldsymbol{\sigma}_{3}\hat{w}_{\mu}\left[\left(G^{+}\right)^{2} - \left(G^{-}\right)^{2} \right]\hat{r}_{\nu}+\boldsymbol{\sigma}_{3}\hat{w}_{\mu}\hat{r}_{\nu}\left[\left(G^{+}\right)^{2} - \left(G^{-}\right)^{2} \right]+2\boldsymbol{\sigma}_{3}\hat{w}_{\mu}\left[G^{-}\hat{r}_{\nu}G^{-}-G^{+}\hat{r}_{\nu}G^{+} \right]\right\rbrace.
\end{align}
Considering the last term, one has
\begin{align}
  &2\boldsymbol{\sigma}_{3}\hat{w}_{\mu}\left[G^{-}\hat{r}_{\nu}G^{-}-G^{+}\hat{r}_{\nu}G^{+} \right] = 2\boldsymbol{\sigma}_{3}i\boldsymbol{\sigma}_{3}\hat{L}^{\alpha}\left[\hat{r}_{\mu},\boldsymbol{\sigma}_{3}\left(G^{\pm}\right)^{-1} \right]\left[G^{-}\hat{r}_{\nu}G^{-}-G^{+}\hat{r}_{\nu}G^{+} \right] \\
  & = 2iC_{\hat{H},\hat{L}^{\alpha}}\hat{r}_{\mu}\delta\left(\eta - \boldsymbol{\sigma}_{3}\hat{H} \right)\hat{r}_{\nu}G^{+} +2iC_{\hat{H},\hat{L}^{\alpha}}\hat{r}_{\mu}G^{-}\hat{r}_{\nu}\delta\left(\eta - \boldsymbol{\sigma}_{3}\hat{H} \right)
\end{align}
where $C_{\hat{H},\hat{L}^{\alpha}}=\boldsymbol{\sigma}_{3}\left[\hat{H},\hat{L}^{\alpha}\boldsymbol{\sigma}_{3} \right]$. We note that for the models considered in this work both the exchange interaction and the Dzyaloshinskii-Moriya interaction preserve the rotational symmetry about the z-axis. Consequently, $C_{\hat{H},\hat{L}^{z}}=\boldsymbol{\sigma}_{3}\left[\hat{H},\hat{L}^{z}\boldsymbol{\sigma}_{3} \right] = 0$. Generally, this particular term $C_{\hat{H},\hat{L}^{\alpha}}$ can also be anticipated to be relatively small compared to the other terms in \eqref{ABmn}. For instance, in systems where magnon-phonon coupling is present, $C_{\hat{H},\hat{L}^{\alpha}}$ scales with the magnon-phonon coupling strength, which tends to be much smaller than the exchange interactions. Hence, it is reasonable to disregard this term in our analysis. Consequently, our final result can be summarized as 
\begin{align}
    A_{\mu\nu}-\frac{1}{2}\frac{dB_{\mu\nu}}{d\eta} = &\frac{1}{4}Tr\left[\boldsymbol{\sigma}_{3}\hat{w}_{\mu}\frac{d}{d \eta}\delta\left(\eta - \boldsymbol{\sigma}_{3}\hat{H} \right)\hat{r}_{\nu}+\boldsymbol{\sigma}_{3}\hat{w}_{\mu}\hat{r}_{\nu}\frac{d}{d \eta}\delta\left(\eta - \boldsymbol{\sigma}_{3}\hat{H} \right)\right].
\end{align}
If we replace $\hat{\boldsymbol{w}}$ by $\hat{\boldsymbol{u}}$, we have
\begin{align}
    \tilde{A}_{\mu\nu}-\frac{1}{2}\frac{d\tilde{B}_{\mu\nu}}{d\eta} = &\frac{1}{4}Tr\left[\boldsymbol{\sigma}_{3}\hat{u}_{\mu}\frac{d}{d \eta}\delta\left(\eta - \boldsymbol{\sigma}_{3}\hat{H} \right)\hat{r}_{\nu}+\boldsymbol{\sigma}_{3}\hat{u}_{\mu}\hat{r}_{\nu}\frac{d}{d \eta}\delta\left(\eta - \boldsymbol{\sigma}_{3}\hat{H} \right)\right].
\end{align}
Note that for a bounded spectrum we have $\delta\left[ \pm \infty - \left( \boldsymbol{\sigma}_{3}E \right)_{nn} \right]=0~ \forall ~n$ leading to
\begin{align}
    \int_{-\infty}^{+\infty}d\eta \left[A_{\mu\nu}\left(\eta \right) -\frac{1}{2}\frac{dB_{\mu\nu}\left(\eta \right)}{d\eta} \right] = 0, ~~~~~~~~\int_{-\infty}^{+\infty}d\eta \left[\tilde{A}_{\mu\nu}\left(\eta \right) -\frac{1}{2}\frac{d\tilde{B}_{\mu\nu}\left(\eta \right)}{d\eta} \right] = 0.
\end{align}
Consequently,
\begin{align}
M_{\mu\nu} &= \frac{1}{2}\sum_{\boldsymbol{k}}\int_{-\infty}^{+\infty}d\eta \rho\left(\eta \right)\int_{-\infty}^{\eta}d\tilde{\eta}\left[A_{\mu\nu}\left(\tilde{\eta} \right) -\frac{1}{2}\frac{dB_{\mu\nu}\left(\tilde{\eta} \right)}{d\tilde{\eta}} + \tilde{A}_{\mu\nu}\left(\tilde{\eta} \right) -\frac{1}{2}\frac{d\tilde{B}_{\mu\nu}\left(\tilde{\eta} \right)}{d\tilde{\eta}} \right].
\end{align}
Using 
\begin{align}
    A_{\mu\nu}\left(\tilde{\eta} \right)-\frac{1}{2}\frac{dB_{\mu\nu}\left( \tilde{\eta}\right)}{d\eta} &=  \frac{i}{4}Tr \left[\boldsymbol{\sigma}_{3}\hat{w}_{\mu}\frac{dG^{+}}{d\tilde{\eta}}\boldsymbol{\sigma}_{3}\hat{v}_{\nu}\delta\left( \tilde{\eta} -\boldsymbol{\sigma}_{3}\hat{H}\right) - \boldsymbol{\sigma}_{3}\hat{w}_{\mu}G^{+}\boldsymbol{\sigma}_{3}\hat{v}_{\nu}\frac{d\delta\left( \tilde{\eta} -\boldsymbol{\sigma}_{3}\hat{H}\right)}{d\tilde{\eta}}\right. \notag \\
    &+ \left.\boldsymbol{\sigma}_{3}\hat{w}_{\mu}\frac{d\delta\left(\tilde{\eta} -\boldsymbol{\sigma}_{3}\hat{H} \right)}{d\tilde{\eta}}\boldsymbol{\sigma}_{3}\hat{v}_{\nu}G^{-} - \boldsymbol{\sigma}_{3}\hat{w}_{\mu}\delta\left(\tilde{\eta} -\boldsymbol{\sigma}_{3}\hat{H} \right)\boldsymbol{\sigma}_{3}\hat{v}_{\nu}\frac{dG^{-}}{d\tilde{\eta}} \right]
\end{align}
and
\begin{align}
    \tilde{A}_{\mu\nu}\left( \tilde{\eta}\right)-\frac{1}{2}\frac{d\tilde{B}_{\mu\nu}\left( \tilde{\eta}\right)}{d\tilde{\eta}} &=  \frac{i}{4}Tr \left[\boldsymbol{\sigma}_{3}\hat{u}_{\mu}\frac{dG^{+}}{d\tilde{\eta}}\boldsymbol{\sigma}_{3}\hat{v}_{\nu}\delta\left( \tilde{\eta} -\boldsymbol{\sigma}_{3}\hat{H}\right) - \boldsymbol{\sigma}_{3}\hat{u}_{\mu}G^{+}\boldsymbol{\sigma}_{3}\hat{v}_{\nu}\frac{d\delta\left( \tilde{\eta} -\boldsymbol{\sigma}_{3}\hat{H}\right)}{d\tilde{\eta}}\right. \notag \\
    &+ \left.\boldsymbol{\sigma}_{3}\hat{u}_{\mu}\frac{d\delta\left(\tilde{\eta} -\boldsymbol{\sigma}_{3}\hat{H} \right)}{d\tilde{\eta}}\boldsymbol{\sigma}_{3}\hat{v}_{\nu}G^{-} - \boldsymbol{\sigma}_{3}\hat{u}_{\mu}\delta\left(\tilde{\eta} -\boldsymbol{\sigma}_{3}\hat{H} \right)\boldsymbol{\sigma}_{3}\hat{v}_{\nu}\frac{dG^{-}}{d\tilde{\eta}} \right]
\end{align}
along with relations \eqref{Greenre} we obtain
\begin{align}
    M_{\mu\nu} &= \frac{i}{4}\sum_{\boldsymbol{k}} \sigma_{3}^{nn}\sigma_{3}^{mm} \frac{\left\langle n\left(\boldsymbol{k} \right) \left\vert \hat{w}_{\boldsymbol{k},\mu}  +  \hat{u}_{\boldsymbol{k},\mu}\right\vert m \left(\boldsymbol{k} \right) \right\rangle \left\langle m \left(\boldsymbol{k} \right)\left\vert \hat{v}_{\nu}\right\vert n \left(\boldsymbol{k} \right) \right\rangle}{\left[\left(\boldsymbol{\sigma}_{3}E\right)_{nn} -\left(\boldsymbol{\sigma}_{3}E\right)_{mm}\right]^{2}} \int_{\left(\boldsymbol{\sigma}_{3}E\right)_{nn}}^{\left(\boldsymbol{\sigma}_{3}E\right)_{mm}}d\eta\rho\left(\eta \right) \notag \\
    & +\frac{i}{8}\sum_{\boldsymbol{k}}\left[\left\langle n \left(\boldsymbol{k} \right) \left\vert \hat{w}_{\boldsymbol{k},\mu}  +  \hat{u}_{\boldsymbol{k},\mu} \right\vert m \left(\boldsymbol{k} \right) \right\rangle   \left\langle m \left(\boldsymbol{k} \right) \left\vert \hat{v}_{\nu}\right\vert n \left(\boldsymbol{k} \right) \right\rangle \times \right. \notag \\
    & \left. \times \frac{\left(\sigma_{3}^{mm}E_{nn}-\sigma_{3}^{nn}E_{mm}  \right)\left\lbrace\rho\left[\left(\boldsymbol{\sigma}_{3}E \right)_{mm} \right]+\rho\left[\left(\boldsymbol{\sigma}_{3}E \right)_{nn} \right]\right\rbrace}{\left[\left(\boldsymbol{\sigma}_{3}E \right)_{nn}-\left(\boldsymbol{\sigma}_{3}E \right)_{mm} \right]^{2}}\right].
\end{align}
Therefore, we have
\begin{align}
    &S_{\mu\nu}+M_{\mu\nu}= \frac{i}{4}\sum_{\boldsymbol{k}} \sigma_{3}^{nn}\sigma_{3}^{mm} \frac{\left\langle n \left(\boldsymbol{k} \right)\left\vert \hat{w}_{\boldsymbol{k},\mu}  +  \hat{u}_{\boldsymbol{k},\mu}\right\vert m \left(\boldsymbol{k} \right)\right\rangle \left\langle m \left(\boldsymbol{k} \right)\left\vert \hat{v}_{\nu}\right\vert n \left(\boldsymbol{k} \right)\right\rangle}{\left[\left(\boldsymbol{\sigma}_{3}E\right)_{nn} -\left(\boldsymbol{\sigma}_{3}E\right)_{mm}\right]^{2}} \int_{\left(\boldsymbol{\sigma}_{3}E\right)_{nn}}^{\left(\boldsymbol{\sigma}_{3}E\right)_{mm}}d\eta\rho\left(\eta \right) \notag \\
    & +\frac{i}{4}\sum_{\boldsymbol{k}}\left[\left\langle n \left(\boldsymbol{k} \right)\left\vert \hat{w}_{\boldsymbol{k},\mu}  +  \hat{u}_{\boldsymbol{k},\mu} \right\vert m\left(\boldsymbol{k} \right) \right\rangle   \left\langle m\left(\boldsymbol{k} \right) \left\vert \hat{v}_{\nu}\right\vert n\left(\boldsymbol{k} \right) \right\rangle\frac{\sigma_{3}^{mm}E_{nn}\rho\left[\left(\boldsymbol{\sigma}_{3}E \right)_{nn} \right]-\sigma_{3}^{nn}E_{mm}\rho\left[\left(\boldsymbol{\sigma}_{3}E \right)_{mm} \right] }{\left[\left(\boldsymbol{\sigma}_{3}E \right)_{mm} - \left(\boldsymbol{\sigma}_{3}E \right)_{nn}\right]^{2}}\right]\\
    &= \frac{i}{4}\sum_{\boldsymbol{k}} \sigma_{3}^{nn}\sigma_{3}^{mm} \frac{\left\langle n\left(\boldsymbol{k} \right) \left\vert \hat{w}_{\boldsymbol{k},\mu}  +  \hat{u}_{\boldsymbol{k},\mu}\right\vert m\left(\boldsymbol{k} \right) \right\rangle \left\langle m\left(\boldsymbol{k} \right) \left\vert \hat{v}_{\nu}\right\vert n\left(\boldsymbol{k} \right) \right\rangle}{\left[\left(\boldsymbol{\sigma}_{3}E\right)_{nn} -\left(\boldsymbol{\sigma}_{3}E\right)_{mm}\right]^{2}} \int_{\left(\boldsymbol{\sigma}_{3}E\right)_{nn}}^{\left(\boldsymbol{\sigma}_{3}E\right)_{mm}}d\eta\rho\left(\eta \right) \notag \\
    & +\frac{i}{4}\sum_{\boldsymbol{k}}\left[\sigma_{3}^{nn}\sigma_{3}^{mm}\left\langle n\left(\boldsymbol{k} \right) \left\vert \hat{w}_{\boldsymbol{k},\mu}  +  \hat{u}_{\boldsymbol{k},\mu} \right\vert m\left(\boldsymbol{k} \right) \right\rangle   \left\langle m\left(\boldsymbol{k} \right) \left\vert \hat{v}_{\nu}\right\vert n \left(\boldsymbol{k} \right)\right\rangle\frac{\sigma_{3}^{nn}E_{nn}\rho\left[\left(\boldsymbol{\sigma}_{3}E \right)_{nn} \right]-\sigma_{3}^{mm}E_{mm}\rho\left[\left(\boldsymbol{\sigma}_{3}E \right)_{mm} \right] }{\left[\left(\boldsymbol{\sigma}_{3}E \right)_{mm} - \left(\boldsymbol{\sigma}_{3}E \right)_{nn}\right]^{2}}\right]\\
    &= \frac{i}{4}\sum_{\boldsymbol{k}} \sigma_{3}^{nn}\sigma_{3}^{mm} \frac{\left\langle n \left(\boldsymbol{k} \right)\left\vert \hat{w}_{\boldsymbol{k},\mu}  +  \hat{u}_{\boldsymbol{k},\mu}\right\vert m\left(\boldsymbol{k} \right) \right\rangle \left\langle m\left(\boldsymbol{k} \right) \left\vert \hat{v}_{\nu}\right\vert n\left(\boldsymbol{k} \right) \right\rangle}{\left[\left(\boldsymbol{\sigma}_{3}E\right)_{nn} -\left(\boldsymbol{\sigma}_{3}E\right)_{mm}\right]^{2}} \int_{\left(\boldsymbol{\sigma}_{3}E\right)_{nn}}^{\left(\boldsymbol{\sigma}_{3}E\right)_{mm}}d\eta\rho\left(\eta \right) \notag \\
    & -\frac{i}{4}\sum_{\boldsymbol{k}}\left[\sigma_{3}^{nn}\sigma_{3}^{mm}\frac{\left\langle n\left(\boldsymbol{k} \right) \left\vert \hat{w}_{\boldsymbol{k},\mu}  +  \hat{u}_{\boldsymbol{k},\mu} \right\vert m\left(\boldsymbol{k} \right) \right\rangle   \left\langle m\left(\boldsymbol{k} \right) \left\vert \hat{v}_{\nu}\right\vert n \left(\boldsymbol{k} \right)\right\rangle}{\left[\left(\boldsymbol{\sigma}_{3}E \right)_{mm} - \left(\boldsymbol{\sigma}_{3}E \right)_{nn}\right]^{2}} \int_{\left(\boldsymbol{\sigma}_{3}E \right)_{nn}}^{\left(\boldsymbol{\sigma}_{3}E \right)_{mm}} d\left[ \eta \rho\left(\eta \right) \right]\right] \\
    &= -\frac{i}{4}\sum_{\boldsymbol{k}}\left[\sigma_{3}^{nn}\sigma_{3}^{mm}\frac{\left\langle n \left(\boldsymbol{k} \right)\left\vert \hat{w}_{\boldsymbol{k},\mu}  +  \hat{u}_{\boldsymbol{k},\mu} \right\vert m\left(\boldsymbol{k} \right) \right\rangle   \left\langle m\left(\boldsymbol{k} \right) \left\vert \hat{v}_{\nu}\right\vert n \left(\boldsymbol{k} \right)\right\rangle}{\left[\left(\boldsymbol{\sigma}_{3}E \right)_{mm} - \left(\boldsymbol{\sigma}_{3}E \right)_{nn}\right]^{2}} \int_{\left(\boldsymbol{\sigma}_{3}E \right)_{nn}}^{\left(\boldsymbol{\sigma}_{3}E \right)_{mm}} \eta d\left[  \rho\left(\eta \right) \right]\right]\\
    &=\frac{i}{4}\sum_{\boldsymbol{k}}\left[\sigma_{3}^{nn}\sigma_{3}^{mm}\frac{\left\langle n\left(\boldsymbol{k} \right) \left\vert \hat{w}_{\boldsymbol{k},\mu}  +  \hat{u}_{\boldsymbol{k},\mu} \right\vert m\left(\boldsymbol{k} \right) \right\rangle   \left\langle m\left(\boldsymbol{k} \right) \left\vert \hat{v}_{\nu}\right\vert n \left(\boldsymbol{k} \right)\right\rangle}{\left[\left(\boldsymbol{\sigma}_{3}E \right)_{mm} - \left(\boldsymbol{\sigma}_{3}E \right)_{nn}\right]^{2}} \int_{\left(\boldsymbol{\sigma}_{3}E \right)_{mm}}^{\left(\boldsymbol{\sigma}_{3}E \right)_{nn}} \eta d\left[  \rho\left(\eta \right) \right]\right]
\end{align}
where we have used $d\left[\eta \rho\left( \eta\right)\right]=\eta d\left[ \rho\left( \eta\right)\right]+d\eta \rho\left( \eta\right)$. Finally, by restoring the sum over the repeated Roman indices and the Planck constant $\hbar$  we arrive at the following expression
\begin{align}
    &S_{\mu\nu}+M_{\mu\nu} \\
    &= \frac{i\hbar}{4}\sum_{\boldsymbol{k}}\sum_{m\neq n}\left[\sigma_{3}^{nn}\sigma_{3}^{mm}\frac{\left\langle n\left(\boldsymbol{k} \right) \left\vert \hat{L}^{\alpha}\boldsymbol{\sigma}_{3}\hat{v}_{\boldsymbol{k},\mu}  +   \hat{v}_{\boldsymbol{k},\mu}\boldsymbol{\sigma}_{3}\hat{L}^{\alpha} \right\vert m\left(\boldsymbol{k} \right) \right\rangle   \left\langle m\left(\boldsymbol{k} \right) \left\vert \hat{v}_{\nu}\right\vert n\left(\boldsymbol{k} \right) \right\rangle}{\left[\left(\boldsymbol{\sigma}_{3}E_{\boldsymbol{k}} \right)_{mm} - \left(\boldsymbol{\sigma}_{3}E_{\boldsymbol{k}} \right)_{nn}\right]^{2}} \int_{\left(\boldsymbol{\sigma}_{3}E_{\boldsymbol{k}} \right)_{mm}}^{\left(\boldsymbol{\sigma}_{3}E_{\boldsymbol{k}} \right)_{nn}} \eta d\left[  \rho\left(\eta \right) \right]\right].
\end{align}
 Inserting the $\eta = k_{B}T \zeta$ and the Orbital angular moment current operator $\hat{j}^{L^{\alpha}}_{\mu}=\frac{1}{4}\left( \hat{L}^{\alpha}\boldsymbol{\sigma}_{3}\hat{v}_{\boldsymbol{k},\mu}  +   \hat{v}_{\boldsymbol{k},\mu}\boldsymbol{\sigma}_{3}\hat{L}^{\alpha}\right)$, we obtain the Orbital Nernst conductivity $\lambda_{\mu\nu}^{L^{\alpha}}=\frac{S_{\mu\nu}+M_{\mu\nu}}{VT}$:

\begin{align}
    \lambda_{\mu\nu}^{L^{\alpha}}  &= \frac{ k_{B}}{\hbar V}\sum_{\boldsymbol{k}} \sum_{m\neq n}\left[ i\hbar^{2}\sigma_{3}^{nn}\sigma_{3}^{mm}\frac{\left\langle n\left(\boldsymbol{k} \right) \left\vert    \hat{j}^{L^{\alpha}}_{\mu}    \right\vert m\left(\boldsymbol{k} \right) \right\rangle   \left\langle m\left(\boldsymbol{k} \right) \left\vert \hat{v}_{\nu}\right\vert n\left(\boldsymbol{k} \right) \right\rangle}{\left[\left(\boldsymbol{\sigma}_{3}E_{\boldsymbol{k}} \right)_{mm} - \left(\boldsymbol{\sigma}_{3}E_{\boldsymbol{k}} \right)_{nn}\right]^{2}} \int_{\left(\boldsymbol{\sigma}_{3}E \right)_{mm}}^{\left(\boldsymbol{\sigma}_{3}E \right)_{nn}} \zeta d\left[  \rho\left(\zeta \right) \right]\right]\\
    &= \frac{ k_{B}}{\hbar V}  \sum_{\boldsymbol{k}}\sum_{m\neq n} i\hbar^{2}\sigma_{3}^{nn}\sigma_{3}^{mm}\frac{\left\langle n\left(\boldsymbol{k} \right) \left\vert    \hat{j}^{L^{\alpha}}_{\mu}    \right\vert m\left(\boldsymbol{k} \right) \right\rangle   \left\langle m\left(\boldsymbol{k} \right) \left\vert \hat{v}_{\nu}\right\vert n\left(\boldsymbol{k} \right) \right\rangle}{\left[\left(\boldsymbol{\sigma}_{3}E_{\boldsymbol{k}} \right)_{mm} - \left(\boldsymbol{\sigma}_{3}E_{\boldsymbol{k}} \right)_{nn}\right]^{2}}\left[ F\left(\rho_{n}  \right) - F\left(\rho_{m}  \right) \right]\\
    & = -\frac{ k_{B}}{\hbar V}  \sum_{\boldsymbol{k}}\sum_{m\neq n} 2\hbar^{2}\sigma_{3}^{nn}\sigma_{3}^{mm}Im\left[\frac{\left\langle n\left(\boldsymbol{k} \right) \left\vert    \hat{j}^{L^{\alpha}}_{\mu}    \right\vert m\left(\boldsymbol{k} \right) \right\rangle   \left\langle m\left(\boldsymbol{k} \right) \left\vert \hat{v}_{\nu}\right\vert n\left(\boldsymbol{k} \right) \right\rangle}{\left[\left(\boldsymbol{\sigma}_{3}E_{\boldsymbol{k}} \right)_{mm} - \left(\boldsymbol{\sigma}_{3}E_{\boldsymbol{k}} \right)_{nn}\right]^{2}}\right] F\left(\rho_{n}  \right),
\end{align}
which leads to
\begin{align}
    \lambda_{\mu\nu}^{L^{\alpha}}  = \frac{ k_{B}}{\hbar V}  \sum_{\boldsymbol{k}}\sum_{n} \Omega^{L^{\alpha},n}_{\mu\nu}\left(\boldsymbol{k} \right) F\left(\rho_{n}  \right) \equiv  \frac{ 2k_{B}}{\hbar V}  \sum_{\boldsymbol{k}}\sum_{n=1}^{N} \Omega^{L^{\alpha},n}_{\mu\nu}\left(\boldsymbol{k} \right) F\left(\rho_{n}  \right)
    \label{EQnlamda}
\end{align}
where $\Omega^{L^{\alpha},n}_{\mu\nu}\left(\boldsymbol{k} \right)=\sum_{m \neq n}\Omega^{L^{\alpha},nm}_{\mu\nu}\left(\boldsymbol{k} \right)$ is the Orbital Berry curvature of the nth band,
\begin{align}
    \Omega^{L^{\alpha},nm}_{\mu\nu}\left(\boldsymbol{k} \right) =-2\hbar^{2}\sigma_{3}^{nn}\sigma_{3}^{mm}Im\left[\frac{\left\langle n\left(\boldsymbol{k} \right) \left\vert    \hat{j}^{L^{\alpha}}_{\mu}    \right\vert m\left(\boldsymbol{k} \right) \right\rangle   \left\langle m\left(\boldsymbol{k} \right) \left\vert \hat{v}_{\nu}\right\vert n\left(\boldsymbol{k} \right) \right\rangle}{\left[\left(\boldsymbol{\sigma}_{3}E_{\boldsymbol{k}} \right)_{mm} - \left(\boldsymbol{\sigma}_{3}E_{\boldsymbol{k}} \right)_{nn}\right]^{2}}\right]
\end{align}
is the projected Orbital Berry curvature of the nth band on the mth band, and  
\begin{equation}\label{F1}
    F\left( \rho_{n}\right) = \left(1+ \rho_{n} \right)ln\left(1+ \rho_{n} \right) -\rho_{n}ln\left(\rho_{n} \right),
\end{equation}

\begin{figure}[h]
\centering
    \includegraphics[width= 1\textwidth]{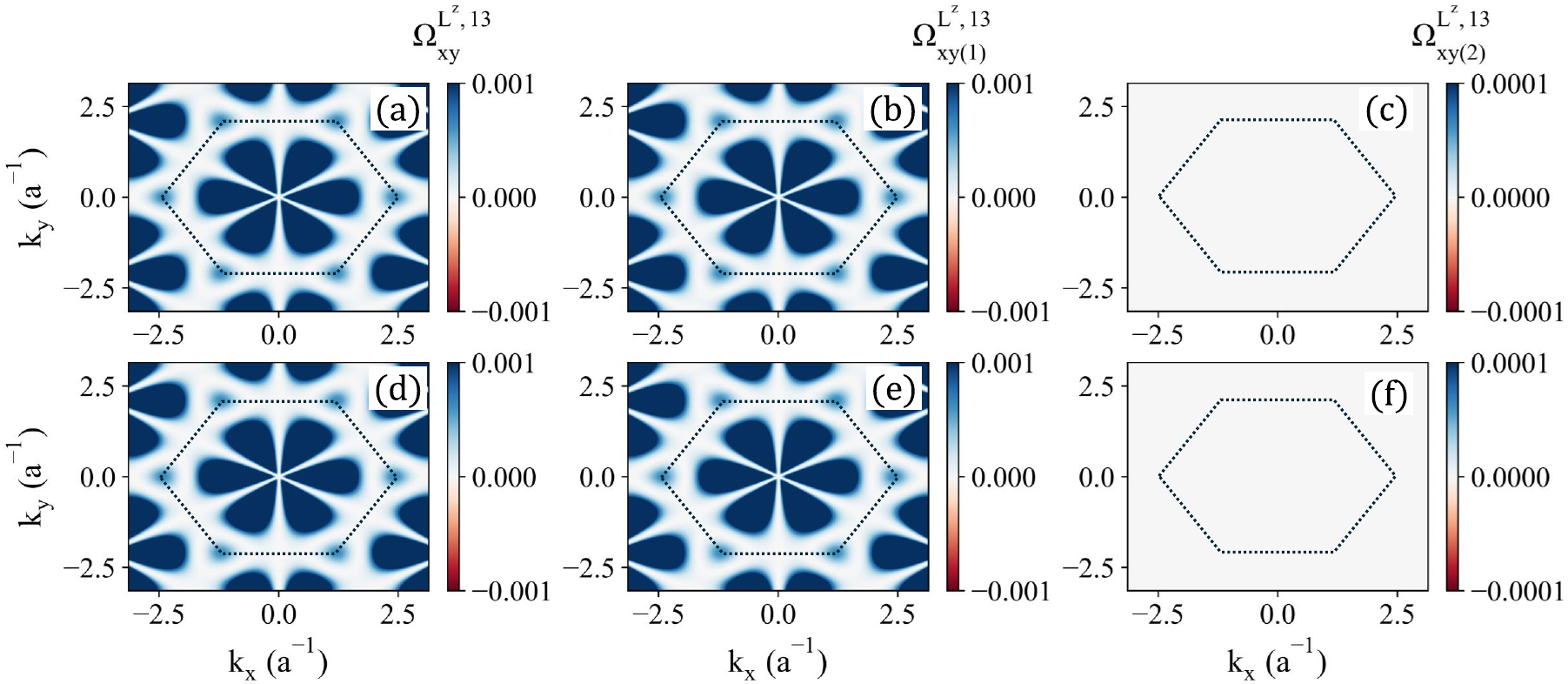}
 \caption{The projected Orbital Berry curvatures from 1st to 3rd band $\Omega^{L^{z},13}_{xy}\left(\boldsymbol{k}\right)$ (a,d), $\Omega^{L^{z},13}_{xy(1)}\left(\boldsymbol{k}\right)$ (b,e) and $\Omega^{L^{z},13}_{xy(2)}\left(\boldsymbol{k}\right)$ (c,f) for MnPS$_3$ as a function of in-plane wave vector $(k_{x},k_{y})$ without (a-c) and with (d-f) the DMI. These calculations were performed with the externally applied magnetic field $B_{z}=1~T$.}
  \label{S7}
\end{figure}
\begin{figure}[h]
\centering
    \includegraphics[width= 1\textwidth]{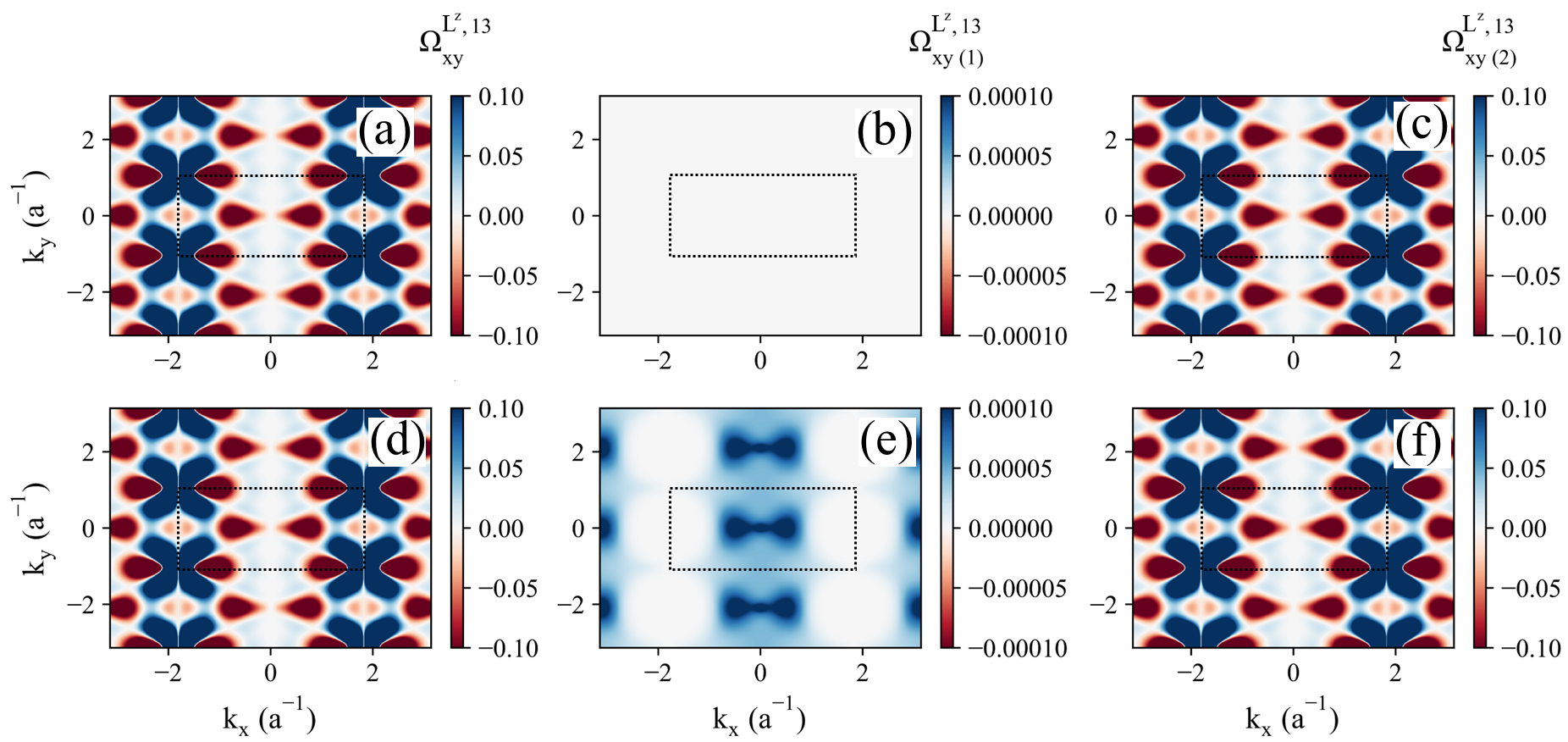}
 \caption{The projected Orbital Berry curvatures from 1st to 3rd band $\Omega^{L^{z},13}_{xy}\left(\boldsymbol{k}\right)$ (a,d), $\Omega^{L^{z},13}_{xy(1)}\left(\boldsymbol{k}\right)$ (b,e) and $\Omega^{L^{z},13}_{xy(2)}\left(\boldsymbol{k}\right)$ (c,f) for NiPSe$_3$ as a function of in-plane wave vector $(k_{x},k_{y})$ without (a-c) and with (d-f) the DMI. These calculations were performed with the externally applied magnetic field $B_{z}=1~T$.}
  \label{S6}
\end{figure}

Overall, the Orbital transverse current of quasiparticle transport underlying the Orbital Nernst effect is computed as:
\begin{equation}
    j^{L^{\alpha}}_{\mu} =  -\lambda^{L^{\alpha}}_{\mu\nu} \partial_{\nu}T
\end{equation}

In Fig.~\ref{S3}, we present the computed orbital Nernst conductivity (ONC) of MnPS$_3$ [Fig.~\ref{S3}(a)] and NiPSe$3$ [Fig.~\ref{S3}(b)] as functions of the external magnetic field $B_{z}$, computed at a constant temperature of 100 K. We observe that the ONC behaves as an even function of $B_z$ for both N\'eel and Zigzag orders. Importantly, the ONC remains finite even without an applied magnetic field, allowing for the probing of this effect in the absence of an external field. Moreover, the ONC for both N\'eel and Zigzag orders exhibits only a slight dependence on the out-of-plane magnetic field $B_z$. This minimal impact results from the magnetic field causing a splitting in the magnon bands corresponding to opposite spins, without inducing any coupling between these distinct bands. Consequently, the weak response of the ONC to the magnetic field $B_z$ arises solely from changes in the magnon population $\rho_{n}$ and hence the function $F\left(\rho_{n} \right)$ in Eq.~\eqref{EQnlamda}, which are due to the shift in the magnon energy band caused by the Zeeman interaction between local spins and external applied magnetic field akin to the behavior of electric polarization under the applied magnetic field as discussed in the main text. 

\subsection{Orbital Berry curvature}
It is worth emphasizing that our linear response theory, as presented here, and our utilization of the Orbital angular moment representation within the Bloch states enable us to treat the intra-band and inter-band orbital angular moment contributions to the orbital Nernst effect equally. In this section, we will distinguish between these two contributions and derive the expression for the topological thermal magnon contribution to the ONE, as utilized in the main text. We begin with the projected Orbital Berry curvature given by:
\begin{align}
    \Omega^{L^{\alpha},nm}_{\mu\nu}\left(\boldsymbol{k} \right) =-\frac{1}{2} \hbar^{2}\sigma_{3}^{nn}\sigma_{3}^{mm}Im\left[\frac{\left\langle n \left(\boldsymbol{k} \right)\left\vert    \hat{L}^{\alpha}\boldsymbol{\sigma}_{3}\hat{v}_{\mu}  +   \hat{v}_{\mu}\boldsymbol{\sigma}_{3}\hat{L}^{\alpha}    \right\vert m \left(\boldsymbol{k} \right) \right\rangle   \left\langle m \left(\boldsymbol{k} \right) \left\vert \hat{v}_{\nu}\right\vert n\left(\boldsymbol{k} \right) \right\rangle}{\left[\left(\boldsymbol{\sigma}_{3}E_{\boldsymbol{k}} \right)_{mm} - \left(\boldsymbol{\sigma}_{3}E_{\boldsymbol{k}} \right)_{nn}\right]^{2}}\right].
\end{align}
Using the completeness relation
\begin{equation}
    \sum_{l}\sigma_{3}^{ll}\left\vert l\left(\boldsymbol{k} \right)\right\rangle \left\langle l\left(\boldsymbol{k} \right)\right\vert \boldsymbol{\sigma}_{3}=\mathcal{I}
\end{equation}
we have
\begin{align}
    &\langle n(\boldsymbol{k})\vert \left( \hat{L}^{\alpha}\boldsymbol{\sigma}_{3}\hat{v}_{\mu}+\hat{v}_{\mu}\boldsymbol{\sigma}_{3}\hat{L}^{\alpha}\right)\vert m(\boldsymbol{k}) \rangle \\
    &= \langle n(\boldsymbol{k})\vert \left( \hat{L}^{\alpha}\sum_{l}\sigma_{3}^{ll}\left\vert l\left(\boldsymbol{k} \right)\right\rangle \left\langle l\left(\boldsymbol{k} \right)\right\vert \boldsymbol{\sigma}_{3}\boldsymbol{\sigma}_{3}\hat{v}_{\mu}+\hat{v}_{\mu}\sum_{q}\sigma_{3}^{qq}\left\vert q\left(\boldsymbol{k} \right)\right\rangle \left\langle q\left(\boldsymbol{k} \right)\right\vert \boldsymbol{\sigma}_{3}\boldsymbol{\sigma}_{3}\hat{L}^{\alpha}\right)\vert m(\boldsymbol{k}) \rangle \\
    &= \langle n(\boldsymbol{k})\vert \left( \hat{L}^{\alpha}\sum_{l}\sigma_{3}^{ll}\left\vert l\left(\boldsymbol{k} \right)\right\rangle \left\langle l\left(\boldsymbol{k} \right)\right\vert \hat{v}_{\mu}+\hat{v}_{\mu}\sum_{q}\sigma_{3}^{qq}\left\vert q\left(\boldsymbol{k} \right)\right\rangle \left\langle q\left(\boldsymbol{k} \right)\right\vert\hat{L}^{\alpha}\right)\vert m(\boldsymbol{k}) \rangle\\
    &= \sum_{l}\sigma_{3}^{ll}\langle n(\boldsymbol{k})\vert \hat{L}^{\alpha}\left\vert l\left(\boldsymbol{k} \right)\right\rangle \left\langle l\left(\boldsymbol{k} \right)\right\vert \hat{v}_{\mu}\vert m(\boldsymbol{k}) \rangle +\sum_{q}\sigma_{3}^{qq}\langle n(\boldsymbol{k})\vert \hat{v}_{\mu}\left\vert q\left(\boldsymbol{k} \right)\right\rangle \left\langle q\left(\boldsymbol{k} \right)\right\vert\hat{L}^{\alpha}\vert m(\boldsymbol{k}) \rangle\\
    &=\left[\sigma_{3}^{nn}L_{nn}^{\alpha}\left(\boldsymbol{k} \right)+\sigma_{3}^{mm}L_{mm}^{\alpha}\left(\boldsymbol{k} \right) \right]\langle n(\boldsymbol{k})\vert \hat{v}_{\mu}\left\vert m\left(\boldsymbol{k} \right)\right\rangle +  \sum_{l\neq n}\sigma_{3}^{ll}\langle n(\boldsymbol{k})\vert \hat{L}^{\alpha}\left\vert l\left(\boldsymbol{k} \right)\right\rangle \left\langle l\left(\boldsymbol{k} \right)\right\vert \hat{v}_{\mu}\vert m(\boldsymbol{k}) \rangle \\
    &+\sum_{q\neq m}\sigma_{3}^{qq}\langle n(\boldsymbol{k})\vert \hat{v}_{\mu}\left\vert q\left(\boldsymbol{k} \right)\right\rangle \left\langle q\left(\boldsymbol{k} \right)\right\vert\hat{L}^{\alpha}\vert m(\boldsymbol{k}) \rangle,
    \end{align}
which leads to
\begin{align}
    &\Omega^{L^{\alpha},nm}_{\mu\nu}\left(\boldsymbol{k} \right)=-\frac{1}{2} \hbar^{2}\sigma_{3}^{nn}\sigma_{3}^{mm}Im\left\lbrace\sum_{l}\frac{\left[ \sigma_{3}^{ll}\langle n(\boldsymbol{k})\vert \hat{L}^{\alpha}\left\vert l\left(\boldsymbol{k} \right)\right\rangle \left\langle l\left(\boldsymbol{k} \right)\right\vert \hat{v}_{\mu}\vert m(\boldsymbol{k}) \rangle \right] \langle m(\boldsymbol{k})\vert \hat{v}_{\nu} \vert n(\boldsymbol{k}) \rangle}{\left[ \sigma_{3}^{nn}\left(E_{\boldsymbol{k}}\right)_{nn}  - \sigma_{3}^{mm}\left(E_{\boldsymbol{k}}\right)_{mm} \right]^{2}}\right\rbrace \\
    &-\frac{1}{2} \hbar^{2}\sigma_{3}^{nn}\sigma_{3}^{mm}Im\left\lbrace\sum_{q}\frac{\left[ \sigma_{3}^{qq}\langle n(\boldsymbol{k})\vert \hat{v}_{\mu}\left\vert q\left(\boldsymbol{k} \right)\right\rangle \left\langle q\left(\boldsymbol{k} \right)\right\vert\hat{L}^{\alpha}\vert m(\boldsymbol{k}) \rangle\right] \langle m(\boldsymbol{k})\vert \hat{v}_{\nu} \vert n(\boldsymbol{k}) \rangle}{\left[ \sigma_{3}^{nn}\left(E_{\boldsymbol{k}}\right)_{nn} - \sigma_{3}^{mm}\left(E_{\boldsymbol{k}}\right)_{mm} \right]^{2}}\right\rbrace\\
    &=-\frac{1}{2} \hbar^{2}\sigma_{3}^{nn}\sigma_{3}^{mm}\left[\sigma_{3}^{nn}L_{nn}^{\alpha}\left(\boldsymbol{k} \right)+\sigma_{3}^{mm}L_{mm}^{\alpha}\left(\boldsymbol{k} \right) \right]Im\left\lbrace\frac{ \left\langle n\left(\boldsymbol{k} \right)\right\vert \hat{v}_{\mu}\vert m(\boldsymbol{k}) \langle m(\boldsymbol{k})\vert \hat{v}_{\nu} \vert n(\boldsymbol{k}) \rangle}{\left[ \sigma_{3}^{nn}\left(E_{\boldsymbol{k}}\right)_{nn} - \sigma_{3}^{mm}\left(E_{\boldsymbol{k}}\right)_{mm} \right]^{2}}\right\rbrace \\
    &-\frac{1}{2} \hbar^{2}\sigma_{3}^{nn}\sigma_{3}^{mm}Im\left\lbrace\sum_{l \neq n}\frac{\left[ \sigma_{3}^{ll}\langle n(\boldsymbol{k})\vert \hat{L}^{\alpha}\left\vert l\left(\boldsymbol{k} \right)\right\rangle \left\langle l\left(\boldsymbol{k} \right)\right\vert \hat{v}_{\mu}\vert m(\boldsymbol{k}) \rangle \right] \langle m(\boldsymbol{k})\vert \hat{v}_{\nu} \vert n(\boldsymbol{k}) \rangle}{\left[ \sigma_{3}^{nn}\left(E_{\boldsymbol{k}}\right)_{nn} - \sigma_{3}^{mm}\left(E_{\boldsymbol{k}}\right)_{mm} \right]^{2}}\right\rbrace\\
    &-\frac{1}{2} \hbar^{2}\sigma_{3}^{nn}\sigma_{3}^{mm}Im\left\lbrace\sum_{q \neq m}\frac{\left[ \sigma_{3}^{qq}\langle n(\boldsymbol{k})\vert \hat{v}_{\mu}\left\vert q\left(\boldsymbol{k} \right)\right\rangle \left\langle q\left(\boldsymbol{k} \right)\right\vert\hat{L}^{\alpha}\vert m(\boldsymbol{k}) \rangle\right] \langle m(\boldsymbol{k})\vert \hat{v}_{\nu} \vert n(\boldsymbol{k}) \rangle}{\left[ \sigma_{3}^{nn}\left(E_{\boldsymbol{k}}\right)_{nn} - \sigma_{3}^{mm}\left(E_{\boldsymbol{k}}\right)_{mm} \right]^{2}}\right\rbrace\\
    &=\Omega^{L^{\alpha},nm}_{\mu\nu(1)}\left(\boldsymbol{k} \right) + \Omega^{L^{\alpha},nm}_{\mu\nu(2)}\left(\boldsymbol{k} \right)
\end{align}
where
\begin{align}
    \Omega^{L^{\alpha},nm}_{\mu\nu(1)}\left(\boldsymbol{k} \right) =-\frac{1}{2} \hbar^{2}\sigma_{3}^{nn}\sigma_{3}^{mm}\left[\sigma_{3}^{nn}L_{nn}^{\alpha}\left(\boldsymbol{k} \right)+\sigma_{3}^{mm}L_{mm}^{\alpha}\left(\boldsymbol{k} \right) \right]Im \left\lbrace\frac{ \left\langle n\left(\boldsymbol{k} \right)\right\vert \hat{v}_{\mu}\vert m(\boldsymbol{k}) \langle m(\boldsymbol{k})\vert \hat{v}_{\nu} \vert n(\boldsymbol{k}) \rangle}{\left[ \sigma_{3}^{nn}\left(E_{\boldsymbol{k}}\right)_{nn} - \sigma_{3}^{mm}\left(E_{\boldsymbol{k}}\right)_{mm} \right]^{2}} \right\rbrace
\end{align}
and
\begin{align}
    &\Omega^{L^{\alpha},nm}_{\mu\nu(2)}\left(\boldsymbol{k} \right) =-\frac{1}{2} \hbar^{2}\sigma_{3}^{nn}\sigma_{3}^{mm}Im\left\lbrace\sum_{l \neq n}\frac{\left[ \sigma_{3}^{ll}\langle n(\boldsymbol{k})\vert \hat{L}^{\alpha}\left\vert l\left(\boldsymbol{k} \right)\right\rangle \left\langle l\left(\boldsymbol{k} \right)\right\vert \hat{v}_{\mu}\vert m(\boldsymbol{k}) \rangle \right] \langle m(\boldsymbol{k})\vert \hat{v}_{\nu} \vert n(\boldsymbol{k}) \rangle}{\left[ \sigma_{3}^{nn}\left(E_{\boldsymbol{k}}\right)_{nn} - \sigma_{3}^{mm}\left(E_{\boldsymbol{k}}\right)_{mm} \right]^{2}}\right\rbrace\\
    &-\frac{1}{2} \hbar^{2}\sigma_{3}^{nn}\sigma_{3}^{mm}Im\left\lbrace\sum_{q \neq m}\frac{\left[ \sigma_{3}^{qq}\langle n(\boldsymbol{k})\vert \hat{v}_{\mu}\left\vert q\left(\boldsymbol{k} \right)\right\rangle \left\langle q\left(\boldsymbol{k} \right)\right\vert\hat{L}^{\alpha}\vert m(\boldsymbol{k}) \rangle\right] \langle m(\boldsymbol{k})\vert \hat{v}_{\nu} \vert n(\boldsymbol{k}) \rangle}{\left[ \sigma_{3}^{nn}\left(E_{\boldsymbol{k}}\right)_{nn} - \sigma_{3}^{mm}\left(E_{\boldsymbol{k}}\right)_{mm} \right]^{2}}\right\rbrace.
\end{align}

Notice that
\begin{align}
    -2 \hbar^{2}\sigma_{3}^{nn}\sigma_{3}^{mm}Im \left\lbrace\frac{ \left\langle n\left(\boldsymbol{k} \right)\right\vert \hat{v}_{\mu}\vert m(\boldsymbol{k}) \langle m(\boldsymbol{k})\vert \hat{v}_{\nu} \vert n(\boldsymbol{k}) \rangle}{\left[ \sigma_{3}^{nn}\left(E_{\boldsymbol{k}}\right)_{nn} - \sigma_{3}^{mm}\left(E_{\boldsymbol{k}}\right)_{mm} \right]^{2}} \right\rbrace \equiv \Omega^{nm}_{\mu\nu}\left(\boldsymbol{k} \right)
\end{align}
is the projected Berry curvature of the nth band on the mth band as defined in Equation (6) in the main text. We can then rewrite the first term as 
\begin{align}
    \Omega^{L^{\alpha},nm}_{\mu\nu(1)}\left(\boldsymbol{k} \right) = \frac{1}{4} \left[\sigma_{3}^{nn}L_{nn}^{\alpha}\left(\boldsymbol{k} \right)+\sigma_{3}^{mm}L_{mm}^{\alpha}\left(\boldsymbol{k} \right) \right]\Omega^{nm}_{\mu\nu}\left(\boldsymbol{k} \right), 
\end{align}
which leads to
\begin{align}
    \Omega^{L^{\alpha},n}_{\mu\nu(1)}\left(\boldsymbol{k} \right) =\sum_{m \neq n}\Omega^{L^{\alpha},nm}_{\mu\nu(1)}\left(\boldsymbol{k} \right) = \sum_{m \neq n}\frac{1}{4} \left[\sigma_{3}^{nn}L_{nn}^{\alpha}\left(\boldsymbol{k} \right)+\sigma_{3}^{mm}L_{mm}^{\alpha}\left(\boldsymbol{k} \right) \right]\Omega^{nm}_{\mu\nu}\left(\boldsymbol{k} \right). 
\end{align}
Consequently, the magnon orbital Nernst conductivity arising from the $\Omega^{L^{\alpha},n}_{\mu\nu(1)}\left(\boldsymbol{k} \right)$ is

\begin{equation}
    \lambda_{\mu\nu(1)}^{L^{\alpha}}=\frac{2k_{B}}{ \hbar V } \sum_{k}\sum_{n=1}^{N} \Omega^{L^{\alpha},n}_{\mu\nu(1)}\left(\boldsymbol{k} \right)F(\rho_{n} ) = \frac{k_{B}}{2 \hbar V} \sum_{k}\sum_{n=1}^{N} \sum_{m\neq n} \left[\sigma_{3}^{nn}L_{nn}^{\alpha}\left(\boldsymbol{k} \right)+\sigma_{3}^{mm}L_{mm}^{\alpha}\left(\boldsymbol{k} \right) \right]\Omega^{nm}_{\mu\nu}\left(\boldsymbol{k} \right)  F(\rho_{n} ).
    \label{ONEtopo}
\end{equation}
Notably, the expression of $\Omega^{L^{\alpha},n}_{\mu\nu(1)}\left(\boldsymbol{k}\right)$ represents the product of intra-band magnon orbital angular moment and Berry curvature. Consequently, Equation~\eqref{ONEtopo} underscores the role of topological thermal magnon bands in inducing the Orbital Nernst effect, specifically via the intra-band magnon OAM. On the other hand, the second term $\Omega^{L^{\alpha},nm}_{\mu\nu(2)}\left(\boldsymbol{k} \right)$ captures the magnon Orbital Nernst effect arising from inter-band magnon orbital angular moment.

We observe that both $\Omega^{L^{\alpha},nm}_{\mu\nu(1)}\left(\boldsymbol{k}\right)$ and $\Omega^{L^{\alpha},nm}_{\mu\nu(2)}\left(\boldsymbol{k} \right)$ depend on the inter-band current density and the energy spacing between two subbands. Because there is no coupling between the two magnons with opposite spin in the systems we consider here, the inter-band current between two-subbands with opposite spin vanishes. Consequently, only the projected orbital Berry curvatures between bands of the same spin remains finite, originating in non-zero interband transitions that adhere to the spin selection rules. Thus, the magnon orbital Nernst conductivities are predominantly influenced by the interband transitions between two magnon subbands with the same spin. This accounts for the weak dependence of the magnon ONCs on the externally applied magnetic field $B_z$ for both MnPS$_3$ and NiPSe$_3$, as shown in Fig.~\ref{S3}: the changing magnetic field does not affect the magnon wavefunction  or change the energy spacing between two subbands of the same spin.
 
\begin{figure}[h]
\centering
    \includegraphics[width= 1\textwidth]{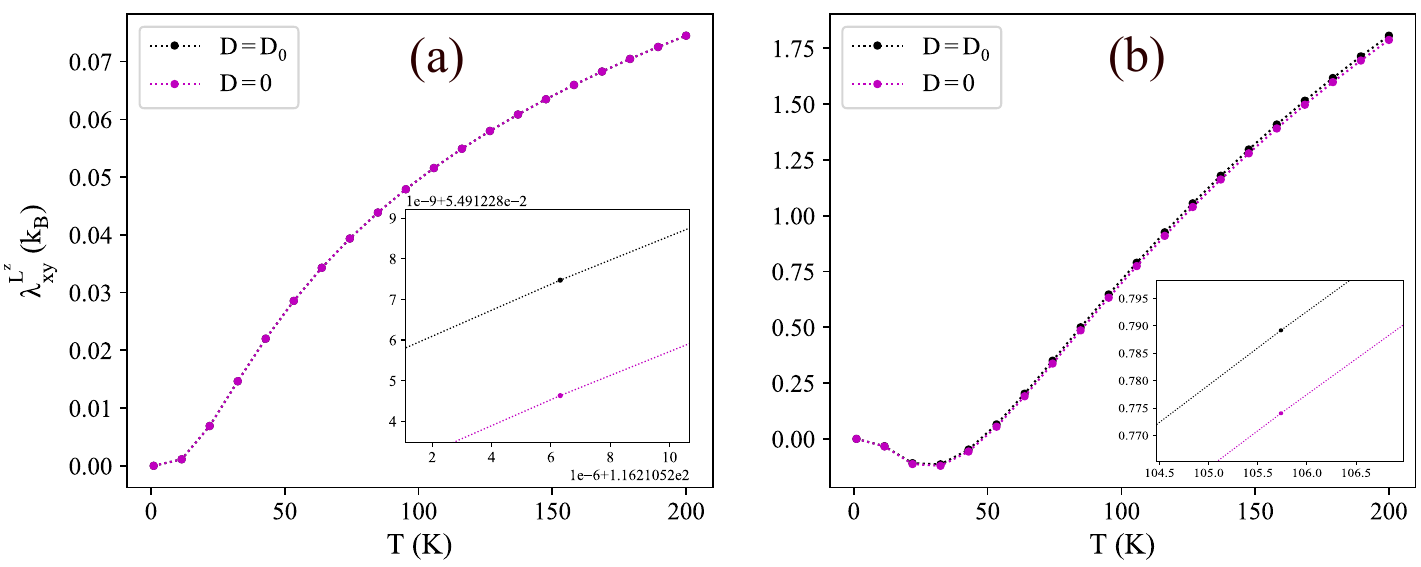}
 \caption{The orbital Nernst conductivity of MnPS$_3$ (a) and NiPSe$_3$ (b) as a function of temperature at fixed applied magnetic field $B_{z}=1~T$. The black and pink colors indicate the results calculated with and without DMI, respectively. The insets present the zoom in at around T = 117 K in figure (a) and T = 105 K in figure (b) showing the vanishing difference between black and pink curve in the case of MnPS$_3$ and weak dependence of ONC of NiPSe$_3$ on the DMI.}
  \label{S8}
\end{figure}

 In Figs.~\ref{S7} and ~\ref{S6} we plot the projected Orbital Berry curvatures of the 1st magnon subband onto the 3rd magnon subband with the same spin up: $\Omega^{L^{z},13}_{xy}\left(\boldsymbol{k}\right)$ (a,d), $\Omega^{L^{z},13}_{xy(1)}\left(\boldsymbol{k}\right)$ (b,e) and $\Omega^{L^{z},13}_{xy(2)}\left(\boldsymbol{k}\right)$ (c,f) for, respectively, MnPS$_3$ (Fig.~\ref{S7}) and NiPSe$_3$ (Fig.~\ref{S6}) as a function of the in-plane wave vector $(k_{x},k_{y})$ without (a-c) and with (d-f) the DMI. As shown in  Fig.~\ref{S7}, the Orbital Berry curvature of MnPS$_3$ is primarily determined by the topological properties of the magnon bands through the first term $\Omega^{L^{z},13}_{xy(1)}\left(\boldsymbol{k}\right)$ because the second term $\Omega^{L^{z},13}_{xy(2)}\left(\boldsymbol{k}\right)$ vanishes regardless of the presence of DMI. As discussed in the main text, due to the symmetry of the N\'eel order, this first term $\Omega^{L^{z},13}_{xy(1)}\left(\boldsymbol{k}\right)$ remains finite even without DMI because both Berry curvature and the intra-band orbital angular moment of the N\'eel phase are finite regardless of the DMI. Additionally, because the wavefunction is unaffected by DMI, the projected orbital Berry curvature $\Omega^{L^{z},13}_{xy(1)}\left(\boldsymbol{k}\right)$ and hence $\Omega^{L^{z},13}_{xy}\left(\boldsymbol{k}\right)$ remains unchanged for MnPS$_3$ when DMI is turned off. Consequently, the orbital Nernst conductivity of MnPS$_3$ exhibits an extremely weak dependence on DMI, as shown in Fig.~\ref{S8}(a). The difference between the black and pink curves shown in the inset of Fig.~\ref{S8}(a) arises solely from changes in the function $F\left( \rho_{n} \right)$ in Eq.~\eqref{ONEtopo} due to the shift in magnon dispersion under the DMI. Since the DMI in MnPS$_3$ is very small (see Table~\ref{tab1}), this difference is vanishingly small.
 
 In contrast, for NiPSe$_3$ in the absence of the DMI the projected Orbital Berry curvatures $\Omega^{L^{z},13}_{xy(1)}\left(\boldsymbol{k}\right)$ vanish simply because of the vanishing intra-band OAM of the magnon in the 2D honeycomb AFM with Zigzag order, as discussed previously. Therefore the projected Berry curvatures $\Omega^{L^{z},13}_{xy}\left(\boldsymbol{k}\right)$ are primarily determined by the $\Omega^{L^{z},13}_{xy(2)}\left(\boldsymbol{k}\right)$. When the DMI is present, the $\Omega^{L^{z},13}_{xy(1)}\left(\boldsymbol{k}\right)$ becomes finite while the $\Omega^{L^{z},13}_{xy(2)}\left(\boldsymbol{k}\right)$ remains unchanged. However, the magnitude of $\Omega^{L^{z},13}_{xy(1)}\left(\boldsymbol{k}\right)$ is approximately three orders smaller than that of $\Omega^{L^{z},13}_{xy(2)}\left(\boldsymbol{k}\right)$, highlighting that the inter-band magnon OAM is the primary contributor to the magnon Orbital Nernst effect. This observation also clarifies why the magnon orbital Nernst conductivity shows a weak dependence on DMI, which is due to the combined effects of changes in the function $F\left( \rho_{n} \right)$ caused by the shift in magnon dispersion and the finiteness of $\Omega^{L^{z},13}_{xy(1)}\left(\boldsymbol{k}\right)$ under the DMI, as shown in Fig.~\ref{S8}(b).

\begin{figure}[h]
\centering
    \includegraphics[width= 1\textwidth]{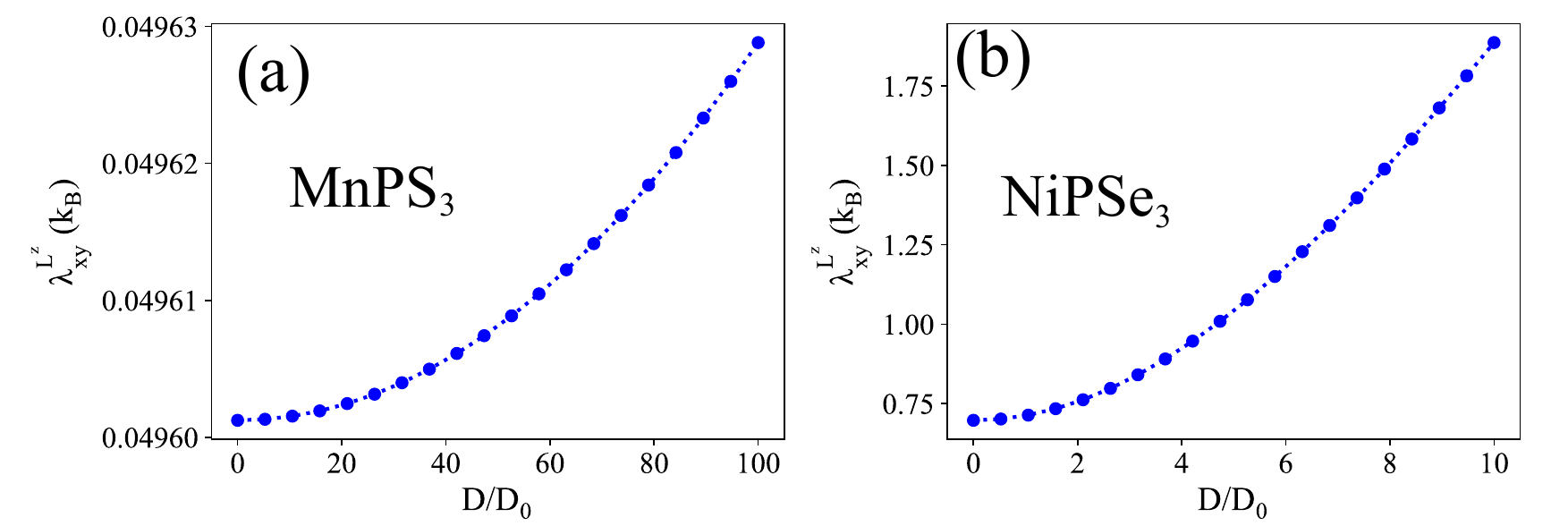}
 \caption{The ONC of MnPS$_3$ (Figure a) and NiPSe$_3$ (Figure b) varies as a function of the Dzyaloshinskii-Moriya interaction strength, parameterized by the ratio $D/D_{0}$. Here $D_{0}$ represents the baseline DMI strength for each material. These calculations were conducted under a constant applied magnetic field of $B_{z}=1~T$ and temperature of 100 K.}
  \label{S10}
\end{figure}

We conclude this section by illustrating the dependence of the orbital Nernst conductivity of both  MnPS$_3$ and NiPSe$_3$ on DMI strength, parameterized by the ratio $D/D_{0}$ in Fig.~\ref{S10}. 

\section{Symmetry and magnon spin current}
We now analyze symmetry constraints on the magnon spin current in the 2D collinear honeycomb AFMs considered in this work. Specifically, we consider the $\mathcal{C}_{S}\mathcal{M}_{x}\mathcal{T}_{a}$ symmetry operation discussed in the main text. The total transverse spin current carried by a magnon along the y-direction under the applied temperature gradient along the x-direction is given by
\begin{equation}
    j_{y}^{S^{z}} = j_{y}^{S^{z}\uparrow}-j_{y}^{S^{z}\downarrow} = -\lambda_{xy}^{S^{z}}\partial_{x} T \label{CMT1}
\end{equation}
where $\lambda_{xy}^{S^{z}}$ is the magon spin conductivity. 

Under the $\mathcal{C}_{S}\mathcal{M}_{x}\mathcal{T}_{a}$ symmetry operation, the temperature gradient changes only its sign:
\begin{equation}
  \left[\mathcal{C}_{S}\mathcal{M}_{x}\mathcal{T}_{a} \right]\partial_{x} T = -\partial_{x} T \label{CMT2}
\end{equation}
while the magnon spin up and down current remain unchanged: 
\begin{align}
    \left[\mathcal{C}_{S}\mathcal{M}_{x}\mathcal{T}_{a} \right]j_{y}^{S^{z}\uparrow} \rightarrow  j_{y}^{S^{z}\uparrow}\\
    \left[\mathcal{C}_{S}\mathcal{M}_{x}\mathcal{T}_{a} \right]j_{y}^{S^{z}\downarrow} \rightarrow  j_{y}^{S^{z}\downarrow}
\end{align}
which leads to
\begin{align}
    \left[\mathcal{C}_{S}\mathcal{M}_{x}\mathcal{T}_{a} \right]j_{y}^{S^{z}} =\left[\mathcal{C}_{S}\mathcal{M}_{x}\mathcal{T}_{a} \right]j_{y}^{S^{z}\uparrow}-\left[\mathcal{C}_{S}\mathcal{M}_{x}\mathcal{T}_{a} \right]j_{y}^{S^{z}\downarrow} = j_{y}^{S^{z}\uparrow}-j_{y}^{S^{z}\downarrow}= j_{y}^{S^{z}} \label{CMT3}.
\end{align}

Because the 2D honecomb AFMs with N\'eel and Zigzag order both preserve $\mathcal{C}_{S}\mathcal{M}_{x}\mathcal{T}_{a}$ symmetry in the absence of DMI, the magnon spin conductivity is unchanged under this operation.  Combining Eqs. \eqref{CMT1}, \eqref{CMT2} and \eqref{CMT3}, one obtains
\begin{equation}
    -\lambda_{xy}^{S^{z}}\partial_{x} T = \lambda_{xy}^{S^{z}}\partial_{x} T.
\end{equation}
Consequently, the magnon spin conductivity must vanish, i.e. $\lambda_{xy}^{S^{z}}=0$, under the $\mathcal{C}_{S}\mathcal{M}_{x}\mathcal{T}_{a}$ symmetry constraint because $\partial_{x} T$ is finite. In other words, there is no transverse spin polarized current carried by magnons in the 2D honeycomb AFMs considered in this work in the absence of DMI regardless of the externally applied magnetic field.

We note that the orbital current remains finite regardless of DMI as discussed in the main text. This suggests that the orbital Nernst effect of magnons is a more universal phenomenon than the magnon spin Nernst effect. Furthermore, a finite transverse spin current carried by magnons under a temperature gradient in the 2D honeycomb antiferromagnets considered here must be identical to the spin current resulting from the spin Nernst effect. This finite spin current arises due to the broken $\mathcal{C}_{S}\mathcal{M}_{x}\mathcal{T}_{a}$ symmetry discussed above, caused by interactions such as the Dzyaloshinskii-Moriya interaction or magnon-phonon coupling that resemble spin-orbit coupling (SOC) in that they couple the spin and orbital angular moment of magnons. In this context, the magnon spin current resulting from the SNE may be viewed as a conversion of orbital current into spin current.

\section{Spin and Orbital accumulations induced by the magnon Nernst effects}
The magnon Nernst effects lead to the accumulation of both orbital and spin moments at the boundaries of finite systems. In this section, we present the formalism for calculating these accumulations at the system’s edges, resulting from the magnon Nernst effect, without delving into detailed computations for specific systems, as that is beyond the scope of this work. To achieve this, we assume the system is periodic along the direction of the temperature gradient but finite in the perpendicular direction to $\nabla T$. The accumulation of magnon spin and orbital moments under an applied temperature gradient can be attributed to two distinct contributions: intrinsic and extrinsic.

\subsection{Intrinsic Contribution: Kubo Formula}
First, we focus on the intrinsic contributions to the spin and orbital density which reads
\begin{align}
    \left\langle \delta \mathcal{M}^{\alpha}\left( \boldsymbol{r} \right)\right\rangle^{in} =  \left\langle  \mathcal{M}^{\alpha}\left( \boldsymbol{r} \right)\right\rangle_{neq}^{in} - \left\langle  \mathcal{M}^{\alpha}\left( \boldsymbol{r} \right)\right\rangle_{eq}^{in}
\end{align}
Using the Kubo formula, we derive these intrinsic contributions to the spin and orbital densities as follows:

\begin{align}
    \left\langle \delta \mathcal{M}^{\alpha}\left( \boldsymbol{r} \right)\right\rangle^{in} = - \lim_{\omega \rightarrow 0}\frac{\partial}{\partial \omega} \int_{0}^{\beta} d\tau e^{i\omega\tau}\left\langle T_{\tau} \mathcal{M}^{\alpha}\left(\boldsymbol{r},\tau \right)J_{\nu}^{Q}\left( 0\right) \right\rangle \nabla_{\nu} \chi
\end{align}
where $\mathcal{M}^{\alpha}\left(\boldsymbol{r} \right)\equiv S^{\alpha}\left(\boldsymbol{r} \right), L^{\alpha}\left(\boldsymbol{r} \right)$ which is the $\alpha$-component of spin and orbital angular moment of at position $\boldsymbol{r}$ respectively. For the finite size, $\boldsymbol{r}$ will become the index $r$ for the unit cells along this dimension which is denoted by y-direction in our consideration. By using the Fourier transformation, one obtains:

\begin{align}
    &\mathcal{M}^{\alpha}\left(\boldsymbol{r} \right) = \frac{1}{2}\sum_{\boldsymbol{k}} \Psi_{\boldsymbol{k}}^{\dagger} M^{\alpha}_{r}\Psi_{\boldsymbol{k}} \\
    &\boldsymbol{J}^{Q} = \frac{1}{4}\int d\boldsymbol{r} \Psi_{\boldsymbol{k}}^{\dagger} \left(\hat{H}_{\boldsymbol{k}}\boldsymbol{\sigma}_{3}\hat{v}_{\boldsymbol{k}}+\hat{v}_{\boldsymbol{k}}\boldsymbol{\sigma}_{3} \hat{H}_{\boldsymbol{k}}\right)\Psi_{\boldsymbol{k}}
\end{align}
where $M^{\alpha}_{r}$ represents the spin and orbital moment density operator of the rth strip. 

In the same maner as the orbital moment current, after evaluating the Kubo terms, one obtains:
\begin{equation}
    \left\langle \delta \mathcal{M}^{\alpha}\left( \boldsymbol{r} \right)\right\rangle^{in} = \kappa_{\nu}^{in}\left(r \right) \nabla_{\nu} T
\end{equation}
where
\begin{align} \label{OAMAin}
    \kappa_{\nu}^{in}\left(r \right) = -\frac{i}{2T}\sum_{\boldsymbol{k},m\neq n}&\langle n\left(\boldsymbol{k} \right) \vert M^{\alpha}_{r}\vert m\left(\boldsymbol{k} \right) \rangle \langle m \left(\boldsymbol{k} \right)\vert \hat{v}_{\boldsymbol{k},\nu}\vert n\left(\boldsymbol{k} \right)\rangle \times \notag \\
    & \times \frac{\left(E_{\boldsymbol{k}}^{nn} \sigma_{3}^{mm} + E_{\boldsymbol{k}}^{mm}\sigma_{3}^{nn} \right)\left\lbrace\rho\left[ \left( \boldsymbol{\sigma}_{3}E_{\boldsymbol{k}}\right)_{mm}\right]-\rho\left[ \left( \boldsymbol{\sigma}_{3}E_{\boldsymbol{k}}\right)_{nn}\right]\right\rbrace}{\left[\left(\boldsymbol{\sigma}_{3}E_{\boldsymbol{k}}\right)_{mm}-\left(\boldsymbol{\sigma}_{3}E_{\boldsymbol{k}}\right)_{nn} \right]^{2}}
\end{align}
where the superscript 'in' denotes the intrinsic contribution.

\subsection{Extrinsic contribution: Boltzmann equation}

We will now assess the extrinsic contribution to the spin and orbital moment density using the Boltzmann equation. Under the relaxation time approximation, denoted by $\tau_{0}$ the Boltzmann equation is expressed as follows:

\begin{equation}
\frac{\rho_{neq}-\rho_{eq}}{\tau_{0}} = -v_{\nu}\nabla_{\nu}T \frac{\partial \rho_{eq}}{\partial T} =   -v_{\nu}\nabla_{\nu}T \frac{E}{k_{B} T^{2}} \frac{e^{E/k_{B}T}}{\left( e^{E/k_{B}T} -1\right)^{2}}  
\end{equation}
where  $\rho_{eq} = \left( e^{E/k_{B}T} -1\right)^{-1}$.

The spin and orbital moment density is given by

\begin{align}
    \left\langle \delta \mathcal{M}^{\alpha}\left( \boldsymbol{r} \right)\right\rangle^{ex} &= \frac{1}{V} \mathcal{M}^{\alpha}\left(\boldsymbol{r} \right) \left( \rho_{neq}-\rho_{eq}\right) = -\frac{1}{V}\mathcal{M}^{\alpha}\left(\boldsymbol{r} \right) v_{\nu} \frac{\tau_{0}E}{k_{B} T^{2}} \frac{e^{E/k_{B}T}}{\left( e^{E/k_{B}T} -1\right)^{2}}\nabla_{\nu}T \\
    &= \kappa_{\nu}^{ex}\left(r \right) \nabla_{\nu} T  
\end{align}
where $V$ is the volume of the system. 

By using the Fourier transformation, one obtains:
\begin{align}\label{OAMAex}
    \kappa_{\nu}^{ex}\left(r \right) = -\frac{\tau_{0}}{2Vk_{B} T^{2}}\sum_{\boldsymbol{k}}\sum_{n=1}^{2N}\langle n\left(\boldsymbol{k} \right) \vert M^{\alpha}_{r}\vert n\left(\boldsymbol{k} \right) \rangle \langle n \left(\boldsymbol{k} \right)\vert \hat{v}_{\boldsymbol{k},\nu}\vert n\left(\boldsymbol{k} \right)\rangle  \frac{\left(\boldsymbol{\sigma}_{3}E_{\boldsymbol{k}}\right)_{nn}e^{\left(\boldsymbol{\sigma}_{3}E_{\boldsymbol{k}}\right)_{nn}/k_{B}T}}{\left( e^{\left(\boldsymbol{\sigma}_{3}E_{\boldsymbol{k}}\right)_{nn}/k_{B}T} -1\right)^{2}}
\end{align}

Equations \ref{OAMAin} and \ref{OAMAex} establish a basic framework for calculating the accumulation of angular moment (both spin and orbital) of magnons at the boundary of a finite system, incorporating intrinsic and extrinsic contributions. An investigation of a finite system, however is beyond the scope of the present paper, and we leave it for future exploration. 
%%%%%%%%%%%%%%%%%%%%%%%%%%%%%%%%%%%%%%%%%%%%%%%%%%%%%%%%%%%%%%%

\section{Magnon-magnon interaction effects}\label{magnon-magnon-interaction}

The topological properties of the magnon bands hosted by 2D honeycomb AFMs, for example Berry curvature and Chern number, are introduced in close analogy with electronic bands in the celebrated Haldane model~\cite{Haldane1988}. In both cases, one considers noninteracting (i.e., infinitely long-lived) quasiparticles as described by the magnon Hamiltonian or Haldane Hamiltonian containing only terms that are quadratic in bosonic or fermionic operators, respectively. Such a quadratic Hamiltonian for magnons is generated by a linearized (i.e., truncated~\cite{Bajpai2021}) HP transformation. We employ the same tranformation [Eqs.~\eqref{eq:hp1} and \eqref{eq:hp2}] with the belief that its success in describing ferromagnets at low temperatures will translate to honeycomb AFMs---that is, conclusions made about the topology of noninteracting magnons produced by the linearized HP transformation can carry over to the full spin Hamiltonian [Eq.~\eqref{Hamilt}]. 

\begin{figure}[h]
    \centering
    \includegraphics[width= 0.4\textwidth]{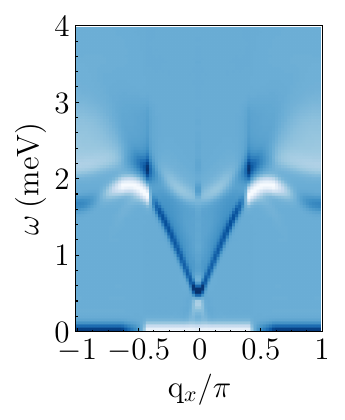}
    \caption{Spin structure factor for a brick ladder, corresponding to a single row of hexagons of the honeycomb lattice,  which hosts quantum spins $S=1/2$ interacting via nearest-neighbor antiferromagnetic exchange interaction and next-nearest-neighbor DMI with $\mathbf{D}_{ij}$ vector in Eq.~\eqref{Hamilt} that is {\em not} parallel to the staggered magnetization.}
    \label{fig:spinstruct}
\end{figure}

In this section we consider the possible consequences of magnon-magnon interactions. We stress that in the work reported in the main manuscript and previous sections of the supplement we consider only DMI oriented strictly out-of-plane, which means that it cannot introduce maganon-magnon interactions. Thus in this section we are considering a more general case in which the DMI has an in-plane component in the full spin Hamiltonian that can lead to important magnon-magnon interaction effects, even at zero temperature. The importance of such effects was revealed by recent perturbative analyses~\cite{Chernyshev2016,Habel2024} of the resulting nonquadratic terms in the bosonic Hamiltonian (e.g.~those containing three bosonic operators). For example, such an effect can lead to~\cite{Habel2024}: spontaneous decay~\cite{Zhitomirsky2013} of edge magnon modes; their hybridization with bulk magnons, thereby delocalizing them away from the edge; and even hybridization of magnons at opposite edges, as mediated by bulk magnons. The three magnon interactions can also lift spurious symmetries~\cite{Gohlke2023} present in the quadratic Hamiltonian, thereby making possible a nonzero transport response~\cite{Hoyer2024} to applied thermal gradient that is absent when one considers only noninteracting magnons. 

To illustrate the potential importance of magnon-magnon interactions, assume that the $\mathbf{D}_{\mathrm{ij}}$ vector has a component in the $x$-direction, which means that the corresponding DMI contains a term 
\begin{equation}
    \sum_{\langle\langle i,j \rangle\rangle}\frac{D_{ij}^x}{2} (\hat{S}_i^++\hat{S}_i^-)\hat{S}_j^z-\hat{S}_i^z(\hat{S}_j^++\hat{S}_j^-).
\end{equation}
The presence of this term in the full spin Hamiltonian [Eq.~\eqref{Hamilt}] generates terms containing three bosonic operators when the square root of the HP transformation is expanded~\cite{Bajpai2021} beyond the lowest order used in Eqs.~\eqref{eq:hp1} and \eqref{eq:hp2}. To understand the importance of these effects, here we employ {\em nonperturbative} calculations for the full spin Hamiltonian using a time-dependent density matrix renormalization group algorithm for the spin structure factor~\cite{White2004}, which we implement via time-dependent variational principle~\cite{Haegeman2016,Chanda2020} calculations within the {\tt ITensor}~\cite{Fishman2022ITensor} package. For purposes of illustrating the effects of magnon-magnon interactions, we use a honeycomb lattice composed of only one row of hexagons hosting an antiferromagnetic nearest-neighbor exchange and next-nearest neighbor DMI. The spin structure factor for this system, which is plotted in Fig.~\ref{fig:spinstruct}, reveals that the degeneracy of the bands of the quadratic Hamiltonian obtained from the linearized HP transformation is lifted when the $\mathbf{D}_{\mathrm{ij}}$ vector has a component in the $x$-direction. In other words, the presence of maganon-magnon interactions breaks the degeneracy of the bands. We stress, however, that our analysis in the main text and other sections of the SI remains valid as long as the DMI vector remains parallel to the staggered magnetization $\mathbf{M}_{\mathrm{stag}}=\mathbf{M}_A-\mathbf{M}_B$.

%%%%%%%%%%%%%%%% SUPPLEMENTARY REFERENCES %%%%%%%%%%%%%%%

% Do NOT include a reference list in the supplement.
% All references must be in a single list at the end of the main text.
% The copyeditors will ensure that the correct reference list appears with each version of the paper
% (print, HTML, PDF, mobile app, metadata for bibliographic databases etc.)

\end{document}